\documentclass[a4paper,12pt,reqno]{amsart}
\pdfoutput=1

\usepackage{a4}
\usepackage{pdflscape}
\usepackage{amsthm}\usepackage{simplewick}
\usepackage[T1]{fontenc}
\usepackage[utf8]{inputenc}
\usepackage[british]{babel}
\usepackage{bbm}
\usepackage{todonotes}
\usepackage{xcolor}
\usepackage[unicode,psdextra]{hyperref}
\usepackage{tikz}
\usepackage{scalerel}
\usetikzlibrary{decorations.pathmorphing,shapes}
\usetikzlibrary{arrows.meta,calc}
\def\centerarc[#1](#2)(#3:#4:#5)% Syntax: [draw options] (center) (initial angle:final angle:radius)
    { \draw[#1] ($(#2)+({#5*cos(#3)},{#5*sin(#3)})$) arc (#3:#4:#5); }

\newcommand{\Z}{\mathcal{Z}}
\newcommand{\T}{\mathcal{T}}
\newcommand{\J}{\mathcal{J}}

\newtheorem{definition}{Definition}
\newtheorem{example}[definition]{Example}
\newtheorem{theorem}[definition]{Theorem}
\newtheorem{lemma}[definition]{Lemma}
\newtheorem{remark}[definition]{Remark}

\newtheorem{corollary}[definition]{Corollary}
\newtheorem{proposition}[definition]{Proposition}

\DeclareMathOperator{\Res}{Res}

\makeatletter
\@addtoreset{definition}{section}
\@addtoreset{equation}{section}
\makeatother

\makeatletter
\let\@wraptoccontribs\wraptoccontribs
\makeatother

\allowdisplaybreaks[4]

\begin{document}

\title[Complete solution of the LSZ Model via Topological Recursion]{Complete solution of the LSZ Model via Topological Recursion}

\author[J. Branahl]{Johannes Branahl\textsuperscript{1}}
\author[A. Hock]{Alexander Hock\textsuperscript{2} \href{https://orcid.org/
		0000-0002-8404-4056}{\scaleto{\includegraphics{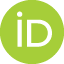}}{10pt}}}
 
\address{\textsuperscript{1}Mathematisches Institut der
  Westfälischen Wilhelms-Universit\"at \hfill \newline
Einsteinstr.\ 62, 48149 M\"unster, Germany \hfill \newline
{\itshape e-mail:} \normalfont
\texttt{j\_bran33@uni-muenster.de}}

\address{\textsuperscript{2}Mathematical Institute, University of Oxford,
  \newline
  Andrew Wiles Building, Woodstock Road, OX2 6GG, Oxford, United Kingdom
  \newline
  {\itshape e-mail:} \normalfont \texttt{alexander.hock@maths.ox.ac.uk}}

\begin{abstract}
We prove that the Langmann-Szabo-Zarembo (LSZ) model with quartic potential, a toy model for a quantum field theory on noncommutative spaces grasped as a complex matrix model, obeys topological recursion of Chekhov, Eynard and Orantin. By introducing two families of correlation functions, one corresponding to the meromorphic differentials $\omega_{g,n}$ of topological recursion, we obtain Dyson-Schwinger equations that eventually lead to the abstract loop equations being, together with their pole structure, the necessary condition for topological recursion. This strategy to show the exact solvability of the LSZ model establishes another approach towards the exceptional property of integrability in some quantum field theories. We compare differences in the loop equations for the LSZ model (with complex fields) and the Grosse-Wulkenhaar model (with hermitian fieldss) and their consequences for the resulting particular type of topological recursion that governs the models.
\end{abstract}


\maketitle
\markboth{\hfill\textsc\shortauthors}{\textsc{{Complete solution of the LSZ Model via Topological Recursion}\hfill}}

\section{Introduction and Main Result}
The property of exact solvability or even integrability of a non-trivial quantum field theory (QFT) in four dimensions is a very active and ongoing discipline in theoretical and mathematical physics. Usually, QFT's turned out to be accessible only by approximate methods, this is why one paid attention to simplified models. An established approach being of interest for us is the Euclidean one \cite{Schwinger:1959} having a proximity to statistical physics, giving the chance of exchanging methods and concepts of both disciplines. New hope arose from the formulation of such models on a suitable noncommutative space. In particular, the fact that this novel geometry developed by A. Connes \cite{Connes94noncommutativegeometry} also bridged the significant gap of how general relativity can be unified with the fundamental laws of quantum mechanics \cite{Doplicher:1994tu} fuelled the debate around this type of toy models. At the turn of the millennium, two models celebrated noteworthy successes towards exact solvability and renormalisability in four dimensions. The quartic model of hermitian fields by Grosse and Wulkenhaar \cite{Grosse:2004yu,Grosse:2012uv} and the model of complex-valued fields with arbitrary interaction by Langmann, Szabo and Zarembo \cite{Langmann:2003if}, both formulated on the noncommutative Moyal space \cite{rieffel1989}. An important fact is that quantum fields on the Moyal space admit an expansion in a matrix base (Moyal algebra).

A suitable approach to the property of exact solvability was given in each case by the finite-dimensional approximation of the quantum fields by matrices. In this finite limit at the so-called self-duality point of the action functional \cite{Langmann:2002cc}, the quantum field theories are reduced to hermitian or complex matrix models with an external field (representing the covariance), an area of research that has its own powerful approaches to solving. This reduction process to an ordinary matrix model starting from an abstract noncommutative algebra can be retraced in great detail in e.g. \cite{SurveyNCG} which is also accessible for beginners in this area of research. By exact solvability we mean the expressibility, at least in principle, of the model's correlation functions by "known" functions. With new methods, it was recently possible to replace the principle expressibility for the hermitian model with concrete expressions that satisfy a remarkable algebraic structure whose investigation is still an ongoing process \cite{Grosse:2019jnv,Branahl:2020yru,Hock:2021tbl}. The key to this is the universal structure of topological recursion (TR) of Chekhov, Eynard and Orantin \cite{Chekhov:2006vd,Eynard:2007kz}.  

In a similar way to the Grosse-Wulkenhaar model, which shrinks in the finite limit to a quartic analogue of the Kontsevich model (QKM) \cite{Kontsevich:1992ti}, we turn to the complex-valued LSZ model, restricted to a quartic interaction. Due to its proximity to the action functional of the Higgs field and the possibility of treating charged quantum fields, this model is quite a bit closer to realistic physical QFTs. Although the complex nature of the fields imposes a more general structure on the partition function, the character of the loop equations for the correlation functions as well as the algebraic structure behind their solution will turn out to be much simpler. In the original work by Langmann, Szabo and Zarembo \cite{Langmann:2003if}, it was proved to be useful to extend the partition function with a second external field. With this following expression of the \textit{partition function}, a more concrete reformulation of exact solvability as well as a more precise notion of integrability based on topological recursion will be possible in this article. The partition function of the LSZ model reads:
\begin{align}
\label{partfunLSZ}
&\mathcal{Z}(J,J^\dagger)\\
&= \int_{C_N} d \Phi d \Phi^\dagger \exp \Big (- N \mathrm{Tr}\bigg [E  \Phi\Phi^\dagger+\tilde{E} \Phi^{\dagger}\Phi + \frac{\lambda}{2} (\Phi^\dagger \Phi)^2\bigg]+N\mathrm{Tr} (\Phi^\dagger J + J^\dagger \Phi) \Big ) \nonumber
\end{align}
Here, the integral is carried out over the space $C_N$ of complex $N \times N$ matrices. $E$ and $\tilde{E}$ are distinct $N \times N$ hermitian matrices with positive eigenvalues that we call \textit{external fields} responsible for the covariance. Their positive and distinct eigenvalues $(E_1,...,E_N)$ and $(\tilde{E}_1,...,\tilde{E}_N)$ are not related. Without loss of generality (see proof of Proposition \ref{prop:ward}), we can assume that both matrices are diagonal due to the invariance of $\mathcal{Z}$ under unitary tranformations. The quantum fields themselves are given by $\Phi$, their interaction strength is encoded in the scalar $\lambda$, the \textit{coupling constant}. We also need to introduce besides the \textit{source} $J\in C_N$ together with its adjoint $J^\dagger$ building together the \textit{source term} that will give rise to the correlation functions by derivatives of $\mathcal{Z}(J,J^\dagger)$ at $J,J^\dagger=0$. 

Note that Langmann, Szabo and Zarembo were in \cite{Langmann:2003if} in particular interested in the case $\tilde{E}=0$, but the same strong properties of exact solvability and integrability (to be precise: integrable in the Kadomtsev-Petviashvili hierarchy) were shown first for this regulated partition function \eqref{partfunLSZ} with $\tilde{E}\neq 0$. Some correlation functions in the case $\tilde{E}=0$ were explicitly solved by the authors, we generalise it to any $\tilde{E}$ and all correlation functions. This extension to any $\tilde{E}$ also became interesting for other applications: In \cite{Santilli:2018} it was shown that the LSZ model computes the probability of atypically large fluctuations in the Stieltjes–Wigert matrix model, which is a $q$-ensemble describing $\mathcal{U}(N)$ Chern–Simons theory on the three-sphere. In their work it is also stated that despite similarities between the LSZ model and the QKM, it \textit{leads to a different analysis and very different set of results}. Using topological recursion, we will show that the two models are indeed deeply related and can be treated with the same techniques. More concretely, the LSZ model is solved by TR and the QKM by blobbed topological recursion (BTR), at least in genus $g=0$, with the same spectral curve $(\overline{\mathbb{C}},x,y,B)$ for $E=\tilde{E}$ \cite{Hock:2021tbl}.

To understand this in detail, we need to recapitulate the framework of topological recursion. This mathematical machinery is \textit{topological} in the sense that it computes an infinite sequence of symmetric meromorphic $n$-forms $\omega_{g,n}$ on $\Sigma^n$, where $g$ can be understood as genus and $n$ to the number of marked points. $\Sigma$ is a Riemann surface independent of $g$ and $n$. It is \textit{recursive} in the sense that the computation is performed recursively in the negative Euler characteristic $-\chi=2g+n-2$ and builds on a set of initial data $(\Sigma,x,y,B)$, where $x,y:\Sigma\to \Sigma_0$ are ramified covering of Riemann surfaces, $\omega_{0,1}=y\, dx$ is a meromorphic differential 1-form on $\Sigma$, regular at the ramification points of $x$, and $\omega_{0,2}=B$ a symmetric bilinear differential form on $\Sigma\times \Sigma$ with double pole on the diagonal and no residue. 

For particular spectral curves, the family of differential forms $\omega_{g,n}$ is directly related for instance to correlation functions (expectation values) of some statistical models. Therefore, the spectral curve recursively encodes the whole information about the model by itself. In the following section we will introduce some mathematical details behind the recursion and a visualisation. 

After its discovery 15 years ago it enabled great success in various, seemingly disconnected fields of mathematical research - as enumerative geometry \cite{Eynard:2016yaa}, intersection numbers of tautological classes of the moduli space of complex curves \cite{Witten:1990hr,Eynard:2011} and integrable hierarchies \cite{Eynard:2017}, and much more. Today, it is almost impossible to give a complete list of applications of this powerful universal procedure. Topological recursion entered the stage of (noncommutative) quantum field theory a few years ago during the attempt to understand the solution structure of the correlation functions of the QKM, formulated by the partition function:
\begin{align}
\label{partfunQKM}
\mathcal{Z}^{(QKM)}(J)= \int_{H_N} d \Phi \exp \bigg (- N \mathrm{Tr}\bigg [E \Phi^2+ \frac{\lambda}{4} \Phi^4\bigg]+N\mathrm{Tr} (\Phi J ) \bigg )
\end{align}
Here, we integrate over the space $H_N$ of hermitian $N \times N$ matrices where $E \in H_N$ has positive eigenvalues $(E_1,...,E_N)$. The exact and concrete solution of the 2-point function of \eqref{partfunQKM}, obtained after complexification and a crucial variable transform \cite{Schurmann:2019mzu,Grosse:2019jnv,Panzer:2018tvy}, gave the possibility to read off the spectral curve for topological recursion \cite{Branahl:2020yru}. For the complex LSZ model, now with two external fields, we will follow a very similar strategy. With established methods as a system of Ward identities (Sec. \ref{ch:setup}) we derive a closed nonlinear equation for the 2-point function of the LSZ model in Sec. \ref{ch:dse}, which is strikingly similar to the QKM \eqref{partfunQKM} in the planar sector. Two variable transforms, occurring in the proof of its solution, yield the corresponding spectral curve. The introduction of a boundary creation operator, creating two families $T$ and $\Omega$ of generalised correlators only building on the knowledge of the 2-point function, is the key step towards topological recursion. Their Dyson-Schwinger equations have exactly the same structure as the loop equations of the hermitian 2-matrix model \cite{Chekhov:2006vd} and the generalised Kontsevich model \cite{Belliard:2021jtj} such that a proof in Sec. \ref{ch:tr} is straightforward. In summary, we will prove the following result:
\begin{theorem}
\label{th:main}
Let  $(\overline{\mathbb{C}},x,y,B= \frac{dz_1\, dz_2}{(z_1-z_2)^2})$ be a spectral curve  with
\begin{align*}
 x(z) = z-\frac{\lambda}{N} \sum_{k=1}^N\frac{1}{y'(\tilde{\varepsilon}_k)(z-\tilde{\varepsilon}_k)} \qquad y(z) =- z + \frac{\lambda}{N}  \sum_{k=1}^{N} \frac{1}{x'(\varepsilon_k)(z-\varepsilon_k)}
\end{align*}
where  $x(\varepsilon_k)=E_k$ and $y(\tilde{\varepsilon}_k)=\tilde{E}_k$. Then, topological recursion generates meromorphic forms $\omega_{g,n}(z_1,...,z_n)$ according to eq. (\ref{eq:TR-intro}), that solve the Langmann-Szabo-Zarembo model for quartic field interaction. In particular, $\omega_{g,n}$ relates to the family of correlators  $\Omega_n^{(g)}$ via $\Omega_n^{(g)}(z_1,...,z_n)dx(z_1)...dx(z_n)=\omega_{g,n}(z_1,...,z_n)$ where
\begin{align*}
 \Omega_n^{(g)}(\varepsilon_{p_1},...,\varepsilon_{p_n}) =  &\bigg [ N^{2-2g-n} \bigg ] (-1)^n \frac{\partial^n}{\partial E_{p_1}...\partial E_{p_n}} \log[\mathcal{Z}(0)] \\
 &+ \frac{\delta_{n,2}\delta_{g,0}}{(E_{p_1}-E_{p_2})^2}
 +\frac{\delta_{n,1}\delta_{g,0}}{\lambda}V'(E_{p_1})
\end{align*}
From the family $\Omega_n^{(g)}$ of meromorphic functions, the exact and explicit solution of all correlation functions $G$ of the LSZ model can be computed recursively via Corollary \ref{cor2+} and Theorem \ref{thm:complete}.
\end{theorem}
\begin{remark}
	The theorems and propositions throughout the article are slightly more general and hold even if we admit multiplicities for the eigenvalues $E_k,\tilde{E}_l$. To avoid technicalities, we stated Theorem \ref{th:main} in this simpler version.
\end{remark}

Interchanging the role of $E$ and $\tilde{E}$ in the LSZ model leads to an interchange between $x$ and $y$ in Theorem \ref{th:main}, which is in general a meaningful phenomenon in TR \cite{Borot:2021thu,Hock:2022wer}, but has in this specific model a trivial consequence (similar to the 2-matrix model).

In summary, topological recursion provides a direct access to the characterisation of the integrable hierarchy behind exact solvability, renewing the discussion of \cite{Langmann:2003if}. Moreover, topological recursion is somehow a necessary starting point towards the representation of all correlation functions in full generality that we derive in Sec. \ref{sec:compl}. We will also compare with results obtained in  \cite{Langmann:2003if} for $\tilde{E}=0$ and point out the aforementioned drastic simplifications that occur in comparison to the QKM in Sec. \ref{ch:cross}. The occurrence to $\omega_{g,n}$ of holomorphic additions in the solutions of QKM, which generate poles away from the ramification points will be explained by the nature of hermitian instead of complex fields. With this comes the necessity to make use of the extended framework of \textit{blobbed topological recursion}, initially developed in \cite{Borot:2015hna} and proved in the planar sector for the QKM in \cite{Hock:2021tbl}. In Sec. \ref{ch:cross}, we also explain TR in in the context of perturbation theory, in particular in the combinatorial limit in which all eigenvalues of the external fields coincide and providing generating functions for the enumeration of bipartite quadrangulations. We will clarify the connection to matrix models tailored for such enumerative problems. In the final discussion and outlook, Sec. \ref{ch:concl}, we give an overview of further connections and possible investigations such as the triviality problem, the integrable hierarchy behind the model and Hurwitz numbers. The technical proofs of the initial data of the recursion as well as of the extended Dyson-Schwinger equations for the correlation functions in full generality are shifted for readability to the appendix.

\section*{ Acknowledgements}
 JB is supported\footnote{``Funded by
  the Deutsche Forschungsgemeinschaft (DFG, German Research
  Foundation) -- Project-ID 427320536 -- SFB 1442, as well as under
  Germany's Excellence Strategy EXC 2044 390685587, Mathematics
  M\"unster: Dynamics -- Geometry -- Structure."} by the Cluster of
Excellence \emph{Mathematics M\"unster}. He would like to thank the University of Oxford for its hospitality. The work of JB at the University of Oxford was additionally financed by the \emph{RTG 2149 Strong and Weak
Interactions – from Hadrons to Dark Matter}. AH is supported by
the Walter-Benjamin fellowship\footnote{``Funded by
  the Deutsche Forschungsgemeinschaft (DFG, German Research
  Foundation) -- Project-ID 465029630}. The authors are grateful to Harald Grosse and Raimar Wulkenhaar, since many of the used techniques are results of our collaboration over the last years.
 
\newpage 

{\footnotesize\tableofcontents}

\section{The Setup}
\label{ch:setup}

Our path towards topological recursion of the LSZ model consists of establishing different families of correlators, becoming more and more simple in their algebraic structure. From the partition function, derivatives with respect to the source terms create the usually considered correlation functions $G_{...}$ from other common quantum field theoretical investigations. 
One remarkable message of this article will be the fact that although they appear as the most natural objects seen from the field theoretical, especially perturbative perspective, another family $\Omega_{...}$ is much more suitable to concretely formulate the exact solvability. 
In this manner, it is almost  impossible to  unravel with usual techniques the common correlation functions $G$, their recursive structure is fairly complicated even after applying TR.

In fact, only the 2-point function will be needed for our proof of TR. We derive its Dyson-Schwinger equation (DSE), which are equations between correlation functions, using a system of Ward identities that are able to decouple the 2-point function in the large $N$-limit from higher correlation function. 
 \newpage
\begin{figure}[h!]
	\centering
	\includegraphics[width= 0.9\textwidth]{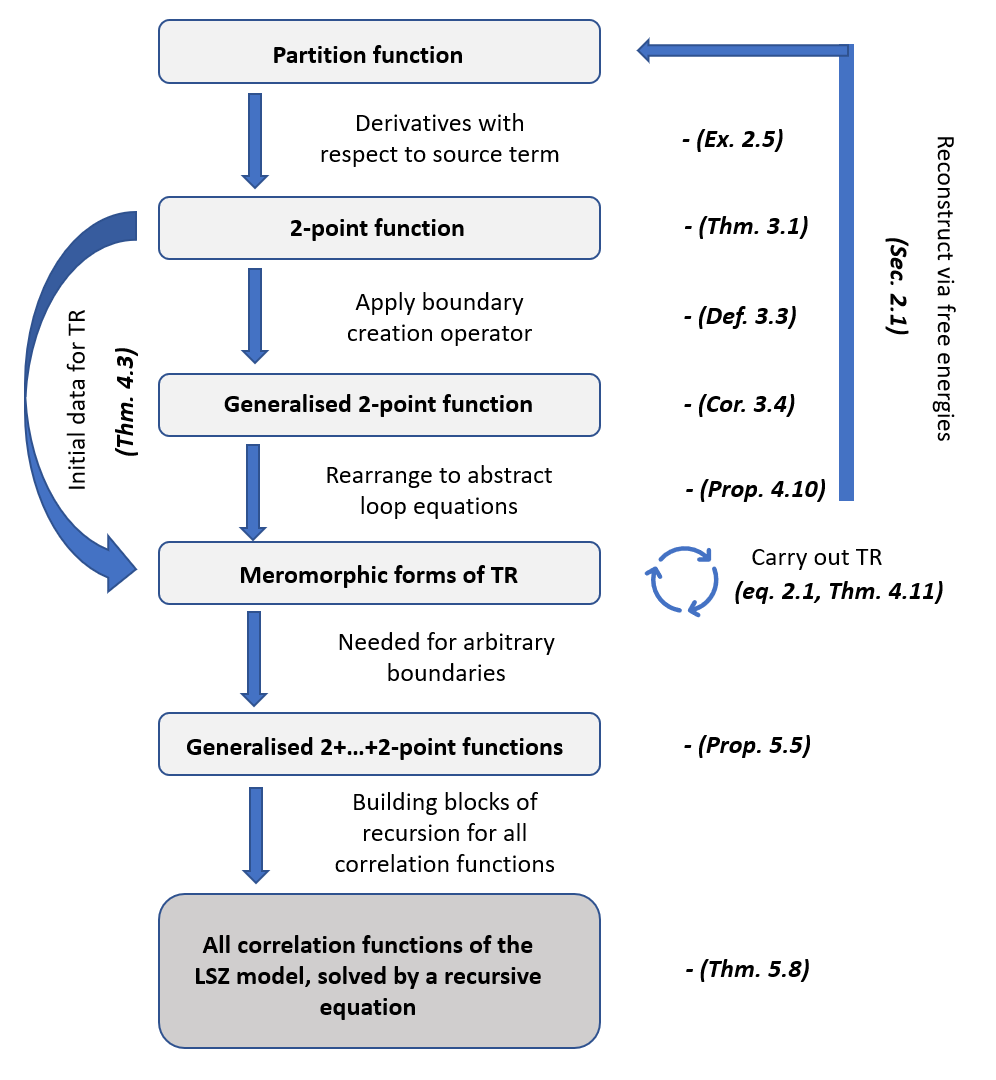} 
	\caption{A guide through the main steps in this article towards the complete solution of the LSZ model. In the sense of matrix models, we arrive there in the fourth step by reconstructing the partition function with the free energies of topological recursion. TR will be however only an intermediate step to formulate the complete solution in the quantum field theoretical sense: Exact solutions as rational functions of all arbitrarily complicated correlation functions.
		\label{fig:flow}}
\end{figure}

 However, this yields for the planar 2-point function to a closed nonlinear equation which is known to be solvable.  In this equation we recognise the appearance of derivatives with respect to the eigenvalues of one of the two external fields $E$, $\partial_{E_p}$ (first observed in \cite{Hock:2020rje}). They play the important role of boundary creation operators in TR, in principle enabling the transition from $\omega_{g,n}$ to $\omega_{g,n+1}$. As intermediate steps, we first need to introduce the family of generalised 2-point functions $T_{...}$ and their (finite) Hilbert transforms. Those are the objects to successfully carry out the final proof of TR via abstract loop equations. Afterwards, TR computes the complete family of $\Omega_{...}$ giving eventually rise to the solution of all generalised correlation functions and thus all objects of quantum field theoretical interest with arbitrarily complicated boundary structure and topology. All these information is visualised in Fig. \ref{fig:flow} also guiding through the main definitions and theorems. But now we will start from scratch in the following subsections, briefly introducing topological recursion and defining the fundamental objects of interest.

\subsection{Topological Recursion}
\begin{figure}[h!t]
\centering
\includegraphics[width= 1\textwidth]{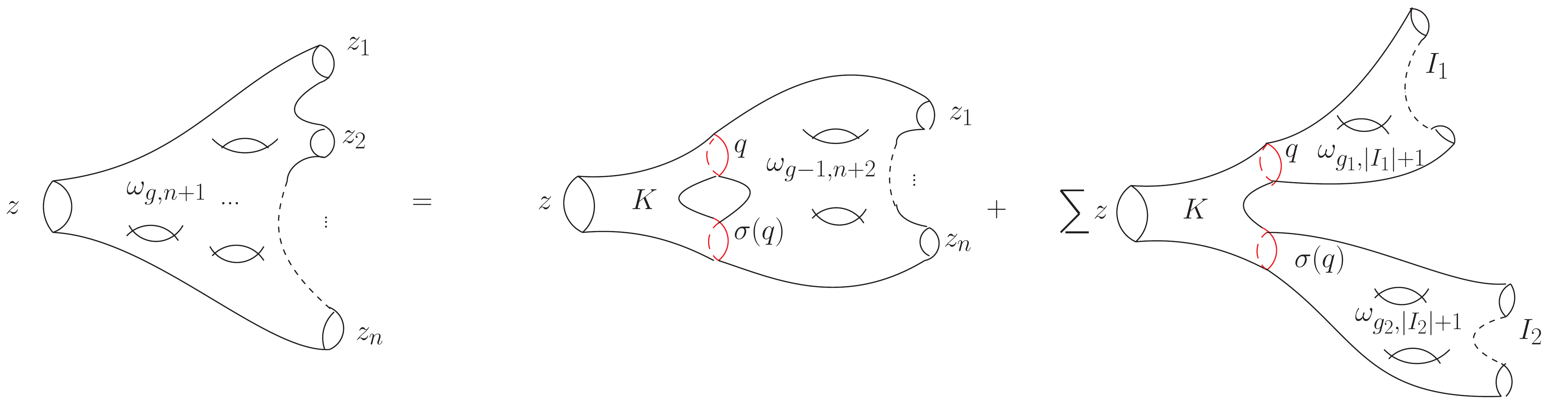} 
\caption{This is a graphical description of the topological recursion formula  eq. (\ref{eq:TR-intro}). There are two different ways to obtain the left hand side
  $\omega_{g,n}$ by gluing the recursion kernel $K$ (a pair of pants, built by the initial data) with something of lower topology: Either one glues
  one object with one genus less and one boundary more $(g-1,n+2)$ along
  two boundaries of the kernel creating the missing genus, or one glues
  two objects with the kernel along its boundaries (causing no genus change): Then one has to
  account any $(g_1,n_1),(g_2,n_2)$ conform with the left hand side --
  the sum over all possible partitions in the master formula
 eq. (\ref{eq:TR-intro}) arises.}
	\label{fig:tr}
\end{figure}

In the introductory section we have already mentioned the initial data $(\Sigma,x,y,B)$ of topological recursion. It contains a Riemann surface $\Sigma$ (which will be the Riemann sphere $\overline{\mathbb{C}}$ for the LSZ model) and the topologically unstable meromorphic forms $\omega_{0,1}(z)=y(z)\,dx(z)$ and $\omega_{0,2}(z_1,z_2)=B(z_1,z_2)=\frac{dz_1\, dz_2}{(z_1-z_2)^2}$ on $\overline{\mathbb{C}}$ and $\overline{\mathbb{C}} \times\overline{\mathbb{C}}$, respectively. They are unstable in the sense that the initial data contains the only $\omega_{g,n}$ with positive Euler characteristic $\chi=2-2g-n$. The recursion is then carried out with techniques of complex analysis \cite{Eynard:2007kz} and yields meromorphic forms $\omega_{g,n}$ on 
\begin{align}
  \label{eq:TR-intro}
&  \omega_{g,n+1}(I,z)
  \\
  & =\sum_{\beta_i}
  \Res\displaylimits_{q\to \beta_i}
  K_i(z,q)\bigg(
  \omega_{g-1,n+2}(I, q,\sigma_i(q))
  +\!\!\!\!\!\!\!\!
   \sum_{\substack{g_1+g_2=g\\ I_1\uplus I_2=I\\
            (g_i,I_i)\neq (0,\emptyset)}}\!\!\!\!\!
   \omega_{g_1,|I_1|+1}(I_1,q)
  \omega_{g_2,|I_2|+1}(I_2,\sigma_i(q))\!\bigg). \nonumber
\end{align}
The notation includes:
\begin{itemize}
\item  $I=\{z_1,\dots,z_n\}$ as a collection of $n$ variables $z_j$
\item  the ramification points
$\beta_i$ of $x$ defined by $dx(\beta_i)=0$
\item  the local Galois involution $\sigma_i\neq \mathrm{id}$ with
$x(q)=x(\sigma_i(q))$ defined in the vicinity of $\beta_i$ with fixed point $\beta_i$
\item the recursion kernel $K_i(z,q)$ is also locally 
defined in the vicinity of $\beta_i$ by
\begin{align*}
K_i(z,q)=\frac{\frac{1}{2}\int^{q}_{\sigma_i(q)}
  B(z,\bullet)}{\omega_{0,1}(q)-\omega_{0,1}(\sigma_i(q))}.
\end{align*}
\end{itemize}
It turns out that $\omega_{g,n}$ are symmetric in its variables and  for $\chi<0$ all
 $\omega_{g,n}$  have poles at the ramification points only, with vanishing residues. A pictorial access to this formula is given in Fig. \ref{fig:tr}.
Finally, we emphasise the existence of quantities belonging to the special case $n=0$. Instead of $\omega_{g,0}$, one usually denotes these quantities by $\mathcal{F}^{(g)}$, the so-called \textit{free energy} of genus $g$. They can be obtained as follows: Define the loop annihilation operator $\Phi(z)$ as a primitive $d_z \Phi(z)=\omega_{0,1}(z)$. Then, for negative Euler characteristic $\chi$, the \textit{Dilaton equation} holds:
\begin{align}
\label{dilaton2}
\sum_{\beta_i} \Res_{z \to \beta_i} \Phi(z)\omega_{g,1}(z)=(2-2g)\mathcal{F}^{(g)} \qquad g>1
\end{align}
The unstable free energies $\mathcal{F}^{(0)}$ and $\mathcal{F}^{(1)}$ obey completely different and much more complicated formulae which will not be of greater interest in this article, see  \cite{Eynard:2007kz}. Altogether, the formal series of $\mathcal{F}^{(g)}$ encodes the complete information of the model due to
\begin{align}
\log \mathcal{Z} = -\sum_{g=0}^\infty N^{2-2g} \mathcal{F}^{(g)}
\end{align}
where $\mathcal{Z}$ is the partition function. Knowing the spectral curve for the LSZ model, we would in principal have its complete solution by topological recursion since the partition function can completely reconstructed from the genus-summed free energies. As announced, the overall goal will be the complete solution in the quantum field theoretical sence (regarding all correlation functions $G$ defined in Defintion \ref{def:FullExpectation} together with \eqref{defG}), where however TR will be the starting point. There is a clear strategy how to reach this: Astonishingly, there is also a universal way to prove that a model is governed by TR (or its extensions). Two relatively mild conditions for the $\omega_{g,n}$ must be fulfilled - they were worked out in \cite{Borot:2013lpa} and read concretely:
\begin{definition}[Abstract loop equations]
\label{def:absloopeq}
 A family of meromorphic differential forms $\omega_{g,n}$ on $\Sigma^n$,
with $g \geq 0$ and $n >0$, fulfils the \textbf{linear loop equation}
if
\begin{align}
\omega_{g,n+1}(u_1,...,u_n,z)+
\omega_{g,n+1}(u_1,...,u_n,\sigma_i(z))=
\mathcal{O}(z-\beta_i)dz
\end{align}
is a holomorphic linear form in $z$ with (at least) a simple zero
at $z \to \beta_i$. The family of $\omega_{g,n}$ fulfils the
\textbf{quadratic loop equation} if
\begin{align}
&\omega_{g-1,n+2}(u_1,...,u_n,z,\sigma_i(z))+ \hspace*{-0.5cm}
\sum_{\substack{g_1+g_2=g \\ I_1\uplus I_2=\{u_1,...,u_n\}}}
\hspace*{-0.8cm}
   \omega_{g_1,|I_1|+1}(I_1,z)
    \omega_{g_2,|I_2|+1}(I_2,\sigma_i(z))
\nonumber \\ 
&=\mathcal{O}((z-\beta_i)^2)(dz)^2
\label{qle}
\end{align}
is a holomorphic quadratic form in $z$ with at least a double zero at $z \to \beta_i$.
\end{definition}
An important subclass of solutions satisfying the abstract loop equations is given by differentials governed by TR
\cite{Borot:2013lpa}, whereas the entirety of solutions is provided
by the so-called blobbed topological recursion that will become of interest when we compare the LSZ model with the case of hermitian matrices. More concretely, we cite:
\begin{theorem}[\cite{Borot:2013lpa}]
\label{thm:ale} Let $\chi=2-2g-n<0$. 
If $\omega_{g,n}$ satisfy the linear and quadratic loop equations and their poles only occur at the ramification points $\beta_i$ of $x(z)$, they are uniquely given by topological recursion eq. (\ref{eq:TR-intro}).
\end{theorem}
It thus solely remains to manipulate the Dyson-Schwinger equations (or loop equations) of suitable quantities in the LSZ model such that it will be possible to show that the two abstract loop equations are fulfilled. To do that, we start with the definition of the natural correlators, seen from the QFT point of view.
 
\subsection{Correlation Functions}
First of all, we bring the partition function \eqref{partfunLSZ} into a convenient form, in order to carry out future calculations simpler. To do that, note that the \textit{fields} $\Phi$, $\Phi^\dagger$ can be related to derivatives of the corresponding adjoint source term: $\Phi^\dagger_{ab}\leftrightarrow \frac{1}{N} \frac{\partial}{\partial J_{ba}}$,  $\Phi_{ab} \leftrightarrow \frac{1}{N} \frac{\partial}{\partial J^\dagger_{ba}}$. With this duality, the \textit{interaction term} can be thus written in a twofold manner:
\begin{align*}
& S_{int}[ \Phi,\Phi^\dagger ] = N\frac{\lambda}{2} \sum_{n_i} \Phi_{n_1n_2}\Phi^\dagger_{n_2n_3}\Phi_{n_3n_4}\Phi^\dagger_{n_4n_1}\\
& S_{int} \bigg [ \frac{1}{N} \frac{\partial}{\partial J^\dagger}, \frac{1}{N} \frac{\partial}{\partial J} \bigg ] =  N\frac{\lambda}{2N^4} \sum_{n_i} \frac{\partial^4}{\partial J^\dagger_{n_2n_1} \partial J_{n_3n_2}\partial J^\dagger_{n_4n_3} \partial J_{n_1n_4}}
\end{align*}
such that it can be pulled out of the integral.
 By symmetrisation and a simple remaining Gaussian integration, which is allowed after the upper rewriting, we can then prove the following rearrangement of the partition function:
\begin{proposition}
\label{prop:partnew}
$\mathcal{Z}(J,J^\dagger)$ can be rewritten as follows:
\begin{align*}
&\mathcal{Z}(J,J^\dagger)=  C \exp \biggl ( - S_{int} \bigg [ \frac{1}{N} \frac{\partial}{\partial J^\dagger}, \frac{1}{N} \frac{\partial}{\partial J} \bigg ] \biggl ) \mathcal{Z}_{free}(J,J^\dagger)
\end{align*}
where
\begin{align*}
C=\prod_{m,n=1}^N \frac{\pi }{N(E_m+\tilde{E}_n)} \qquad \mathrm{and} \qquad  \mathcal{Z}_{free}(J,J^\dagger)=\exp \biggl ( N \sum_{m,n=1}^N \frac{J_{nm}J^\dagger_{mn}}{E_n+\tilde{E}_m} \biggl ) 
\end{align*}
\begin{proof}
	First, we symmetrise the kinetic term. The external fields can be diagonalised, $E= \delta_{mn}E_m$ and $\tilde{E}= \delta_{mn}\tilde{E}_m$, as the partition function is invariant under unitary transformations $U \in \mathcal{U}(N)$ (see later in the discussion of the Ward identity, Proposition \ref{prop:ward}). Therefore, the trace can be rewritten to
	\begin{align*}
		\mathrm{Tr}[E  \Phi\Phi^\dagger+\tilde{E}  \Phi^{\dagger}\Phi ] = \sum_{m,n=1}^N (E_n \Phi_{nm} \Phi^\dagger_{mn}+ \tilde{E}_n  \Phi^\dagger_{nm} \Phi_{mn}) =  \sum_{m,n=1}^N (E_n+\tilde{E}_m) \Phi_{nm} \Phi^\dagger_{mn}
	\end{align*}
	Next, we rewrite also the kinetic term combined with the source term:
	\begin{align*}
		& (E_n+\tilde{E}_m) \Phi_{nm} \Phi^\dagger_{mn} -J_{nm}\Phi^\dagger_{mn}-\Phi_{nm}J^\dagger_{mn}\\
		&=  (E_n+\tilde{E}_m)  \bigg ( \Phi_{nm}- \frac{J_{nm}}{ (E_n+\tilde{E}_m)} \bigg )\bigg (\Phi^\dagger_{mn}- \frac{J^\dagger_{mn}}{ (E_n+\tilde{E}_m)} \bigg) - \frac{J_{nm}J^\dagger_{mn}}{ (E_n+\tilde{E}_m)}
	\end{align*}
	Now, we use the invariance of the measure under the transformations $\Phi_{nm} \to \Phi_{nm}' =\Phi_{nm}- \frac{J_{nm}}{ (E_n+\tilde{E}_m)}$, implying $\Phi_{mn}^\dagger \to (\Phi_{mn}^\dagger)'=\Phi_{mn}^\dagger- \frac{J^\dagger_{mn}}{ (E_n+\tilde{E}_m)}$.
	Finally, one carries out the $2N^2$ Gaussian integrations.
\end{proof}
\end{proposition}

In order to define the correlation functions, let us first define the expectation value:
\begin{definition}\label{def:FullExpectation}
We define moments
\begin{align}\label{eq:FullExpectation}
 \langle \Phi^\dagger_{q_1p_1}\Phi_{p_2q_2}..\Phi_{p_Kq_K}\rangle :=\frac{\int d\Phi d\Phi^\dagger
 \, \Phi^\dagger_{q_1p_1}\Phi_{p_2q_2}..\Phi_{p_Kq_K} e^{-S[\Phi,\Phi^\dagger]}}{\int d\Phi d\Phi^\dagger
 \,  e^{-S[\Phi,\Phi^\dagger]}}.
\end{align}
where $K$ even and the action functional reads $S[\Phi,\Phi^\dagger]=- N \mathrm{Tr}\big [E \Phi \Phi^\dagger+\tilde{E} \Phi^{\dagger} \Phi+ \frac{\lambda}{2} (\Phi^\dagger \Phi)^2\big]$. The connected part of this expression is denoted by $ \langle \Phi^\dagger_{q_1p_1}\Phi_{p_2q_2}..\Phi_{p_Kq_K}\rangle_c$. The full expectation value is given in terms of the connected ones by
\begin{align*}
 \langle \Phi^\dagger_{q_1p_1}\Phi_{p_2q_2}..\Phi_{p_Kq_K}\rangle=\sum_{\text{partitions}}
 \langle \Phi^\dagger_{q_{i^1_1}p_{i^1_1}}..\Phi_{p_{i^1_{k^1}}q_{i^1_{k^1}}}\rangle_c
 ..\langle \Phi^\dagger_{q_{i^j_1}p_{i^j_1}}..\Phi_{p_{i^j_{k^j}}q_{{i^j_{k^j}}}}\rangle_c,
\end{align*}
where the sum over partitions is understood as a sum over all possible 
decompositions.
\end{definition}
\begin{remark}
	It is quite easy to see that if the number $K$ in Defintion \ref{def:FullExpectation} is odd or the number of $\Phi$ and $\Phi^\dagger$ is different that the expectation value vanishes.
\end{remark}
 As usual in quantum field theory and probability theory, the connected expectation value is equivalently obtained by 
derivatives with respect to to the source terms, applied on the logarithm of the partition function. In this article we will follow on this connected part of the expectation values
\begin{align*}
 \frac{1}{N^K}\frac{\partial^K}{\partial J_{p_1q_1}\partial J^\dagger_{q_2p_2}..\partial J^\dagger_{q_Kp_K}}\log \frac{\mathcal{Z}(J,J^\dagger)}{\mathcal{Z}(0)}\bigg\vert_{J,J^\dagger=0}=
 \langle \Phi^\dagger_{q_1p_1}\Phi_{p_2q_2}..\Phi_{p_Kq_K}\rangle_c.
\end{align*}
The indices $\{p_i,q_i\} \in \{1,N\}$ may build cycles, otherwise the expectation value vanishes (see \cite{Schurmann:2019mzu}). A correlation function $G$ built of $b$ cycles  of indices $\{p^\beta_{1},q_1^\beta,p^\beta_{2},q_2^\beta,...,p_{N_\beta}^\beta, q_{N_\beta}^\beta\}$ with $\beta\in \{1,...,b\}$  having each cycle length $2N_\beta$ will correspond to the following connected expectation value as follows:
\begin{align}\label{defG}
  G_{|p_1^1q_1^1...p_{N_1}^1q_{N_1}^1|...|p_1^bq_1^b...p_{N_b}^bq_{N_b}^b|}  :=N^{b-2} \frac{\partial^{2N_1+...+2N_b} \log [\mathcal{Z}(J,J^\dagger)]}{\partial J_{p_1^1q_1^1}...\partial J^\dagger_{q^1_{N_1}p^1_{1}}...\, \partial J_{p_1^bq_1^b}...\partial J^\dagger_{q^b_{N_b}p^b_{1}}} |_{J,J^\dagger =0}.
\end{align}
We call this type of correlation function a  $2N_1+...+2N_b$-\textit{point functions}. This will  be taken as definition for the correlation function which we are interested in from QFT perspective.
Thus, the logarithm of the partition function can be expressed formally in terms of correlation functions:
\begin{align}\label{loginG}
&  \log [\mathcal{Z}(J,J^\dagger)] \\\nonumber
&= \log[\mathcal{Z}(0)] +\sum_{b=1}^\infty \sum_{N_1,...,N_b=1}^\infty \sum_{p_1^1...q_{N_b}^b=1}^N \sum_{g=0}^\infty \frac{N^{2-b-2g}}{b!} G^{(g)}_{|p_1^1q_1^1...p_{N_1}^1q_{N_1}^1|...|p_1^bq_1^b...p_{N_b}^bq_{N_b}^b|} \prod_{\beta=1} ^b \frac{\mathbb{J}_\beta}{N_\beta}
\end{align}
with $\mathbb{J}_\beta= J_{p_1^\beta q_1^\beta}...J^\dagger_{q^\beta_{N_\beta}p^\beta_{1}}$. $J$ and $J^\dagger$ show always up alternating, giving rise to pairs $\{p_i^j,q_i^j\}$.  Furthermore, the common genus expansion $G = \sum_{g=0}^\infty N^{-2g} G^{(g)}$ is inserted. For the purpose of this article, it is sufficient for concrete calculations to focus on the 2-point function $G_{|pq|}$. 
\begin{example}
\label{ex:2p}
The naive derivatives with respect to the (adjoint) source terms gives, making use of Proposition \ref{prop:partnew}:
\begin{align}
G_{|pq|}& =\frac{1}{N} \frac{\partial^2 \log [\mathcal{Z}(J,J^\dagger)]}{\partial J_{pq} \partial J^\dagger_{qp}}|_{J,J^\dagger =0} \nonumber \\
&=  C \frac{\partial}{\partial J_{qp}^\dagger} \frac{1}{\mathcal{Z}(J,J^\dagger)} e^{-S_{int}[\partial_J,\partial_{J^\dagger}]} \frac{J^\dagger_{qp}}{E_p+\tilde{E}_q} \mathcal{Z}_{free}(J,J^\dagger)_{J,J^\dagger =0} \nonumber \\\nonumber
& = \frac{1}{E_p+\tilde{E}_q} \bigg [1-\frac{\partial}{\partial J_{qp}^\dagger} \frac{1}{\mathcal{Z}(J,J^\dagger)} \frac{\partial S_{int}(\partial_J,\partial_{J^\dagger})}{\partial \Phi_{qp}^\dagger}  \mathcal{Z}(J,J^\dagger) \bigg ]_{J,J^\dagger =0}\\\nonumber
&=\frac{1}{E_p+\tilde{E}_q} \bigg [1-\lambda \bigg(\frac{1}{N^2}\sum_{n,m=1}^NG_{|pnmq|}+G_{|pq|}\frac{1}{N}\sum_{n=1}^N(G_{|pn|}+G_{|nq|})\\
&\qquad\qquad\qquad +\frac{1}{N^3}\sum_{n=1}^N(G_{|pq|pn|}+G_{|pq|nq|})\bigg)\bigg]
\label{eq:2p}
\end{align}
where we used the general chain rule:
\begin{align*}
e^{f(\partial_x)}(x \cdot g(x)) = xe^{f(\partial_x)}g(x)+f'(\partial_x)e^{f(\partial_x)}g(x) \qquad e^{f(\partial_x)} := \sum_{k=0} ^\infty \frac{[f(\partial_x)]^k}{k!},
\end{align*}
where $f(x)$ denotes some polynomial in $x$, $g(x)$ is smooth. We have applied in the last step the expansion \eqref{loginG}. We observe that we need further relations between correlation functions to turn this ansatz into a closed Dyson-Schwinger equation, since the 2-point function depends on the 4-point function and on the $2+2$-point function which are a priori not known. In this way, Dyson-Schwinger equations are building a tower of coupled equations iteratively depending on higher topologies. 
\end{example}

\subsection{Ward Identities}
An exceptional phenomenon in the treated class of QFT toy models is the decoupling of the correlation function of interest from all higher topologies. It was already observed in the early 2000's that a system of \textit{Ward identities} offers the possibility to obtain closed Dyson-Schwinger equations. In this class of models the identities rely on the fact of unitary invariance of the partition function, first developed in \cite{Disertori:2006nq} for complex quantum fields, but one external field. This procedure can be adapted to the LSZ model, we are going to prove:
\begin{proposition}
\label{prop:ward}
Let $E_l\neq E_k$. 
The partition function of the LSZ model satisfies the following system of Ward-identities:
\begin{align*}
\sum_{n=1}^N\bigg ( \frac{E_k-E_l}{N} \frac{\partial^2}{\partial J_{ln} \partial J^\dagger_{nk}} - J^\dagger_{nl} \frac{\partial}{\partial J^\dagger_{nk}}+J_{kn} \frac{\partial}{\partial J_{ln}}\bigg )\mathcal{Z}(J,J^\dagger)=0
\end{align*}
\begin{proof}
	We transform the fields under $U\in \mathcal{U}(N)$ as $\Phi^{(U)} =U\Phi $ and thus $(\Phi^\dagger)^{(U)} =  \Phi^\dagger U^\dagger$. First of all, we recognise that there is a $U \in \mathcal{U}(N)$ such that $ (\Phi^\dagger)^{(U)} E (\Phi)^{(U)} $ diagonalises $E$, $E_{mn} = E_m\delta_{mn}$. \footnote{Introducing another unitary transformation $\Phi^{(U)} = \Phi U$ and thus $(\Phi^\dagger)^{(U)} = U^\dagger  \Phi^\dagger $ guarantees the claimed diagonalisability of $\tilde{E}$ in the same manner.} Let now an other $U=e^{i A}$ a unitary transform on the space of complex $N \times N$ matrices with $i A$ in the $\mathfrak{u}(N)$ algebra. Assume $A$ is a small deviation from the unit element such that we expand the expansion around the identity $U = \mathbbm{1} + i A + \mathcal{O}(A^2)$. Noticing  that the partition function is invariant under unitary transformations, we have the condition $\mathcal{Z}(\Phi^{(U)},(\Phi^\dagger)^{(U)})-\mathcal{Z}(\Phi,\Phi^\dagger)=0$.
	Insert the field transformations into the partition function and consider the small variation in $A$. The interaction term is untouched by $U$ and the source term behaves at order $A^1$ like $i\, \mathrm{Tr}(A(\Phi J^\dagger -J \Phi^\dagger ))$. We finally discuss the kinetic term. Despite the occurrence of an additional matrix $\tilde{E}$, the Ward identity does not lose its original shape of \cite{Disertori:2006nq}, as inserting the unitary transforms gives:
	\begin{align*}
		& \mathrm{Tr}(\Phi^\dagger E \Phi + \Phi \tilde{E} \Phi^\dagger) - \mathrm{Tr}(  (\Phi^\dagger)^{(U)} E\Phi^{(U)} + \Phi^{(U)} \tilde{E}(\Phi^\dagger)^{(U)}) \\
		& =  \mathrm{Tr}(\Phi^\dagger E \Phi) - \mathrm{Tr}( \Phi^\dagger U^\dagger E U \Phi  ) \\
		& =   \mathrm{Tr}(\Phi^\dagger E \Phi) - \mathrm{Tr}( \Phi^\dagger ( \mathbbm{1} - i A) E ( \mathbbm{1} + i A) \Phi  ) + \mathcal{O}(A^2) \\
		& = - i \mathrm{Tr}( A\Phi \Phi^\dagger E  - A E\Phi \Phi^\dagger)+ \mathcal{O}(A^2)
	\end{align*}
	Exploiting $U^\dagger U = \mathbbm{1}$, the additional external field $\tilde{E}$ cancels and does not show up in the identity. We also used invariance of the trace under cyclic permutations. Substituting $\Phi^\dagger_{ab}\leftrightarrow \frac{1}{N} \frac{\partial}{\partial J_{ba}}$,  $\Phi_{ab} \leftrightarrow \frac{1}{N} \frac{\partial}{\partial J^\dagger_{ba}}$ and taking the derivative $\frac{\partial}{\partial A_{lk}}$ at $A=0$ finishes the proof.
	
	A second way to deduce the Ward identity is by partial integration similar to the derivation of the Dyson-Schwinger equation (see \cite{Hock:2018wup} for details).
\end{proof}
\end{proposition}
The most important tool to deduce closed Dyson-Schwinger equations shows up during the discussion of coinciding matrix indices. It is a fact that the cycle decomposition persists for such a coincidence, meaning that a 2+2-point function always stays topologically different from a 4-point function. As the correlation functions possess a perturbative expansion into rational functions of the eigenvalues $E_m,\tilde{E}_n$ being continuously differentiable, one may interpret the correlation function as $\mathcal{C}^1$-functions of all their indices. With this continuity argument, we prove:
\begin{proposition}
\label{prop:ward2}
For  distinct $E_m$, the partition function $\mathcal{Z}(J,J^\dagger)$ \eqref{partfunLSZ} of the LSZ model satisfies the system of Ward identities
 \begin{align}
\label{eq:Ward}
\sum_{n=1}^N \frac{\partial^2 \mathcal{Z}(J,J^\dagger)}{\partial J_{ln} \partial J^\dagger_{nk}} = \delta_{k,l} W_k[J,J^\dagger]\mathcal{Z}(J,J^\dagger) + \frac{N}{E_k-E_l}\sum_{n=1}^N \biggl ( J^\dagger_{nl} \frac{\partial}{\partial J^\dagger_{nk}}-J_{kn} \frac{\partial}{\partial J_{ln}}\bigg )\mathcal{Z}(J,J^\dagger)
 \end{align}
where
 \begin{align*}
  W_k[J,J^\dagger]:=&\sum_{b=1}^\infty \sum_{N_1,..,N_b=1}^\infty \sum_{p_1^1,q_1^1..,q_{N_b}^b=1}^N
 \sum_{g=0}^\infty \frac{N^{2-b-2g}}{b!}\prod_{\beta=1}^b\frac{\mathbb{J}_{\beta}}{N_\beta}\times\\
 &\bigg(  \frac{1}{N}\sum_{n=1}^N G^{(g)}_{|kn|
 p_1^1q_1^1...p_{N_1}^1q_{N_1}^1|...|p_1^bq_1^b...p_{N_b}^bq_{N_b}^b|}\\
 &+\sum_{M=1}^\infty\sum_{n,k_1,..,k_{2M-1}=1}^N G^{(g)}_{|knk_1..k_{2M-1}n|
p_1^1q_1^1...p_{N_1}^1q_{N_1}^1|...|p_1^bq_1^b...p_{N_b}^bq_{N_b}^b|}J_{nk_1}J^\dagger_{k_1k_2}..J^\dagger_{k_{2M-1}n}\bigg) 
 \end{align*}
and $\mathbb{J}_\beta= J_{p_1^\beta q_1^\beta}...J^\dagger_{q^\beta_{N_\beta}p^\beta_{1}}$.
\begin{proof}
	We rewrite the second derivative as
	\begin{align*}
		&\frac{\partial^2 \mathcal{Z}(J,J^\dagger)}{\partial J_{ln}\partial J^\dagger_{nk}} = \mathcal{Z}(J,J^\dagger) \bigg [ \frac{\partial^2 \log[\mathcal{Z}(J,J^\dagger)]}{\partial J_{ln}\partial J^\dagger_{nk}} +  \frac{\partial \log[\mathcal{Z}(J,J^\dagger)]}{\partial J^\dagger_{nk}} \frac{\partial \log[\mathcal{Z}(J,J^\dagger)]}{\partial J_{ln}} \bigg ]\\
		&=  (\delta_{kl} W_k[J,J^\dagger]+ W^{reg}_{k,l}[J,J^\dagger])\mathcal{Z}(J,J^\dagger) 
	\end{align*}
	Note that the product of logarithms will not contribute. $W^{reg}_{k,l}[J,J^\dagger]$ is regular and comes from the Ward identity of Proposition \ref{prop:ward}, which is safely divide by $\frac{E_k-E_l}{N}$ yielding
	\begin{align*}
		 W^{reg}_{k,l}[J,J^\dagger] \mathcal{Z}(J,J^\dagger) = \frac{N}{E_k-E_l}\sum_{n=1}^N \biggl ( J^\dagger_{nl} \frac{\partial}{\partial J^\dagger_{nk}}-J_{kn} \frac{\partial}{\partial J_{ln}}\bigg )\mathcal{Z}(J,J^\dagger).
	\end{align*}
	By l'Hospital's rule, the coincidence of indices $k \to l$ yields a continuous transition and a finite (regular) result.
	For the term proportional to $\delta_{k,l} $, we have to look at the action of
	 $\frac{\partial^2}{\partial J_{kn}\partial J^\dagger_{nk}}$ in a way which can not produced by the Ward identity. There are two cases, first $\frac{\partial^2}{\partial J_{kn}\partial J^\dagger_{nk}}$ can act on a cycle of length 2 and produces a closed cycle. The second possibility is a cycle of length $2M+2 \geq 4$, partially hit by the these derivatives. Expanding the $\log \mathcal{Z}$ as in \eqref{loginG} leads to all terms appearing in $W_k$.
\end{proof}
\end{proposition}

\section{Dyson-Schwinger Equations}
\label{ch:dse}
\subsection{The 2-Point Function}
With the results of the previous sections we are prepared to easily derive the closed Dyson-Schwinger equation for the 2-point function.
\begin{proposition}
\label{thm:2p}
The 2-point function of the LSZ model satisfies
 \begin{align*}
  G_{|pq|}&=
\frac{1}{E_p+\tilde{E}_q}-\frac{\lambda}{E_p+\tilde{E}_q} 
\bigg\{  G_{|pq|}\cdot \frac{1}{N}\sum_{n=1}^N G_{|pn|} 
 \\
&+\frac{1}{N^3}\sum_{n=1}^N G_{|pn|pq|}
+\frac{1}{N^2} G_{|pqpq|}
+\frac{1}{N}\sum_{n=1}^N\frac{G_{|pq|}- G_{|nq|}}{E_n-E_p} 
\bigg\}\;.
 \end{align*}
\begin{proof}
	Eq. \ref{eq:2p} turns with the upper Ward identity eq. \ref{eq:Ward} into
	\begin{align*}
		G_{|pq|}& =\frac{1}{E_p+\tilde{E}_q} \bigg [1- \frac{\lambda}{\mathcal{Z}(0)} \sum_{m,n=1}^N\frac{\partial^4}{\partial J_{pm}\partial J_{mn}^\dagger\partial J_{nq}\partial J_{qp}^\dagger}   \mathcal{Z}(J,J^\dagger) \bigg ]_{J,J^\dagger =0}\\
		& = \frac{1}{E_p+\tilde{E}_q} \bigg [1-\frac{\lambda}{N^3\mathcal{Z}(0)} \frac{\partial^2  W_p[J,J^\dagger] \mathcal{Z}(J,J^\dagger)}{\partial J_{pq} \partial J^\dagger_{qp}} \\
		&- \frac{\lambda}{N^2\mathcal{Z}(0)} \sum_{m,n=1}^N \frac{1}{E_n-E_p} \frac{\partial^2}{\partial J_{nq} \partial J^\dagger_{qp}}  \bigg (  J_{nm} \frac{\partial}{\partial J_{pm}}-J^\dagger_{mp} \frac{\partial}{\partial^\dagger J_{mn}} \bigg ) \mathcal{Z}(J,J^\dagger)\bigg ]_{J,J^\dagger =0}
	\end{align*}
	Now, all surviving correlation functions after setting $J,J^\dagger =0$ have to be identified. From $W_k$ we easily read off all the terms in Theorem \ref{thm:2p} except  $\frac{1}{N}\sum_{n=1}^N\frac{G_{|pq|}- G_{|nq|}}{E_n-E_p}$. This term comes from the last line if $J_{nm}$ is hit by $\frac{\partial}{\partial J_{nq}}$ fixing $m=q$, and if $J_{mp}^\dagger$ is hit by $\frac{\partial}{\partial J^\dagger_{qp}}$ fixing $m=p$.
\end{proof}
\end{proposition} 
A formal genus expansion $G = \sum_{g=0}^\infty N^{-2g} G^{(g)}$ yields at order $N^{-2g}$:
\begin{align}
\label{genusexp}
 G^{(g)}_{|pq|}&=
\frac{\delta_{0,g}}{E_p+\tilde{E}_q}-\frac{\lambda}{E_p+\tilde{E}_q} 
\bigg\{  \sum_{h=0}^{g} G^{(h)}_{|pq|}
\frac{1}{N}\sum_{n=1}^N G^{(g-h)}_{|pn|}
\\
&+\frac{1}{N}\sum_{n=1}^N G^{(g-1)}_{|pn|pq|}
+G^{(g-1)}_{|pqpq|}
+\frac{1}{N}\sum_{n=1}^N \frac{G^{(g)}_{|pq|}- G^{(g)}_{|nq|}}{E_n-E_p} 
 \bigg\}\nonumber.
\end{align}
 Note that in contrast to hermitian fields, we do not have symmetry under exchanging indices, $G_{|pq|}\neq G_{|qp|}$. For the planar 2-point function we thus have structurally the same nonlinear  equation as in the QKM with hermitian fields (compare with \cite[eq. (9)]{Schurmann:2019mzu}) except for $\tilde{E}_q$. We will see that also their exact solutions are very similar.

\subsection{Generalised Correlation Functions}
 So far, we carried out derivatives with respect to the source term. A crucial step towards TR, however, is keeping the eyes on derivatives with respect to the eigenvalues of the external field $E$ - for example:
\begin{align*}
&-N\frac{\partial}{\partial E_{p}} \int_{C_N}  d \Phi d \Phi^\dagger \exp \bigg (- N \mathrm{Tr}\bigg [E \Phi \Phi^\dagger+\tilde{E} \Phi^{\dagger} \Phi+ \frac{\lambda}{2} (\Phi^\dagger \Phi)^2 -  (\Phi^\dagger J+J^\dagger \Phi)\bigg]   \bigg )\bigg\vert_{J=0} \\
=&N^2\sum_{n=1}^N\int_{C_N}  d \Phi d \Phi^\dagger\,  \Phi_{pn}\Phi^\dagger_{np} \exp \bigg (- N \mathrm{Tr}\bigg [E \Phi \Phi^\dagger+\tilde{E} \Phi^{\dagger} \Phi+ \frac{\lambda}{2} (\Phi^\dagger \Phi)^2 -  (\Phi^\dagger J+J^\dagger \Phi)\bigg]   \bigg )\bigg\vert_{J=0}
\\
 =& \sum_n \frac{\partial^2}{\partial J_{pn}\partial J^\dagger_{np}}  \int_{C_N} d \Phi d \Phi^\dagger \exp \bigg (- N \mathrm{Tr}\bigg [E \Phi^\dagger \Phi+\tilde{E} \Phi \Phi^{\dagger}\\
&\qquad\qquad\qquad\qquad\qquad\qquad\qquad\qquad + \frac{\lambda}{2} (\Phi^\dagger \Phi)^2 -  (\Phi^\dagger J+J^\dagger \Phi) \bigg]    \bigg )\bigg\vert_{J=0} \\
=& \frac{1}{N} \sum_{n=1}^N G_{|pn|}=:\Omega_{p}
\end{align*}
With this simple calculation we defined the first member of the family $\Omega$, namely $\Omega_{p}$, via derivative with respect to $E_p$  that will be later related to topological recursion. In particular, we read off that in some sense the initial data of this TR must be encoded in the 2-point function. This appearance of derivatives also allows for a convenient new form of the Dyson-Schwinger equation for the 2-point function:
\begin{align}
\label{2pnew}
  G_{|pq|}&=
\frac{1}{E_p+\tilde{E}_q} 
\bigg (1-  \lambda G_{|pq|}\cdot \Omega_p
 +\frac{\lambda}{N} \frac{\partial G_{|pq|}}{\partial E_p}
-\frac{\lambda}{N}\sum_{\substack{{n=1}\\{n\neq p}}}^N\frac{G_{|pq|}- G_{|nq|}}{E_n-E_p} 
\bigg )
\end{align}
\begin{remark}
\label{rem:xy}
By exchanging the role of the two external fields $E, \tilde{E}$, it is possible to rewrite this equation in an equivalent form:
\begin{align}
\label{2pnew2}
  G_{|pq|}&=
\frac{1}{E_p+\tilde{E}_q} 
\bigg (1-  \lambda G_{|pq|}\cdot \frac{1}{N}\sum_{n=1}^NG_{|nq|}
 +\frac{\lambda}{N} \frac{\partial G_{|pq|}}{\partial \tilde{E}_q}
-\frac{\lambda}{N}\sum_{\substack{{n=1}\\{n\neq q}}}^N\frac{G_{|pq|}- G_{|pn|}}{\tilde{E}_n-\tilde{E}_q} 
\bigg )
\end{align}
\end{remark}
The repeated appearance of the derivative suggests to introduce the following objects:
\begin{definition}[Generalised correlation functions]
\label{def:gDSE}
For pairwise different $ E_{p_i}$ we define the \textit{generalised correlation functions} by repetitive action of the boundary creation operator $-N\frac{\partial}{\partial E_{p_i}}$:
\begin{align*}
  T_{p_1,...,p_m\|pq}
  &:=\frac{(-N)^m\partial^m}{\partial E_{p_1}...\partial 
    E_{p_m}}G_{|pq|} 
\end{align*}
In the same manner, we introduce a family $\Omega$ defined as
\begin{align*}
\Omega_{p_1,...,p_m} &:= 
\frac{(-N)^{m-1}\partial^{m-1}\Omega_{p_1}}{
\partial E_{p_2}...\partial 
E_{p_{m}}} +\frac{\delta_{m,2}}{(E_{p_1}-E_{p_2})^2}\;,\qquad
m\geq 2\;,
\end{align*}
where $ \Omega_{p_1}=\frac{1}{N} \sum_{k=1}^N G_{|p_1k|}$. 
\end{definition}
For these objects we will establish similar Dyson-Schwinger equations. For coinciding indices a derivative with respect to one of the indices has to be taken into account, e.g.
\begin{align}
	-N\frac{\partial}{\partial E_p}G_{pq}=T_{p\|pq}-N\lim_{n\to p}\frac{G_{nq}-G_{pq}}{E_n-E_q}.
\end{align}
The latter term is well-defined since, as argued earlier, any correlation function has a unique perturbative expansion in $\lambda$ which is a rational function in $E_i$ (and $\tilde{E}_i$) at each order in $\lambda$. Using the representation \eqref{2pnew} for the DSE of the 2-point function, it is straightforward to show:
\begin{corollary}
	Let $I=\{p_1,...,p_m\}$ be a index set and we abbreviate $T_{p_1,...,p_m\|pq}=T_{I\|pq}$ and $\Omega_{p_1,...,p_m}=\Omega_{I}$. Then the generalised correlation function $T_{I\|pq}$ and $\Omega_{I}$ satisfy the DSE's
\begin{align}
&\Big(\tilde{E}_q+E_p+\frac{\lambda}{N}
\sum_{\substack{l=1\\l\neq p}}^N  \frac{1}{E_l-E_p}\Big)T_{I\|pq}
- \frac{\lambda}{N}\sum_{\substack{l=1 \\ l\notin I,p}}^N
 \frac{T_{I\|lq}}{E_l-E_p}
\label{DSE-T2}
\\
&= \delta_{0,|I|}
-
\lambda\bigg\{
\sum_{I'\uplus I''=I}
\Omega_{I',p} T_{I''\|pq}
-\frac{1}{N}\frac{\partial T_{I\|pq}}{\partial E_p}
\nonumber
+ 
\sum_{j=1}^m \frac{\partial}{\partial E_{q_j}}
\Big(\frac{T_{I\setminus q_j\|q_jq}}{E_{q_j}-E_p}\Big)
\bigg\}\;.
\nonumber
\end{align}
and 
\begin{align}
&\Omega_{I,p}
= \frac{\delta_{|I|,1}}{
(E_{p_1}-E_p)^2}
+
\frac{1}{N}\sum_{k=1}^NT_{I\|pk}
\label{DSE-Om}
\end{align}
\begin{proof}
	Follows immediately by careful application of Definition \ref{def:gDSE} on eq. (\ref{2pnew}). It is in particular noteworthy that the second matrix index of the 2-point function is never hit by the action of the derivative $\partial_{E_{p_i}}$ - this is responsible for the major differences to the QKM (hermitian fields).
\end{proof}
\end{corollary}
As a final step, we have to compute one further DSE resulting from \eqref{DSE-T2} by multiplying with $\frac{\lambda}{N}\frac{1}{v-\tilde{E}_q}$ for some complex variable $v$ and summing over $q$, which yields
\begin{align}\nonumber
	&\big(v+V'(E_p)\big)\bigg(-\delta_{|I|,0}+\frac{\lambda}{N}\sum_q \frac{T_{I\|pq}}{v-\tilde{E}_q}\bigg)+\frac{\lambda^2}{N^2}\sum_q\sum_{n\notin I,p}\frac{T_{I\|nq}}{(E_p-E_n)(v-\tilde{E}_q)}\\\label{H2P}
	&=-\delta_{|I|,0}(\tilde{V}'(v)+V'(E_p))-\lambda \sum_{I_1\uplus I_2=I} \bigg(-\delta_{|I_1|,0}+\frac{\lambda}{N}\sum_q \frac{T_{I_1\|pq}}{v-\tilde{E}_q}\bigg) \Omega_{I_2, p}\\\nonumber
	&+\lambda \sum_{p_i\in I}\frac{\partial}{\partial E_{p_i}}\bigg(\frac{\frac{\lambda }{N}\sum_{q}\frac{T_{I\setminus p_i\| p_iq}}{v-\tilde{E}_q}}{E_p-E_{p_i}}\bigg)+\frac{\lambda }{N}\frac{\partial }{\partial E_p}\bigg(\frac{\lambda }{N}\sum_{q} \frac{T_{I\|pq}}{v-\tilde{E}_q}\bigg),
\end{align}
where $V'(E_p)=E_p-\frac{\lambda}{N}\sum_{n\neq p}\frac{1}{E_p-E_n}$ and $\tilde{V}'(x)=x-\frac{\lambda}{N}\sum_{k}\frac{1}{x-\tilde{E}_k}$.

\begin{remark}
	Notice that the DSE \eqref{DSE-T2} of $T_{I\|pq}$ and \eqref{H2P} of $\big(-\delta_{|I|,0}+\frac{\lambda}{N}\sum_q \frac{T_{I\|pq}}{v-\tilde{E}_q}\big)$ are of very similar form. However, by construction the solution of latter one is necessarily a rational function in $v$ with poles only at $\tilde{E_q}$, or equivalenty by multiply with $\prod_{q} (v-\tilde{E}_q)$ this would be a polynomial of degree $N$.
\end{remark}

In the next section, we will solve the DSE \eqref{H2P} and start with the first one $I=\emptyset$, which will provide the initial data. It  is  well-known that solutions of this type of models are achieved by complex analytic techniques. Therefore, we have to choose an analytic continuation of the correlations functions in the neighbourhood of $E_i$ (and $\tilde{E}_j$, respectively).

\section{Topological Recursion of the LSZ Model}
\label{ch:tr}

\subsection{Analytic Continuation and Initial Data}

To determine the solution of a correlation function, we will choose an analytic continuation imposed by the DSE's. The correlation functions are not uniquely continued in this way, but very canonical. Evaluating the analytic functions at the points $E_i$ (and/or $\tilde{E}_j$) yields however the unique result of the DSE's (\ref{DSE-T2}) and  (\ref{DSE-Om}). The same kind of continuation is performed in the Kontsevich model (see e.g. \cite{Eynard:2016yaa}) or in the quartic analogue of the Kontsevich model \cite{Branahl:2020yru}.

\begin{definition}[Analytic continuation]\label{def:complexT}
	DSE \ref{DSE-T2} (or equivalently DSE \eqref{H2P}) suggests the following extension:
	\begin{itemize}
		
		\item[\textup{(a)}] Introduce holomorphic functions 
		$G,T,\tilde{\Omega}$ in several complex variables, 
		defined on Cartesian products of a connected neighbourhood 
		$\mathcal{V}$ of $\{E_1,...,E_N\}$ (and $\tilde{\mathcal{V}}$ of $\{\tilde{E}_1,...,\tilde{E}_N\}$) in $\mathbb{C}$ ,  
		which at $E_1,\dots,E_N$ (and $\tilde{E}_1,...,\tilde{E}_N$) agree with the previous correlation functions:
		\begin{align*}
			G(E_p,\tilde{E}_q)&\equiv G_{|pq|}\;,
			\\
			T(E_{p_1},...,E_{p_n}\| E_p,\tilde{E}_q|)
			&\equiv T_{p_1,\dots,p_n\|pq}\;,\qquad
			\\
			\tilde{\Omega}(E_{p_1},...,E_{p_n})
			&\equiv \Omega_{p_1,\dots,p_n}\;.
		\end{align*}
		
		\item[\textup{(b)}] Write eq. \eqref{DSE-T2} (or equivalently \eqref{H2P})
		in terms of $G,T,\tilde{\Omega}$ and postulate that they extend to
		pairwise different points
		$\{E_p\mapsto\zeta,\tilde{E}_q\mapsto \eta,E_{p_j} \mapsto \zeta_j\}$ of
		$\mathcal{V}$ and $\tilde{\mathcal{V}}$.

		\item[\textup{(c)}] Complexify the derivative by 
		\begin{align*}
			\frac{\partial }{\partial E_p} f(E_p)\mapsto
			\frac{f(\zeta)-f(E_p)}{\zeta-E_p}
			+\frac{\partial }{\partial E_p}\Big\vert_{E_p\mapsto \zeta} f(\zeta)
		\end{align*}
		such that the
		$\frac{\partial }{\partial E_p}\big\vert_{E_p\mapsto \zeta}$-derivative
		acts in the sense of Definition \ref{def:gDSE} with extension to
		$E_p\mapsto \zeta$, and a difference quotient which tends for
		$\zeta\to E_p$ to the derivative on the argument of $f$.
		
		\item[\textup{(d)}] Keep the $E_n$ and $\tilde{E}_k$ in summations over $k,n \in \{1,\dots,N\}$ and 
		complete the $n$-summation of $E_n$ with the difference quotient term 
		of \textup{(c)}.
		Consider the equations for
		$\zeta,\zeta_i \in \mathcal{V}\setminus
		\{E_1,\dots,E_N\}$ and $\eta\in \tilde{\mathcal{V}}\setminus \{\tilde{E}_1,...,\tilde{E}_N \} $.
		
		\item[\textup{(e)}] Define the values of $G,T,\tilde{\Omega}$ at 
		$\zeta=E_p,\eta=\tilde{E}_q,\zeta_i =E_{p_i}$ and 
		at coinciding points by a limit procedure.
	\end{itemize}
\end{definition}

The holomorphic extension gives us more freedom for the correlation functions. The property of having pairwise distinct eigenvalues $E_1,...,E_N$ (and $\tilde{E}_1,...,\tilde{E}_N$) can be relaxed to have coinciding eigenvalues. We will deal in the following with the more general situation that $e_1,...,e_d$ are the $d$ pairwise distinct eigenvalues of $E$ with multiplicities $r_1,...,r_d$, i.e. $\sum_{n=1}^dr_n=N$  and $\tilde{e}_1,...,\tilde{e}_{\tilde{d}}$ are the $\tilde{d}$ pairwise distinct eigenvalues of $\tilde{E}$ with multiplicities $\tilde{r}_1,...,\tilde{r}_{\tilde{d}}$, i.e. $\sum_{k=1}^{\tilde{d}}\tilde{r}_k=N$).

Together with the complexification defined and described in Definition \ref{def:complexT}, the analytically continued DSE's \eqref{DSE-T2} and \eqref{H2P} spell out in a straightforward way. We illustrate the procedure with the following example of $I=\emptyset$:
\begin{example}\label{ex:Gen2P}
	The analytically continued DSE's \eqref{DSE-T2} and \eqref{H2P} read for  $I=\emptyset$:
	\begin{align}\label{Gen2Pcompl1}
		&\bigg(\eta+V'(\zeta)+\lambda \tilde{\Omega}(\zeta)\bigg)T(\|\zeta,\eta|)+\lambda\frac{1}{N^2}T(\zeta\|\zeta,\eta|)
		=1 -\frac{\lambda}{N}\sum_{n=1}^dr_n\frac{T(\|e_n,\eta|)}{\zeta-e_n}
	\end{align}
and
\begin{align}\label{H2Pcompl1}
	&\big(v+V'(\zeta)+\lambda \tilde{\Omega}(\zeta)\big)\bigg(-1+\frac{\lambda}{N}\sum_{q=1}^{\tilde{d}}\tilde{r}_q \frac{T(\|\zeta,\tilde{e}_q|)}{v-\tilde{e}_q}\bigg)+\frac{\lambda^2 }{N^3}\sum_{q=1}^{\tilde{d}}\tilde{r}_q \frac{T(\zeta\|\zeta,\tilde{e}_q|)}{v-\tilde{e}_q}\\\nonumber
	=&-\tilde{V}'(v)-V'(\zeta)-\frac{\lambda^2}{N^2}\sum_{q,n=1}^{\tilde{d},d}\frac{T(\|e_n,\tilde{e}_q|)}{(\zeta-e_n)(v-\tilde{e}_q)}.
\end{align}
The potentials are given by $V'(\zeta)=\zeta-\frac{\lambda}{N}\sum_{n=1}^d\frac{r_n}{\zeta-e_n}$ and $\tilde{V}'(x)=x-\frac{\lambda}{N}\sum_{k=1}^{\tilde{d}}\frac{\tilde{r}_k}{x-\tilde{e}_k}$ and the $n$-summation got unrestricted due to \textup{(c)} of Definition \ref{def:complexT}. We emphasise the crucial fact that the rhs (and therefore the lhs as well) of \eqref{Gen2Pcompl1} is a rational function in $\zeta$ with simple poles at the points $e_n$. Multiplying by $\prod_{n=1}^d (\zeta-e_n)$ generates a polynomial of degree $d$. Furthermore,  the rhs (and therefore the lhs) of \eqref{H2Pcompl1} is a rational function in $\zeta$ and $v$ with simple poles at the points $e_n$ and $\tilde{e}_q$. Multiplying by $\prod_{n=1}^d (\zeta-e_n)\prod_{q=1}^{\tilde{d}} (v-\tilde{e}_q)$ generates a polynomial of degree $d+1$ in $\zeta$ and $\tilde{d}+1$ in $v$.
\end{example}

As mentioned earlier, the correlation functions are graded by the genus $g$ and have a formal genus expansion in $N$. In case of $g=0$, this decouples \eqref{Gen2Pcompl1} and \eqref{H2Pcompl1} from $T(\zeta\|\zeta,.|)$ and yields a nonlinear equation for $T(\|\zeta,\eta|)$. Together with the rationality property underlined in Example \ref{ex:Gen2P}, these two equations can be solved simultaneously. The solution is based on some natural, weak assumptions.
\begin{theorem}[Initial data for topological recursion]\label{Thm:InitialData}
		Let $\lambda,e_n,\tilde{e}_k>0$. Assume two rational functions $x,y:\overline{\mathbb{C}}\to \overline{\mathbb{C}}$ exist with $x$ of rational degree $\tilde{d}+1$ and $y$ of rational degree $d+1$, both normalised such that $ x(z)=z+\mathcal{O}(1)$ and $y(z)=-z+\mathcal{O}(1)$. 
	Let further $G(x(z),y(w))=T(\|x(z),y(w)|)=:\mathcal{G}(z,w)$, and let $x,y$ satisfy the functional relations
	\begin{align}\label{yGsum}
		y(z)=&-V'(x(z))-\frac{\lambda}{N}\sum_{k=1}^{\tilde{d}}\tilde{r}_k\mathcal{G}(z,\tilde{\varepsilon}_k)\\
		x(z)=&\tilde{V}'(y(z))+\frac{\lambda}{N}\sum_{n=1}^{d}r_n\mathcal{G}(\varepsilon_n,z),
	\end{align}
	where the points $\varepsilon_n$ and $\tilde{\varepsilon}_k$ are implicitly defined by $e_n=:x(\varepsilon_n)$ and $\tilde{e}_k=y(\tilde{\varepsilon}_k)$.  We denote the $\tilde{d}+1$ solutions of $x(v)=x(z)$ by $v\in\{z,\hat{z}^1,...,\hat{z}^{\tilde{d}}\}$ and the $d+1$ solutions of $y(v)=y(w)$ by $v\in\{w,\hat{\tilde{w}}^1,...,\hat{\tilde{w}}^{d}\}$.  Assume that $y(\hat{z}^{l})$ and $x(\hat{\tilde{w}}^{l})$ as well as $\mathcal{G}(z,\hat{z}^{l})$ and $\mathcal{G}(\hat{\tilde{w}}^{l},w)$ are finite $\forall l$. Then, $x,y$ and $\mathcal{G}$ are uniquely determined by
	\begin{align}\label{xy}
		x(z)=z-\frac{\lambda}{N}\sum_{k=1}^{\tilde{d}}&\frac{\tilde{r}_k}{y'(\tilde{\varepsilon}_k)(z-\tilde{\varepsilon}_k)},\qquad y(z)=-z+\frac{\lambda}{N}\sum_{n=1}^{d}\frac{r_n}{x'(\varepsilon_n)(z-\varepsilon_n)}\\
		\mathcal{G}(z,w)=&\frac{E(x(z),y(w))}{(x(z)-x(w))(y(z)-y(w))},\\ \text{where}\qquad 
		 E(x(z),y(w))=&(y(w)-y(z))\prod_{k=1}^{\tilde{d}}\frac{y(w)-y(\hat{z}^k)}{y(w)-y(\tilde{\varepsilon}_k)}\\
		 =&(x(z)-x(w))\prod_{n=1}^{d}\frac{x(z)-x(\hat{\tilde{w}}^n)}{x(z)-x(\varepsilon_n)}.
	\end{align}
 \begin{proof}
 	See Appendix \ref{sec:2p}
 \end{proof}
\end{theorem}
Note again that $\mathcal{G}(z,w) \neq \mathcal{G}(w,z)$ for unrelated $E$ and $\tilde{E}$ in contrast to the hermitian model.
Here we encounter a general phenomenon during the solution strategy of a certain matrix model using TR. The correlator of the most basic topology is a sophisticated (nonlinear) problem, often related to a Riemann-Hilbert problem.  Its solution already reveals the geometry of the spectral curve, implicitly containing all the information about higher topological sectors. Indeed, the same holds for the LSZ model: The solution given in Theorem \ref{Thm:InitialData} provides most of the initial data for topological recursion $(\Sigma,x,y,B)$ as we will show in the rest of this section, where $\Sigma=\overline{\mathbb{C}}$ and $x,y$ given by \eqref{xy}. The rational parametrisation of $x,y$ implies to a genus 0 spectral curve, which uniquely determines $B(z,w)=\frac{dz\,dw}{(z-w)^2}$. We will show that the Bergman kernel is directly related to $\Omega$ with topology $(g,n)=(0,2)$ in Appendix \ref{sec:om02}. 

It still remains to show how the correlators $\omega_{g,n}$ computed by TR, eq. (\ref{eq:TR-intro}), are related to the correlation functions of the LSZ model. It turns out that $\omega_{g,n}$ are in direct relation to $\Omega_{q_1,...,q_n}$ defined in Definition \ref{def:gDSE}. To prove this, we will need to perform a decisive variable transformation implied by the solution of $x(z)$ and $y(z)$.

\subsection{Complexified Dyson-Schwinger Equations}
The hidden algebraic structure  behind the LSZ model becomes visible after a final variable transformation imposed by $x(z)$ and $y(z)$, which was achieved by the solution of the 2-point function in Theorem \ref{Thm:InitialData}. The variable transformation via $x(z)$ and $y(z)$ is a transformation depending on the coupling constant $\lambda$. We define the correlation functions on the (usually called) $z$-plane:
\begin{definition}[Correlation functions on the $z$-plane]\label{def:compl}
	Let $G,T,\tilde{\Omega}$ be the functions in several 
	complex variables obtained by the complexification of
	Definition~\ref{def:complexT} and by 
	admitting multiplicities $r_n,\tilde{r}_k$ of $e_n,\tilde{e}_k$. Then functions 
	$\mathcal{G},\mathcal{T},\Omega_n$ of several 
	complex variables are introduced by
	\begin{align*}
		\mathcal{G}(u,w)&:=G(x(u),y(w))\;,
		\\
		\mathcal{T}(z_1,...,z_n\|u,w|)
		&:=
		T(x(z_1),...,x(z_n)\|x(u),y(w)|)\;,
		\\
		\Omega_{n}(z_1,....,z_n)&:=
		\tilde{\Omega}(x(z_1),...,x(z_n))+\frac{\delta_{n,1}}{\lambda}V'(x(z_1)) \;.
	\end{align*}
	We let $\mathcal{T}(\emptyset\|u,w|):=
	\mathcal{G}(u,w)$.
\end{definition}
We write the genus-expanded correlation functions as
\begin{align}
	\mathcal{T}(z_1,...,z_n\|u,w|)=&:\sum_{g=0}^\infty N^{-2g} \mathcal{T}^{(g)}(z_1,...,z_n\|u,w|)\\
	\Omega_{n}(z_1,....,z_n)=&:\sum_{g=0}^\infty N^{-2g}\Omega_{g,n}(z_1,....,z_n).
\end{align}
Combining all of the previous constructions and definitions, we finally get a complexified DSE on the $z$-plane, which takes a well-known form.
\begin{corollary}[DSE of topological recursion type]\label{Cor:DSE}
	Let $I=\{z_2,...,z_n\}$ and 
	\begin{align*}
		H_{g,n}(v;z;I):=&-\delta_{|I|,0}\delta_{g,0}+\frac{\lambda}{N}\sum_{k=1}^{\tilde{d}} \tilde{r}_k\frac{\mathcal{T}^{(g)}(I\|z,\tilde{\varepsilon}_k)}{v-\tilde{e}_k},\\
		P_{g,n}(v;x;I):=&\delta_{|I|,0}\delta_{g,0}(\tilde{V}'(v)+V'(x))+\frac{\lambda^2}{N^2}\sum_{k,n=1}^{\tilde{d},d} \tilde{r}_kr_n\frac{\mathcal{T}^{(g)}(I\|\varepsilon_n,\tilde{\varepsilon}_k)}{(v-\tilde{e}_k)(x-e_n))}\\
		&-\sum_{z_i\in I}\frac{\partial}{\partial x(z_i)}\frac{\lambda^2}{N}\sum_{k=1}^{\tilde{d}} \tilde{r}_k\frac{\mathcal{T}^{(g)}(I\setminus z_i\|z_i,\tilde{\varepsilon}_k)}{(v-\tilde{e}_k)(x-x(z_i))}+\frac{\lambda \delta_{g,0}\delta_{n,2}}{(x-x(z_2))^2},
	\end{align*}
	where $H$ is by definition rational in $v$ and $P$ rational in $v$ and $x$. Then, $H$ and $P$ satisfy the following DSE
	\begin{align}\label{DSEHP}
		&(v-y(z))H_{g,n}(v;z;I)+P_{g,n}(v;x(z);I)\\\nonumber
		=&-\lambda \sum_{\substack{g_1+g_2=g\\ I_1\uplus I_2=I}}^\prime H_{g_1,|I_1|+1}(v;z;I_1)\Omega_{g_2,|I_2|+1}(I_2,z)-\lambda H_{g-1,n+1}(v;z;z,I),
	\end{align}
	where the primed sum excludes $(g_2,I_2)=(0,\emptyset)$.
	\begin{proof}
		The DSE of $H$ and $P$ arises from DSE \eqref{H2P}. First, the complexification of Definition \ref{def:complexT} has to be considered to formulate the DSE in terms of $T$ and $\tilde{\Omega}$ in the variables $\zeta_j, \zeta$ and $\eta$ similar to Example \ref{ex:Gen2P}. Then, Theorem \ref{Thm:InitialData} is applied with \eqref{yGsum} to get $y(u)$ on the lhs of the equation. Following the variable transformation given in Definition \ref{def:compl} via $x(z)$ and $y(z)$, and the definition of $H$ and $P$  yields the assertion.
	\end{proof}
\end{corollary}

Exactly the same structure of the DSE of Corollary \ref{Cor:DSE} appears in the 2-matrix model \cite[eq. (2-18)]{Chekhov:2006vd} and the generalised Kontsevich model \cite[eq. (3.16)]{Belliard:2021jtj}. We will follow the same steps to proof TR. From the definitions, we have for the asymptotic expansion in $v$ the following leading order terms: 
\begin{align}\label{HOmega}
	H_{g,n}(v;z;I)=-\delta_{|I|,0}\delta_{g,0}+&\frac{1}{v}\bigg(\lambda\Omega_{g,n}(z,I)
	-\delta_{|I|,0}\delta_{g,0}V'(x(z))-\frac{\lambda \delta_{|I|,1}\delta_{g,0}}{(x(z)-x(z_2))^2}\bigg)\\\nonumber
	&+\mathcal{O}(v^{-2})
\end{align}

We recall that $H_{g,n}(v;z;I)$ is a rational function in $v$ with simple poles located at $\tilde{e}_k$. Consequently, after multiplying by $\prod_{k=1}^{\tilde{d}}(v-y(\tilde{\varepsilon}_k))$ we obtain a polynomial in $v$ of degree $\tilde{d}$ for $(g,n)=(0,0)$ and degree $\tilde{d}-1$ for any other $(g,n)$. Furthermore, $P_{g,n}(v;x(z);I)$ of \eqref{DSEHP}  is a rational function in $v$ and $x(z)$, where all poles are known as well. After multiplying with $\prod_{k=1}^{\tilde{d}}(v-y(\tilde{\varepsilon}_k))$ it turns  (for $2g+n-2\geq 0$)  into a polynomial of degree $\tilde{d}$ in $v$. A polynomial of degree $\tilde{d}$ is uniquely determined by $\tilde{d}+1$ points.
In particular, we already know the lowest topology in $H$ and $P$:
\begin{align}\label{H01P01}	H_{0,1}(v;z;\emptyset)=-\prod_{k=1}^{\tilde{d}}\frac{v-y(\hat{z}^k)}{v-\tilde{e}_k},
\qquad P_{0,1}(v;x(z);\emptyset)=(v-y(z))\prod_{k=1}^{\tilde{d}}\frac{v-y(\hat{z}^k)}{v-\tilde{e}_k}.
\end{align}
These expressions were used in the proof of Theorem \ref{Thm:InitialData}, see App. \ref{sec:2p}. Additionally, the fact that $P_{g,n}(v;x(z);I)$ depends on $x(z)$ rather than on $z$ yields several identities for DSE \eqref{DSEHP} by
\begin{align*}
	P_{g,n}(v;x(z);I)=P_{g,n}(v;x(\hat{z}^k);I),
\end{align*}
where the $\hat{z}^k$'s were defined as the  preimages under $x$, i.e. $x(z)=x(\hat{z}^k)$ for $k=1,...,\tilde{d}$.
These properties of the functions $H$ and $P$ will be crucial in the proof of TR. Before going into more details, we will finish this subsection with the solution for the second ingredient of the initial data, $\Omega_{0,2}$.
\begin{proposition}\label{pro:om02}
	The cylinder amplitude of the LSZ model reads
	\begin{align}
		\Omega_{0,2}(z_1,z_2)=\frac{1}{x'(z_1)x'(z_2)(z_1-z_2)^2}.
	\end{align}
	\begin{proof}
		See Appendix \ref{sec:om02}
	\end{proof}
\end{proposition}

\subsection{Proof of Topological Recursion}
The proof will proceed in three steps, in complete analogy to the proof that the generalised Kontsevich model is governed by topological recursion \cite{Belliard:2021jtj}:
\begin{itemize}
	\item[\textbf{I.}] The solution of $H_{g,n}$ and $P_{g,n}$ is derived in terms of sums of the $\Omega_{g',n'}$ where the arguments are the preimages $\hat{z}^k$ with $2g+n-2\geq 2g'+n'-2$
	\item[\textbf{II.}] The linear and quadratic loop equations are extracted from the exact solution obtained in \textbf{I.}
	\item[\textbf{III.}] TR is concluded from the linear and quadratic loop equations of \textbf{II.}, see Theorem \ref{thm:ale}
\end{itemize}
\textbf{I.}
The solutions of $H$ and $P$ are constructed by the following functions (first defined in \cite{Bouchard:2012yg}, but we follow the notation of \cite{Bouchard:2016obz,Belliard:2021jtj})
\begin{definition}
	Let $\underline{t}:=\{t_1,...,t_k\}$ and $I=\{z_2,...,z_n\}$, then we define
	\begin{align}
		\mathcal{E}^{(k)}\Omega_{g,n}(\underline{t};I):=\sum_{\mu \vdash \underline{t}}\sum_{I_1\uplus ...\uplus I_{l(\mu)}=I}\sum_{\substack{h_1+...h_{l(\mu)}\\
		=g+l(\mu)-k}}\bigg(\prod_{i=1}^{l(\mu)}\Omega_{h_i,|\mu_i\uplus I_i|}(\mu_i,I_i)\bigg),
	\end{align}
where $\mu \vdash \underline{t}$ means that $\mu$ is a set partition of $\underline{t}$, which consists of $l(\mu)$ parts $\mu_1,...,\mu_{l(\mu)}$. In particular, we set $\mathcal{E}^{(0)}\Omega_{g,n}(\emptyset;I)=\delta_{g,0}\delta_{n,1}$.
\end{definition}
In the next step, we introduce the functions $\check{H}_{g,n}$ and $\check{P}_{g,n}$, which are designed to coincide with $H_{g,n}$ and $P_{g,n}$:
\begin{definition}\label{Def:Hcheck}
	Let $\tau(z)=\{z,\hat{z}^1,...,\hat{z}^{\tilde{d}}\}$ be the set of solutions of the equation $x(\bullet)=x(z)$, and let $\tau_0(z)=\tau(z)\setminus \{z\}$. For $I=\{z_2,...,z_n\}$, define
	\begin{align}
		\check{H}_{g,n}(v;z;I)=&-\frac{1}{\prod_{k=1}^{\tilde{d}}(v-\tilde{e}_k)}\sum_{i=0}^{\tilde{d}}\lambda^iv^{\tilde{d}-i}\sum_{\underline{t}\subset_i\tau_0(z) }\mathcal{E}^{(i)}\Omega_{g,n}(\underline{t};I)\\
		\check{P}_{g,n}(v;x(z);I)=&\frac{1}{\prod_{k=1}^{\tilde{d}}(v-\tilde{e}_k)}\sum_{i=0}^{\tilde{d}+1}\lambda^iv^{\tilde{d}+1-i}\sum_{\underline{t}\subset_i\tau(z) }\mathcal{E}^{(i)}\Omega_{g,n}(\underline{t};I),
	\end{align}
	where $\subset_i$ takes a subset of $i$ elements. The main difference in the definition of $\check{H}$ and $\check{P}$ is the sum of $\underline{t}$ which  either lies in $\tau_0(z)$ or $\tau(z)=\tau_0(z)\cup \{z\}$. By definition, $\check{P}$ is a rational function in $x(z)$ due to the symmetric representation in all preimages $\hat{z}^i$, including $z$.
\end{definition}
For $g=0$ and $I=\emptyset$ the expressions considerably simplify.
It is straightforward to show their lowest topology $(g,n)=(0,1)$ with $y(z)=-\lambda \Omega_{0,1}(z)$ by their definitions: 
\begin{align}\label{Hcheck01}
	\check{H}_{0,1}(v;z;\emptyset)=-\prod_{k=1}^{\tilde{d}}\frac{v-y(\hat{z}^k)}{v-\tilde{e}_k}\quad \text{and}\quad \check{P}_{0,1}(v;x(z);\emptyset)=(v-y(z))\prod_{k=1}^{\tilde{d}}\frac{v-y(\hat{z}^k)}{v-\tilde{e}_k}
\end{align}
These expressions coincide with \eqref{H01P01}. The coincidence holds for any $(g,n)$:
\begin{proposition}\label{Prop:HHcheck}
	Let $I=\{z_2,...,z_n\}$. Then,
	the following holds:
	\begin{align*}
		\check{H}_{g,n}(v;z;I)=H_{g,n}(v;z;I)\qquad \text{and}\qquad \check{P}_{g,n}(v;x;I)=P_{g,n}(v;x;I).
	\end{align*}
\begin{proof}
	Decompose $\check{P}_{g,n}$ of Definition \ref{Def:Hcheck} in terms of $\check{H}_{g',n'}$ by separating $z\in \tau(z)$ yields an equation between $\check{P}_{g,n}$ and $\check{H}_{g',n'}$ with $2g+n-2\geq 2g'+n'-2$. Applying \cite[Lemma 3.17]{Bouchard:2016obz} 
	\begin{align}
		\mathcal{E}^{(k+1)}\Omega_{g,n}(z,\underline{t};I)=\mathcal{E}^{(k)}\Omega_{g-1,n+1}(\underline{t};z,I)+\sum_{\substack{g_1+g_2=g\\I_1\uplus I_2=I}}\Omega_{g_1,|I_1|+1}(z,I_1)\mathcal{E}^{(k)}\Omega_{g_2,|I_2|+1}(\underline{t};I_2),
	\end{align}
reduces the previous equation to 
\begin{align}\label{DSEHPcheck}
	&(v-y(z))\check{H}_{g,n}(v;z;I)+\check{P}_{g,n}(v;x(z);I)\\\nonumber
	=&-\lambda \sum_{\substack{g_1+g_2=g\\ I_1\uplus I_2=I}}^\prime \check{H}_{g_1,|I_1|+1}(v;z;I_1)\Omega_{g_2,|I_2|+1}(I_2,z)-\lambda \check{H}_{g-1,n+1}(v;z;z,I),
\end{align}
which coincides with \eqref{DSEHP} of Corollary \ref{Cor:DSE} (we refer to \cite{Belliard:2021jtj} for more details). Now, we proceed by induction in the negative Euler characteristic $-\chi=2g+n-2$, where the initial step was already shown in \eqref{Hcheck01}. Assume that Proposition \ref{Prop:HHcheck} holds for all $2g'+n'-2<2g+n-2$, then we obtain by subtracting \eqref{DSEHP} from \eqref{DSEHPcheck}
\begin{align}
	P_{g,n}(v;x(z);I)-\check{P}_{g,n}(v;x(z);I)=(v-y(z))(\check{H}_{g,n}(v;z;I)-H_{g,n}(v;z;I)).
\end{align}
Since the lhs is a rational function of $x(z)$, it does not change by $z\mapsto \hat{z}^j$
\begin{align}
	P_{g,n}(v;x(z);I)-\check{P}_{g,n}(v;x(z);I)=(v-y(\hat{z}^j))(\check{H}_{g,n}(v;\hat{z}^j;I)-H_{g,n}(v;\hat{z}^j;I)).
\end{align}
Setting $v=y(\hat{z}^j)$ for $j=0,...,\tilde{d}$, the rhs vanishes  at $\tilde{d}+1$ points, where $P$ and $\check{P}$ coincide:
\begin{align}\label{Pd+1}
	P_{g,n}(y(\hat{z}^j);x(z);I)=\check{P}_{g,n}(y(\hat{z}^j);x(z);I),\qquad \forall j=0,...,\tilde{d}
\end{align}
Recall that after multiplying $\check{P}$ by $\prod_{k=1}^{\tilde{d}}(v-\tilde{e}_k)$ it becomes a polynomial of degree $\tilde{d}$, which is uniquely determined by the $\tilde{d}+1$ points in \eqref{Pd+1}. This gives rise to the equality between $P$ and $\check{P}$. Furthermore, the DSE \eqref{DSEHPcheck} directly leads to the equality between $H$ and $\check{H}$.
\end{proof}
\end{proposition}

\textbf{II.}
The equality given in Proposition \ref{Prop:HHcheck} is more involved than it seems. Expanding it around $v$, we obtain equations for $\Omega_{g,n}$ which are equivalent to the linear and quadratic loop equations (see Definition \ref{def:absloopeq}):
\begin{proposition}\label{Prop:linquad}
	Let $I=\{z_2,...,z_n\}$ and $2g+n-2 > 0$, then
	\begin{align}\label{linearloop}
		\sum_{k=0}^{\tilde{d}}\Omega_{g,n}(\hat{z}^k,I)=0
	\end{align}
and 
\begin{align}\label{quadloop}
	&\frac{1}{2}\sum_{k=0}^{\tilde{d}}\bigg(\sum_{\substack{I_1\uplus I_2=I\\g_1+g_2=g}}\Omega_{g_1,|I_1|+1}(\hat{z}^k,I_1)\Omega_{g_2,|I_2|+1}(\hat{z}^k,I_2)+\Omega_{g-1,n+1}(\hat{z}^k,\hat{z}^k,I)\bigg)\\\nonumber
	=&\sum_{z_i\in I}\frac{\partial}{\partial x(z_i)}\frac{\Omega_{g,n-1}(I)}{x(z)-x(z_i)}-\frac{1}{N}\sum_{k=1}^{d}r_k\frac{\Omega_{g,n}(\varepsilon_k,I)}{x(z)-e_k}
\end{align}

\begin{proof}
	The first equation is obtained by identifying (due to Proposition \ref{Prop:HHcheck}) the expressions
	\begin{align*}
		[v^{\tilde{d}-1}]\bigg(\check{H}_{g,n}(v;z;I)\prod_{k=1}^{\tilde{d}}(v-\tilde{e}_k)\bigg)=&-\lambda \sum_{k=1}^{\tilde{d}}\Omega_{g,n}(\hat{z}^k,I)\\
		[v^{\tilde{d}-1}]\bigg(H_{g,n}(v;z;I)\prod_{k=1}^{\tilde{d}}(v-\tilde{e}_k)\bigg)=&\lambda\Omega_{g,n}(z,I),
	\end{align*}
	which follows directly by their definitions for $2g+n-2\geq 0$. Going to the next order in $v$, we obtain
	\begin{align*}
		&[v^{\tilde{d}-2}]\bigg(\check{H}_{g,n}(v;z;I)\prod_{k=1}^{\tilde{d}}(v-\tilde{e}_k)\bigg)\\
		=&-\frac{\lambda^2}{2}\bigg(\sum_{\substack{I_1\uplus I_2=I\\g_1+g_2=g}}\Omega_{g_1,|I_1|+1}(z,I_1)\Omega_{g_2,|I_2|+1}(z,I_2)+\Omega_{g-1,n+1}(z,z,I)\bigg)\\
		&+\frac{\lambda^2}{2}\sum_{k=1}^{\tilde{d}}\bigg(\sum_{\substack{I_1\uplus I_2=I\\g_1+g_2=g}}\Omega_{g_1,|I_1|+1}(\hat{z}^k,I_1)\Omega_{g_2,|I_2|+1}(\hat{z}^k,I_2)+\Omega_{g-1,n+1}(\hat{z}^k,\hat{z}^k,I)\bigg)\\
		&+\lambda^2\sum_{z_i\in I}\frac{\partial}{\partial x(z_i)}\frac{\Omega_{g,n-1}(I\setminus z_i,z)}{x(z)-x(z_i)}-\frac{\lambda^2\delta_{g,0} \delta_{|I|,2}}{(x(z)-x(z_2))^2(x(z)-x(z_3))^2}\\
		&+V'(x(z))\bigg(\Omega_{g,n}(z,I)-\frac{\lambda \delta_{g,0}\delta_{|I|,1}}{(x(z)-x(z_2))^2}\bigg)
	\end{align*}
	where $\mathcal{E}^{(2)}$ was written out explicitly, the sum over the preimages was anti-symmetrised and the linear loop equation \eqref{linearloop} applied.
	
	On the other hand, the expansion of $H_{g,n}(v;z;I)$ is achieved by expanding the DSE \eqref{DSEHP} of Corollary \ref{Cor:DSE}
	\begin{align*}
		&[v^{\tilde{d}-2}]\bigg(H_{g,n}(v;z;I)\prod_{k=1}^{\tilde{d}}(v-\tilde{e}_k)\bigg)\\
		=&-\lambda^2\bigg(\sum_{\substack{I_1\uplus I_2=I\\g_1+g_2=g}}\Omega_{g_1,|I_1|+1}(z,I_1)\Omega_{g_2,|I_2|+1}(z,I_2)+\Omega_{g-1,n+1}(z,z,I)\bigg)
		\\
		&-\frac{\lambda^2}{N}\sum_{k=1}^{d}r_k\frac{\Omega_{g,n}(\varepsilon_k,I)}{x(z)-e_k}+\lambda^2\sum_{z_i\in I}\frac{\partial}{\partial x(z_i)}\frac{\Omega_{g,n-1}(I\setminus z_i,z)+\Omega_{g,n-1}(I)}{x(z)-x(z_i)}\\
		&-\frac{\lambda^2\delta_{g,0} \delta_{|I|,2}}{(x(z)-x(z_2))^2(x(z)-x(z_3))^2}
		+V'(x(z))\bigg(\Omega_{g,n}(z,I)-\frac{\lambda \delta_{g,0}\delta_{|I|,1}}{(x(z)-x(z_2))^2}\bigg).
	\end{align*}
	Comparing both expressions, we confirm the assertion.
\end{proof}
\end{proposition}

\textbf{III.}
First, we comment on the pole structure of $\Omega_{g,n}$. From the exact solutions
\begin{align*}
	\Omega_{0,1}(z)=-\frac{1}{\lambda}y(z),\qquad \Omega_{0,2}(z_1,z_2)=\frac{1}{x'(z_1)x'(z_2)(z_1-z_2)^2},
\end{align*}
we can deduce inductively that all $\Omega_{g,n}(z,...)$ have for $2g+n-2>0$ poles only located at the ramification point $z$ of $x$, i.e. $x'(z)=0$. This is observed by expanding the DSE \eqref{DSEHP} in $v$ with 
\begin{align*}
	H_{g,n}(v;z;I)=\frac{\lambda \Omega_{g,n}(z,I)}{v}+\mathcal{O}(v^{-2}).
\end{align*}
Comparing the poles on both sides inductively, we see that
some additional possible poles could arise at $z=z_i$ or $z=\varepsilon_n$. However, these poles cancel exactly due to the exact solution of $\Omega_{0,1}$ and $\Omega_{0,2}$ having also poles at the diagonal $z=z_i$ and $z=\varepsilon_n$. Therefore, $\Omega_{g,n}(z,I)$ has for $2g+n-2>0$ only poles at the ramification points of $x(z)$.
 
Let the simple ramification points of $x(z)$, where $x(z)$ is given by Theorem \ref{Thm:InitialData}, be denoted by $\{\beta_1,...,\beta_{2\tilde{d}}\}$ with $x'(\beta_i)=0$. At a ramification point two branches meet, i.e. for $z=\beta_j$ exists an $i\in \{1,...,\tilde{d}\}$ with $\hat{z}^i=\beta_j$. We will denote in the vicinity of $\beta_j$ this specific preimages $\hat{z}^i$ around $\beta_j$ by $\sigma_j(z):=\hat{z}^i$ being the local Galois involution introduced in Sec. \ref{ch:setup}. 

\begin{theorem}\label{thm:TROm}
	Let $I=\{z_2,...,z_n\}$ and $2g+n-2>0$. Then, one can compute recursively in $2g+n-2$ all $\Omega_{g,n}$ by topological recursion
	\begin{align}\label{TRTheorem}
		\Omega_{g,n}(z,I)&x'(z)=\lambda \sum_{i=1}^{2 \tilde{d}}\Res\displaylimits_{q\to \beta_i} \frac{\frac{1}{z-q}-\frac{1}{z-\sigma_i(q)}}{2(y(q)-y(\sigma_i(q)))}dx(q)\\\nonumber
		&\times\bigg\{\sum_{\substack{I_1\uplus I_2=I\\g_1+g_2=g}}^\prime \Omega_{g_1,|I_1|+1}(q,I_1)\Omega_{g_2,|I_2|+1}(\sigma_i(q),I_2)+\Omega_{g-1,n+1}(q,\sigma_i(q),I)\bigg\},
	\end{align}
where the sum is taken over all the residues are at the ramification points $\beta_i$ of $x(z)$, i.e. $x'(\beta_i)=0$. The primed sum $\sum^\prime$ excludes the cases $(g_i,I_i)=(0,\emptyset)$.

\begin{proof}
	We deduce from the linear loop equation \eqref{linearloop} of Proposition \ref{Prop:linquad} for the behavior around $\beta_i$
	\begin{align}
		\Omega_{g,n}(z,I)+\Omega_{g,n}(\sigma_i(z),I)=\mathcal{O}((z-\beta_i)^0).
	\end{align}
	Combining the quadratic loop equation \eqref{quadloop} with the linear, we obtain in the vicinity of $\beta_i$
	\begin{align}\label{quadproof}
		\frac{1}{\lambda}(y(z)-y(\sigma_i(z)))\Omega_{g,n}(z,I)+&\sum_{\substack{I_1\uplus I_2=I\\g_1+g_2=g}}^\prime\Omega_{g_1,|I_1|+1}(z,I_1)\Omega_{g_2,|I_2|+1}(\sigma_i(z),I_2)\\
		+&\Omega_{g-1,n+1}(z,\sigma_i(z),I)=\mathcal{O}((z-\beta_i)^0).
	\end{align}
	Finally, we compute by Cauchy's formula and by shifting the contour $\Omega_{g,n}$ as
	\begin{align*}
		\Omega_{g,n}(z,I)x'(z)=&  \sum_{i=1}^{2 \tilde{d}} \Res\displaylimits_{q\to z}\frac{\Omega_{g,n}(q,I)dx(q)}{z-q}=-\sum_{i=1}^{2 \tilde{d}}\Res\displaylimits_{q\to \beta_i}\frac{\Omega_{g,n}(q,I)dx(q)}{z-q}\\
		=&-\lambda \sum_{i=1}^{2 \tilde{d}} \Res\displaylimits_{q\to \beta_i}\frac{\sum_{\substack{I_1\uplus I_2=I\\g_1+g_2=g}}^\prime\Omega_{g_1,|I_1|+1}(q,I_1)\Omega_{g_2,|I_2|+1}(\sigma_i(q),I_2)}{(y(q)-y(\sigma_i(q)))(z-q)}dx(q),
	\end{align*}
	where we took into account that $\Omega_{g,n}$ has only poles at the ramification points $\beta_i$ and the quadratic loop equation \eqref{quadproof} after dividing by $\frac{1}{\lambda}(y(z)-y(\sigma_i(z)))$. Since the residue is invariant under the local Galois involution $q\mapsto \sigma_i(q)$ around $\beta_i$, we derived \eqref{TRTheorem}.
\end{proof}
\end{theorem}
Relating the correlation function $\Omega_{g,n}$ to the differential forms $\omega_{g,n}$ of TR by 
\begin{align*}
	\Omega_{g,n}(z_1,...,z_n)\prod_{i=1}^ndx(z_i)=\omega_{g,n}(z_1,...,z_n)
\end{align*}
gives the most common representation of TR already given in \eqref{eq:TR-intro}. This finishes the proof that the Langmann-Szabo-Zarembo model is governed by topological recursion.

\section{Complete Solution of the LSZ Model}
\label{sec:compl}
In some sense, we have already written down a complete solution of the LSZ model: Knowing all meromorphic forms $\omega_{g,n}$ by the universal topological recursion formula, it is a feasible task to reconstruct the partition function eq. (\ref{partfunLSZ}) using the genus-summed free energies $\mathcal{F}^{(g)}$ as described in Sec. \ref{ch:setup}. From the matrix model perspective, this is a satisfying answer. However, the LSZ model was designed as a quantum field theoretical toy model and thus we are eventually interested in the explicit solutions of the $2N_1+...+2N_b$-point function - in other words the shape of all correlation functions $G$ of arbitrary boundary structures. 

In this section we will give a simple recursion formula for these most general correlation functions. There basic building blocks are, as indicated already in Fig. \ref{fig:flow}, the generalised 2+...+2-point functions. We thus have to extend the procedure to obtain the Dyson-Schwinger equation for the (generalised) 2-point function - which is straightforward, but lengthy. The complexification also works in exactly the same manner and gives rise to a relatively simple residue formula for their computation that has a clear pictorial interpretation (see \cite{Eynard:2007gw,Branahl:2020yru}). With this formula, all building blocks of aforementioned recursion can be calculated. This recursion will be the main result of this section as it extends the knowledge about the particular form of loop equations also showing up in the 2-matrix model and the generalised Kontsevich model. Due to a plethora of indices and little learning effects, the laborious proofs are shifted into Appendix \ref{sec:appproof} for the sake of readability.

As announced, we start with a more general Dyson-Schwinger equation:

\begin{proposition}
\label{prop:GpqJ}
Let $\mathcal{J}=\{J^2,...,J^b\}$ for $J^s=[p^s,q^s]$.
Then for pairwise different $p,q,p^s,q^s$ one has 
the Dyson-Schwinger equation for the $2+...+2$-point function:
\begin{align}
G_{|pq|\mathcal{J}|}
&= \frac{\delta_{|\mathcal{J}|,0}}{E_p+\tilde{E}_q}
+
\frac{\lambda}{E_p+\tilde{E}_q}\Big\{
-\Omega_p G_{|pq|\mathcal{J}|}
+\frac{1}{N}\frac{\partial G_{|pq|\mathcal{J}|}}{\partial E_p} 
\nonumber
\\
& 
-
\sum_{\substack{\mathcal{J}'\uplus \mathcal{J}''=\mathcal{J}\\
\mathcal{J}'\neq \emptyset}}
T_{p\|\mathcal{J}'|} G_{|pq|\mathcal{J}''|}+ \frac{1}{N}\sum_{\substack{l=1 \\ l\neq p}}^N 
\frac{G_{|lq|\mathcal{J}|}-G_{|pq|\mathcal{J}|}}{E_l-E_p}
\nonumber
\\
&+ \sum_{s=2}^b 
\frac{G_{|p^sq^sp^sq|\mathcal{J}\setminus J^s|}
-G_{|pq^sp^sq|\mathcal{J}\setminus J^s|}
}{E_{p^s}-E_p}  
\Big\}\;.
\label{GpqJ}
\end{align}
 \end{proposition}
In the same manner as for the 2-point function, the repeated application of the boundary creation operator $-N\frac{\partial }{\partial E_{p_i}}$ gives without further effort the following corollary.
\begin{corollary}
	Let $\mathcal{J}=\{J^2,...,J^b\}$ for $J^s=[p^s,q^s]$ and $I=\{p_1,...,p_m\}$.
	Then for pairwise different $p,q,p^s,q^s$ one has 
	the Dyson-Schwinger equation for the generalised $2+...+2$-point function:
\begin{align}
&\Big(\tilde{E}_q+E_p+\frac{\lambda}{N}
\sum_{\substack{l=1\\l\neq p}}^N  \frac{1}{E_l-E_p}\Big)
T_{I\|pq|\mathcal{J}|}
- \frac{\lambda}{N}
\sum_{\substack{l=1 \\ l\notin I,p}}^N
 \frac{T_{I\|lq|\mathcal{J}|}}{E_l-E_p}
\nonumber
\\
&= \delta_{0,|\mathcal{J}|}\delta_{0,|I|}
-
\lambda\bigg\{
\sum_{I'\uplus I''=I}
\Omega_{I',p} T_{I''\|pq|\mathcal{J}|}
-\frac{1}{N}\frac{\partial T_{I\|pq|\mathcal{J}|}}{\partial E_p}
\nonumber
\\
&+
\sum_{\substack{I'\uplus I''=I\\
\mathcal{J}'\uplus \mathcal{J}''=\mathcal{J},~\mathcal{J}'\neq \emptyset}}
\hspace*{-1.5em}
T_{I',p\|\mathcal{J}'|} T_{I''\|pq|\mathcal{J}''|}
+ 
\sum_{i=1}^m \frac{\partial}{\partial E_{p_i}}
\Big(\frac{T_{I\setminus p_i\|p_iq|\mathcal{J}|}}{E_{p_i}-E_p}\Big) \nonumber
\\
&+ \sum_{s=2}^b 
\frac{T_{I\|p^sq^sp^sq|\mathcal{J}\setminus J^s|}
-T_{I\|pq^sp^sq|\mathcal{J}\setminus J^s|}
}{E_{p^s}-E_p}  \Big\}.
\label{DSE-T2long}
\end{align}
\end{corollary}

In analogy to the previous strategy of analytic continuation, we rewrite the (generalised) correlation functions for arbitrary boundary components as
\begin{align}
\label{eq:complexi}
G(E_{p_1^1},\tilde{E}_{q_{1}^{1}},...,\tilde{E}_{q_{N_1}^{1}}|\dots|E_{p_1^b},\tilde{E}_{q_{1}^{b}},...,\tilde{E}_{q^b_{N_b}}) 
&\equiv G_{|p_1^1q_1^1...q^1_{N_1}|\dots|p_1^bq_1^b...q^b_{N_b}|}\;,
\\
T(E_{p_1},...,E_{p_m}\| 
E_{p_1^1},...,\tilde{E}_{q^1_{N_1}}|\dots|E_{p_1^b},...,\tilde{E}_{q^b_{N_b}}) 
&\equiv T_{p_1,\dots,p_m\|p_1^1...q^1_{N_1}|\dots|p_1^b...q^b_{N_b}|}\;. \nonumber
\end{align}
Moreover, we use again the change of variables suggested by the solution of the 2-point function in Theorem~\ref{Thm:InitialData}, giving us again meromorphic functions $\mathcal{G}$ and $\mathcal{T}$ of several complex variables. We thus define:
\begin{definition}
\label{def:complexification} 
We define functions  $\mathcal{G}$ and $\mathcal{T}$ in several complex variables by the variable transforms $x(z)$ and $y(w)$ and the functions $G$ and $T$ complexified in eq. (\ref{eq:complexi}). They have the known formal genus expansion as well:
\begin{align*}
&\mathcal{G}^{(g)}(z_1^1,w_1^1,...,z^1_{N_1},w^1_{N_1}|\dots|
z_1^b,w_1^b,...,z^b_{N_b},w^b_{N_b})\\
&:=
G^{(g)}(x(z_1^1),y(w^1_{1}),...,x(z^1_{N_1}),y(w^1_{N_1})|\dots|
x(z_1^b),y(w^b_{1}),...,x(z^b_{N_b}),y(w^b_{N_b}))\;,
\end{align*}
and
\begin{align*}
&\mathcal{T}^{(g)}(u_1,...,u_m\|z_1^1,...,w^1_{N_1}|\dots|
z_1^b,...,w^b_{N_b}|) \\
&:=
T^{(g)}(x(u_1),...,x(u_m)\|x(z_1^1),...,y(w^1_{N_1})|\dots|
x(z_1^b),...,y(w^b_{N_b}))
\end{align*}
We let
\begin{align*}
&\mathcal{T}^{(g)}(\emptyset\|z_1^1,...,w^1_{N_1}|\dots|
z_1^b,...,w^b_{N_b}|)
:=
\mathcal{G}^{(g)}(z_1^1,...,w^1_{N_1}|\dots|
z_1^b,...,w^b_{N_b})
\end{align*}
 \end{definition}
With this preparation, we can formulate the complexified version of eq. (\ref{DSE-T2long}):
\begin{corollary}
\label{corr:DSEcomplex}
Let $\mathcal{J}=\{J^2,...,J^b\}$ for $J^s=[z^s,w^s]$ 
and $I=\{u_1,...,u_m\}$.
The complexification of Definition~\ref{def:complexification} 
turns after formal genus expansion, inclusion 
of multiplicities $r_i$ of $e_i$ as well as $\tilde{r}_j$ of $\tilde{e}_j$ and the change of variables, which involves 
the rational functions $x$ and $y$ of Theorem~\ref{Thm:InitialData},
the Dyson-Schwinger equation for the generalised $2+...+2$-point function eq. (\ref{DSE-T2long}) into the following equation:
\begin{align}
&(y(w)-y(z)) \mathcal{T}^{(g)}(I\|z,w|\mathcal{J}|)
- \frac{\lambda}{N}
\sum_{k=1}^d
 \frac{r_k \mathcal{T}^{(g)}(I\|\varepsilon_k,w|\mathcal{J}|)}{
x(\varepsilon_k)-x(z)}
\label{DSE-cT2}
\\
&= \delta_{0,|\mathcal{J}|}\delta_{0,|I|}\delta_{g,0}
-
\lambda\bigg\{
\sum_{\substack{I_1\uplus I_2=I\\
g_1+g_2=g,~(g_1,I_1)\neq (0,\emptyset)}} \hspace*{-2em}
\Omega^{(g_1)}_{|I_1|+1}(I_1,z) 
\mathcal{T}^{(g_2)}(I_2\|z,w|\mathcal{J}|)
\nonumber
\\
&+\T^{(g-1)}(I,z\|z,w|\mathcal{J}|)
+
\sum_{\substack{I_1\uplus I_2=I\\
\mathcal{J}_1\uplus \mathcal{J}_2=\mathcal{J},~
\mathcal{J}_1\neq \emptyset\\
g_1+g_2=g}}
\hspace*{-1.5em}
\mathcal{T}^{(g_1)}(I_1,z\|\mathcal{J}_1|) 
\mathcal{T}^{(g_2)}(I_2\|z,w|\mathcal{J}_2|)
\nonumber
\\
& 
+ 
\sum_{i=1}^m \frac{\partial}{\partial x(u_i)}
\Big(\frac{\mathcal{T}^{(g)}(I{\setminus} u_i\|u_i,w|\mathcal{J}|)}{
x(u_i)-x(z)}\Big)
\nonumber
\\
&- \sum_{s=2}^b\frac{
\mathcal{T}^{(g)}(I\|z^s,w^s,z^s,w|\mathcal{J}{\setminus} J^s|)
-
\mathcal{T}^{(g)}(I\|z,w^s,z^s,w|\mathcal{J}{\setminus} J^s|)
}{x(z^s)-x(z)}
\nonumber
\bigg\},
\nonumber
\end{align}
\end{corollary}

The Lagrange interpolation formula (see Lemma \ref{lem:interpol}) together with the solution of the 2-point function allows to solve this equation by a residue formula:

 \begin{proposition} \label{prop2}
 	Let $\big\{\hat{\tilde{w}}^1,...,\hat{\tilde{w}}^d\big\}$ be the solutions of $y(\bullet)=y(w)$. Then, the Dyson-Schwinger equation of Corollary \ref{corr:DSEcomplex} is solved by
 \begin{align*}
 \mathcal{T}^{(g)}&(I\|z,w|\mathcal{J}|)
 =\lambda\mathcal{G}^{(0)}(z,w)\sum_{j=1}^d \Res\displaylimits_{t\to z,\hat{\tilde{w}}^j}\frac{x'(t)\, dt}{(x(z)-x(t))(y(w)-y(t))\mathcal{G}^{(0)}(t,w)} \\
&\times\bigg[\sum_{\substack{I_1\uplus I_2=I\\
g_1+g_2=g,~(g_1,I_1)\neq (0,\emptyset)}} \hspace*{-2em}
\Omega^{(g_1)}_{|I_1|+1}(I_1,t) 
\mathcal{T}^{(g_2)}(I_2\|t,w|\mathcal{J}|)
\nonumber
\\
&+\T^{(g-1)}(I,t\|t,w|\mathcal{J}|)
+
\sum_{\substack{I_1\uplus I_2=I\\
\mathcal{J}_1\uplus \mathcal{J}_2=\mathcal{J},~
\mathcal{J}_1\neq \emptyset\\
g_1+g_2=g}}
\hspace*{-1.5em}
\mathcal{T}^{(g_1)}(I_1,t\|\mathcal{J}_1|) 
\mathcal{T}^{(g_2)}(I_2\|t,w|\mathcal{J}_2|)
\nonumber
\\
& 
+ 
\sum_{i=1}^m \frac{\partial}{\partial x(u_i)}
\Big(\frac{\mathcal{T}^{(g)}(I{\setminus} u_i\|u_i,w|\mathcal{J}|)}{
x(u_i)-x(t)}\Big)
\nonumber
\\
&- \sum_{s=2}^b\frac{
\mathcal{T}^{(g)}(I\|z^s,w^s,z^s,w|\mathcal{J}{\setminus} J^s|)
-
\mathcal{T}^{(g)}(I\|t,w^s,z^s,w|\mathcal{J}{\setminus} J^s|)
}{x(z^s)-x(t)}\bigg]
\nonumber
 \end{align*}
 \end{proposition}

The result of Proposition \ref{prop2} can be grasped in a more compact way by rewriting two terms:
\begin{corollary}\label{cor2+} 	 
Proposition \ref{prop2} is equivalent to
\begin{align*}
\mathcal{T}^{(g)}(I\|z,w|\J)
&= \sum_{s=2}^b \sum_{j=1}^d \sum_{i=1}^m\Res\displaylimits_{t\to z,\hat{\tilde{w}}^j,u_i,z^s}
  \frac{\lambda\mathcal{G}^{(0)}(z,w)  x'(t)\, dt}{(x(z)-x(t))(y(w)-y(t))\mathcal{G}^{(0)}(t,w)} 
\\
&\times\bigg[\sum_{\substack{I_1\uplus I_2=I\\
\mathcal{J}_1\uplus \mathcal{J}_2=\mathcal{J}\\
g_1+g_2=g}}^\prime
\mathcal{T}^{(g_1)}(I_1,t\|\mathcal{J}_1|) 
\mathcal{T}^{(g_2)}(I_2\|t,w|\mathcal{J}_2|)
\nonumber
\\
&\qquad +\T^{(g-1)}(I,t\|t,w|\mathcal{J}|)
+\sum_{s=2}^b\frac{
\mathcal{T}^{(g)}(I\|t,w^s,z^s,w|\mathcal{J}{\setminus} J^s|)
}{x(z^s)-x(t)}
		\bigg],
	\end{align*}
where the primed sum excludes $(g_1,I_1,\mathcal{J}_1)=(0,\emptyset,\emptyset)$.
\end{corollary}
\begin{remark}\label{rmk:2MMH}
	This residue formula has an intuitive geometric interpretation that the interested reader can find in \cite[eq. (4-1) and (4-2)]{Eynard:2007gw}. We have a one-to-one correspondence to the 2-matrix model. Due to our restricitions to boundary lengths of 2, the first two visualised summands in \cite[eq. (4-2)]{Eynard:2007gw} do not show up here, but have to be added when going to arbitrary $N_s$. The generalised correlation functions $\mathcal{T}^{(g)}(I\|z,w|\J)$ correspond to the $H^{(g)}_{k_L;m;n}(S_1,...,S_l;p_1,...,p_m;q_1,...,q_n)$, where the boundaries in $\J$ correspond to $\{S_2,...,S_l\}$ and the set of marked points $u_i \in I$ to $p_i$. Another set of marked points $q_j$ could be generated by acting with $\frac{\partial}{\partial \tilde{E}_q}$on $G_{...}$. We especially refer to eq. (3-9) in \cite{Eynard:2007gw} to compare the complete structural equivalence to our 2-point function: $\mathcal{G}^{(0)}(z,w) \equiv H^{(0)}_{1;0;0}(z,w)$.
\end{remark}
We now have collected all components necessary to write the $2N_1+...+2N_b$-point function in terms of polynomials in (2+...+2)-point functions as well as denominators $\frac{1}{E_k-E_l}$. The complete solution of the LSZ model in the quantum field theoretical perspective reads:
\begin{theorem}
\label{thm:complete}
Let  $\J=\{J^2,...,J^b\}$, $J^\beta=\{p_1^\beta, q_1^\beta,...p_{N_\beta}^\beta, q_{N_\beta}^\beta \}$ and $\beta \in \{2,...,b\}$. The generalised $2N_1+...+2N_b$-point function satisfies the following recursive equation:
 \begin{align*}
  &T^{(g)}_{I\|p_1^1 q^1_1..q^1_{N_1}|\mathcal{J}|}=-\frac{\lambda}{\tilde{E}_{q^1_1}-\tilde{E}_{q^1_{N_1}}}\bigg\{
  \sum_{k=2}^{N_1}\frac{T^{(g-1)}_{I\|p^1_kq^1_{1}..q^1_{k-1}|p^1_1q^1_k..q^1_{N_1}|\mathcal{J}|}
 -T^{(g-1)}_{I\|p^1_1q^1_1p^1_2..q^1_{k-1}|p^1_{k}..q^1_{N_1}|\mathcal{J}|}}{E_{p^1_k}-E_{p^1_1}}\\ 
 +&\sum_{\beta=2}^{b}\sum_{k=1}^{N_\beta}\frac{T^{(g)}_{I\|p_1^\beta q_1^\beta..p_k^\beta q^1_1p^1_2..q^1_{N_1}
 p^1_1 q_{k}^\beta..q^\beta_{N_\beta}|\mathcal{J}\backslash \{J^\beta\}|}-
 T^{(g)}_{I\|p_1^\beta q_1^\beta..q_{k-1}^\beta p^1_1 q^1_1..q^1_{N_1}
  p_{k}^\beta..q^\beta_{N_\beta}|\mathcal{J}\backslash \{J^\beta\}|}}{E_{p^\beta_k}-E_{p^1_1}}\\
  +&\sum_{k=2}^{N_1}
  \sum_{\substack{\mathcal{J}'\uplus \mathcal{J}''=\mathcal{J}\\I_1\uplus I_2=I\\h+h'=g}}   \frac{T^{(h)}_{I_1\|p^1_kq^1_{1}..q^1_{k-1}|\mathcal{J}'|} 
 T^{(h')}_{I_2\|p^1_1q^1_k..q^1_{N_1}|\mathcal{J}''|}-
 T^{(h)}_{I_1\|p^1_1 p^1_2..q^1_{k-1}|\mathcal{J}'|} T^{(h')}_{I_2\|p^1_{k}..q^1_{N_1}|\mathcal{J}''|}}{E_{p^1_k}-E_{p^1_1}}\bigg\}
 \end{align*}
\end{theorem}

\begin{remark}
	As mentioned before, we have a direct correspondence of all correlation functions for the LSZ model in terms of $\mathcal{T}$ to the hermitian 2-matrix model (see Remark \ref{rmk:2MMH}). However,  the algebraic recursive equation of Theorem \ref{thm:complete} was never proved for the 2-matrix model. It is straightforward to write the theorem in terms of meromorphic functions $\mathcal{T}$ with $x(z), y(w)$. Therefore, this algebraic formula should also hold in the 2-matrix model. Furthermore, cyclic symmetry within a boundary, i.e. 
	\begin{align*}
		T^{(g)}_{I\|p_1^1 q^1_1p_2^1q_2^1..q^1_{N_1}|\mathcal{J}|}=T^{(g)}_{I\|p_2^1q_2^1..q^1_{N_1}p_1^1 q^1_1|\mathcal{J}|}
	\end{align*}
	is a direct consequence of the equation of Theorem \ref{thm:complete}, which was also never proved for general correlation functions in the 2-matrix model, but should obviously hold by definition.
\end{remark}

Thus, we have a concrete algorithm to derive all correlation functions building on topological recursion. The algorithm is recursive in the negative Euler characteristic $-\chi=2g+n-2$. For any $2N_1+...+2N_b$-point function, Theorem \ref{thm:complete} has to be applied until one ends up just with $2+...+2$-point functions. For these, Corollary \ref{cor2+} applies, which includes recursively further $2+...+2$-point functions and $\Omega_{g,n}$ of less topology or equal Euler characteristic. The $\Omega_{g,n}$ are derived by topological recursion, Theorem \ref{thm:TROm}, where the initial data is provided by Theorem \ref{Thm:InitialData}.

We end with an example of the above recursion formula being in particular interesting from combinatorial perspective:
\begin{example}
There is a particular example in the planar sector which analogously occurs for hermitian fields \cite{DeJong} or in the 2-matrix model \cite{Eynard:2005iq}: The recursion boils down for the planar $2N$-point function to:
\begin{align*}
  G^{(0)}_{|p_1 q_1..q_{N}|}=&-\lambda
  \sum_{k=1}^{N-2}
  \frac{G^{(0)}_{|p_1q_{k+1}..q_{N}|} 
  G^{(0)}_{|p_{k+1}q_1..q_k|}-
 G^{(0)}_{|p_{k+1}..q_{N}|} G^{(0)}_{|p_1 q_1..q_{k}|}}{(E_{p_{k+1}}-E_{p_1})(\tilde{E}_{q_1}-\tilde{E}_{q_{N}})},
	\end{align*}
The explicit structure of the produced polynomials follow a remarkable combinatorial pattern based on nested Catalan tuples that was unraveled in \cite{DeJong}. This combinatorial pattern might by generalised to higher genus and more boundaries by finding an explicit structure for the recursion of Theorem \ref{thm:complete}.
\end{example}
We have now reached all solutions of correlators in the LSZ model in their full complexity. Some simple topologies shall now be compared with known limits of this model.

\section{Cross Checks}
\label{ch:cross}

In this section we take the obtained results from the LSZ model and compare them with known results from other models that can be reached by certain limits. Setting $\tilde{E}=0$, we get back the original model the authors of \cite{Langmann:2003if} were interested in - since they introduced the additional external field $\tilde{E}$ only as an auxiliary quantity.  Setting $\tilde{E}=E$, the Quartic Kontsevich Model (QKM) is obtained for the first simple topologies (for higher topologies the combinatorics changes drastically). Finally, we perform a perturbative analysis for both complex and hermitian fields $\Phi$ and compare the different classes of ribbon graphs that are generated. In particular, we relate them to known results from enumerative geometry for the complex and hermitian 1-matrix model, arising in the \textit{combinatorial limit} of a single $N$-fold degenerate eigenvalue $e$ and $\tilde{e}$, respectively (in other words $d=\tilde{d}=1$).

\subsection{Recovering the Results of Langmann-Szabo-Zarembo \cite{Langmann:2003if}}
As already mentioned, the original work of Langmann, Szabo and Zarembo used the external matrix $\tilde{E}$ as a regulator, which was switched off for evaluating the exact solution of the correlators. In \cite{Langmann:2003if},  Langmann, Szabo and Zarembo have computed the solution of $\Omega^{(0)}_p$ with classical techniques like the  Harish-Chandra-Itzykson-Zuber formula \cite{Harish57,Itz1980} together with the Riemann-Hilbert equation. An important observation was that their ''master equation'' was of the same type as in the Kontsevich-Penner model, which had a universal procedure to solve it. This procedure relies essentially on the fact of having a one-cut solution. 

We will see that this is just possible if all eigenvalues of $\tilde{E}$ coincide. For the specific case of $\tilde{E}=0$, the function $x(z)$ of Theorem \ref{Thm:InitialData} breaks down to a degree-two function in $z$. Its inverse has just two branches so that only one cut occurs. In particular, we have in this case from our theorem with $\tilde{d}=1$
\begin{align*}
	x(z)=z-\frac{\lambda}{y'(\tilde{\varepsilon})(z-\tilde{\varepsilon})},\qquad y(z)=-z+\frac{\lambda}{N}\sum_{n=1}^{d}\frac{r_n}{x'(\varepsilon_n)(z-\varepsilon_n)},
\end{align*}
where $0=y(\tilde{\varepsilon})$ and $e_n=x(\varepsilon_n)$. Let now 
\begin{align*}
	b_1+b_2=2\tilde{\varepsilon},\qquad b_1-b_2=4\sqrt{-\frac{\lambda}{ y'(\tilde{\varepsilon})}},
\end{align*}
then we can invert $x(z)$ to
\begin{align*}
	z(x)=\frac{x+\frac{b_1+b_2}{2}\pm \sqrt{(x-b_1)(x-b_2)}}{2}.
\end{align*}
It is straightforward to check the following relations
\begin{align*}
	x'(z)^2(\tilde{\varepsilon}-z)^2=&(x(z)-b_1)(x(z)-b_2) \\  \bigg(\frac{2 (x(z)-e_n)}{x'(\varepsilon_n)(z-\varepsilon_n)}-\frac{x(z)-e_n}{x'(\varepsilon_n)(\tilde{\varepsilon}-\varepsilon_n)}-1\bigg)^2=&\frac{(x(z)-b_1)(x(z)-b_2)}{(e_n-b_1)(e_n-b_2)}
\end{align*}
which implies (by taking the right square root branch) for $\Omega^{(0)}_1$
\begin{align*}
	&-\frac{y(z)+V'(x(z))}{\lambda}=\frac{z-x(z)}{\lambda}+\frac{1}{N}\sum_{n=1}^d r_n\bigg(\frac{1}{x(z)-e_n}-\frac{1}{x'(\varepsilon_n)(z-\varepsilon_n)}\bigg)\\
	=&\frac{\sqrt{(x-b_1)(x-b_2)}-x}{2\lambda}+\frac{1}{2N}\sum_{n=1}^d r_n\bigg(\frac{1}{x-e_n}-\frac{\sqrt{(x-b_1)(x-b_2)}}{(x-e_n)\sqrt{(e_n-b_1)(e_n-b_2)}}\bigg).
\end{align*}
This result is in perfect coincidence with the exact solution provided by \cite[eq. (D.14)]{Langmann:2003if} (up to a global sign). The two implicit constraints on $b_1$ and $b_2$ in \cite[eq. (D.15)]{Langmann:2003if} are in one-to-one correspondence to $0=y(\tilde{\varepsilon})$ and $b_1-b_2=4\sqrt{-\frac{\lambda}{ y'(\tilde{\varepsilon})}}$.
From this analysis, we see that the solution of Langmann, Szabo and Zarembo is a very specific case of the result provided by Theorem \ref{Thm:InitialData}.

\subsection{Analytical Comparison with the Hermitian Model (\ref{partfunQKM})}
\begin{figure}[h!]
  \centering
    \includegraphics[width= 0.99\textwidth]{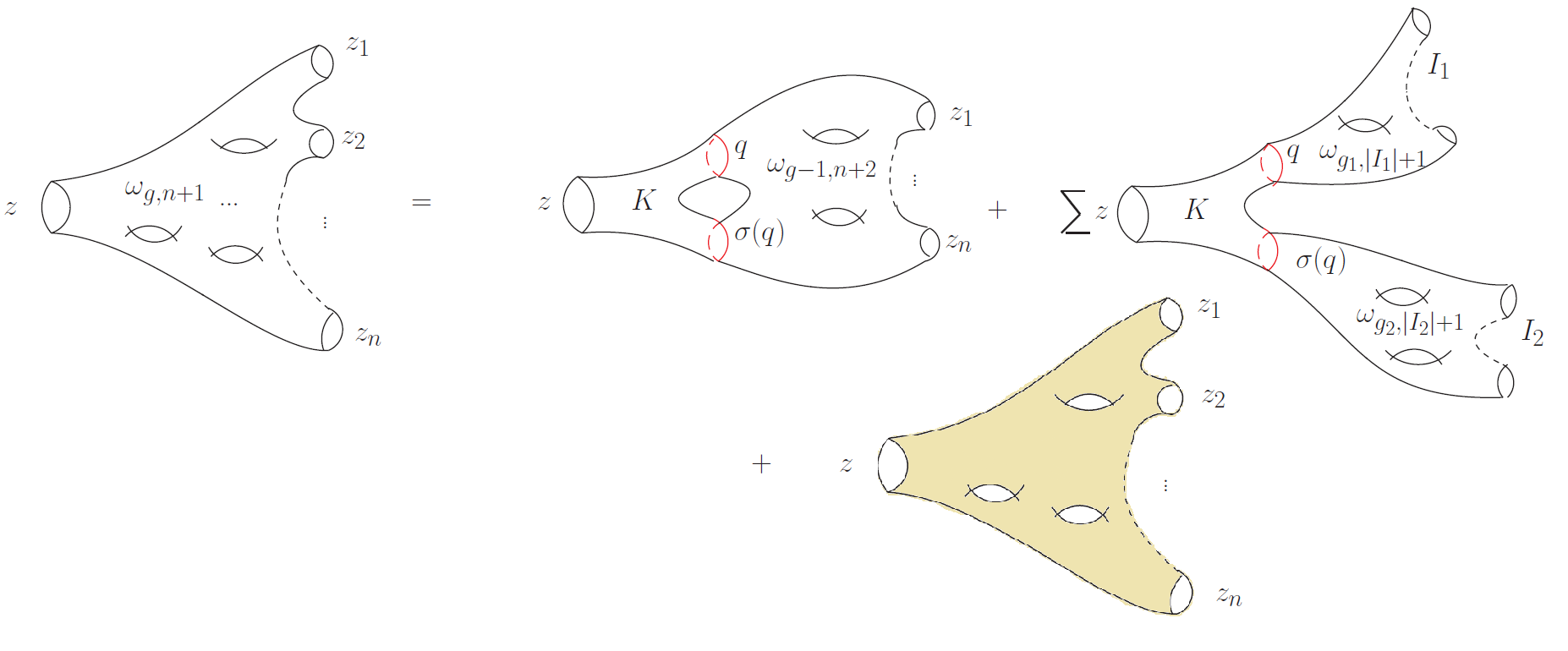} 
    \caption{Graphical interpretation of blobbed topological recursion: The usual recursion formula is enriched by a holomorphic (at ramification points) term  $\mathcal{H}_z\omega_{g,n+1}$ (coloured) that appears as a surplus structure in the solution of the loop equations. It has to be seen as additional data that has to be taken into account at each further recursion step. 
      \label{diagrams}}
\end{figure}

This subsection is dedicated to a structural comparison of the different types of topological recursion solving the models defined by eq. (\ref{partfunLSZ}) and  eq. (\ref{partfunQKM}) with complex and hermitian fields, respectively. Although looking at hermitian matrices $\Phi=\Phi^\dagger$ at first sight seems to reduce the complexity of the model, the contrary is the case. Performing very similar steps as for the solution of the LSZ model, we end up in the hermitian model (QKM) of  eq. (\ref{partfunQKM})  with meromorphic forms which has additional  properties. These properties can be described by three facts:
\begin{itemize}
\item having only one external field $E$, $x(z)$ and $y(z)$ boil down to \cite[Thm. 4.1]{Branahl:2021}:
\begin{align*}
x(z)=R(z)=z-\sum_{k=1}^d \frac{r_k}{x'(\varepsilon_k)(z+\varepsilon_k)} \, , \qquad y(z)=-R(-z)\; .
\end{align*}
This limit is reached by $\varepsilon_k = -\tilde{\varepsilon}_k$ implied by $E=\tilde{E}$. It is decisive that both $x$ and $y$ can be expressed by the same change of variables $R(z)$, such that we have a global reflection symmetry $x(z)=-y(-z)$. In particular, this implies e.g. $\mathcal{G}^{(0)}(z,w)=\mathcal{G}^{(0)}(w,z)$
\item $\omega_{0,2}$ is enlarged by a "reflected" Bergman kernel: $\omega_{0,2}(z_1,z_2)=B(z_1,z_2)-B(z_1,-z_2)$
\item $\omega_{g,n+1}(z,z_1,...,z_n)$ has poles on the antidiagonals $z=-z_i$, for $g>0$,  $\omega_{g,n+1}(z,z_1,...,z_n)$ has additionally poles at $z=0$
\end{itemize}
Altogether, these three facts suggest to turn to the more general framework of \textit{blobbed topological recursion} (BTR), developed in \cite{Borot:2015hna}. This framework finds application in the solution of models in which the abstract loop equations are fulfilled, but additional terms with poles away from the ramification points show up. This suggests to decompose every $\omega_{g,n}$ into a holomorphic (at the ramification points $\beta_i$) part  $\mathcal{H}\omega_{g,n}$ and a polar part $\mathcal{P}\omega_{g,n}$. The holomorphic parts may be interpreted as an infinite tower of surplus initial data, that cannot be handled with usual topological recursion. Instead, they successively contribute at each recursion step in the Euler characteristic and mix with polar contributions in the recursion formula. A visualisation of this phenomenon is depicted in Fig. \ref{diagrams}.

Historically, the treatment of so-called \textit{stuffed maps} first stressed the necessity of this enlarged framework \cite{Borot:2013fla}. The detailed analysis of the loop equations of (so far) the planar sector of the QKM allowed for a concrete model-specific recursion, having astonishing similarities to the known formula generating the polar part \cite{Hock:2021tbl}. The much more complicated non-planar sector is still work in progress, only some first examples were explicitly worked out. We do not want to review all these accomplishments and rather refer to the article \cite{Branahl:2021} being readable also for beginners in both noncommutative quantum field theory and topological recursion. Instead, we want to analyse the differences in the Dyson-Schwinger equations for the $\Omega_{q_1,...,q_n}^{(g)}$ and the reasons for it. The three above-mentioned facts correspond to the following phenomenological differences:
\begin{itemize}
\item Having only one external field $E$, the boundary creation operator $-N \frac{\partial}{\partial E_p}$ additionally hits the second argument of the 2-point function $G_{|ak|}$ in $\Omega_{a}=\frac{1}{N}\sum_{k} G_{|ak|}$. This has obviously an impact on the exact solution given by the ramified cover $y(z)_{\tilde{E}=E}=-R(-z)$ - adding a second term to eq. (\ref{DSE-Om}) and thus giving rise to the extended Bergman kernel. This does not happen in the LSZ model, where we have distinguished dependencies $E_k,\tilde{E}_l$.
\item The hermitian model allows objects like a 1+1-point function (first solved in \cite{Schurmann:2019mzu}), and in general any partition function with odd boundary lengths, as long as their sum is even, again -  adding a third term to eq. (\ref{DSE-Om}). We will see that also perturbatively, these objects are forbidden in the complex model.
\end{itemize}
These two observations lead to the following Dyson-Schwinger equation \cite{Branahl:2020yru}:
\begin{align}
&\Omega^{(QKM)}_{I,q}
= \frac{\delta_{|I|,1}}{
(E_{q_1}-E_q)^2}
+
\frac{1}{N}\sum_{\substack{l=1\\l\notin I}}^NT_{I\|ql|}
\textcolor{red}{- \sum_{j=1}^m 
\frac{\partial T_{I\setminus q_j\|qq_j|}}{\partial E_{q_j}} 
+\frac{1}{N^2} T_{I\|q|q|}\;,}
\label{DSE-OmQKM}
\end{align}
Comparing with the proof of the Bergman kernel $B(z_1,z_2)$ in the LSZ model in App. \ref{sec:om02}, the first new term compared to eq. (\ref{DSE-Om}) gives rise to the pole on the antidiagonal being responsible for the additional $-B(z_1,-z_2)$, the first blob in blobbed topological recursion of the QKM:
\begin{align}
&\Omega^{(QKM)}_{0,2}(z_1,z_2)
=\frac{1}{x'(z_1)x'(z_2)}\bigg( \frac{1}{(z_1-z_2)^2}+\frac{1}{(z_1+z_2)^2}\bigg).
\end{align}
 This structure inherits to the whole planar sector with more holomorphic parts with poles of  $\omega_{0,n+1}(z,z_1,...,z_n)$ on the antidiagonals $z=-z_i$. We see that the second new term is suppressed by a factor $\frac{1}{N^2}$, thus it only contributes for $g>0$. The evaluation of the 1+1-point function on the diagonal then gives rise to poles at $z=0$, the fixed point of the global reflection symmetry in the spectral curve, $x(z)=R(z)=-y(-z)$ (see above). Note that these two kinds of poles emerge from two important differences between the LSZ model and the QKM: The coexistence of one single external field together with the property of the quantum fields to be hermitian leads to the recursive structure of the QKM in its full complexity. We will see in the coming subsection that these consequences have clear interpretation in perturbation theory.

\subsection{Perturbation Theory}
Where in physical quantum field theory, perturbative expansions are often the only way to produce results, they become in toy models a powerful tool to justify the calculations for the exact solutions. Comparing hermitian and complex fields seen from the perturbative perspective is worth an investigation as well.

Matrix models of the considered type are perturbatively (or combinatorially) expanded into ribbon graphs, which is known since the 70's \cite{tHooft:1973alw,Brezin:1977sv}. The external matrix $E$ (and $\tilde{E}$) associates an additional weight for the free propagator. We refer to \cite{Branahl:2020uxs} for some more rigorous details on the hermitian model and try to summarize the most important facts. 

\begin{figure}[h!]
	\centering
	\includegraphics[width= 0.99\textwidth]{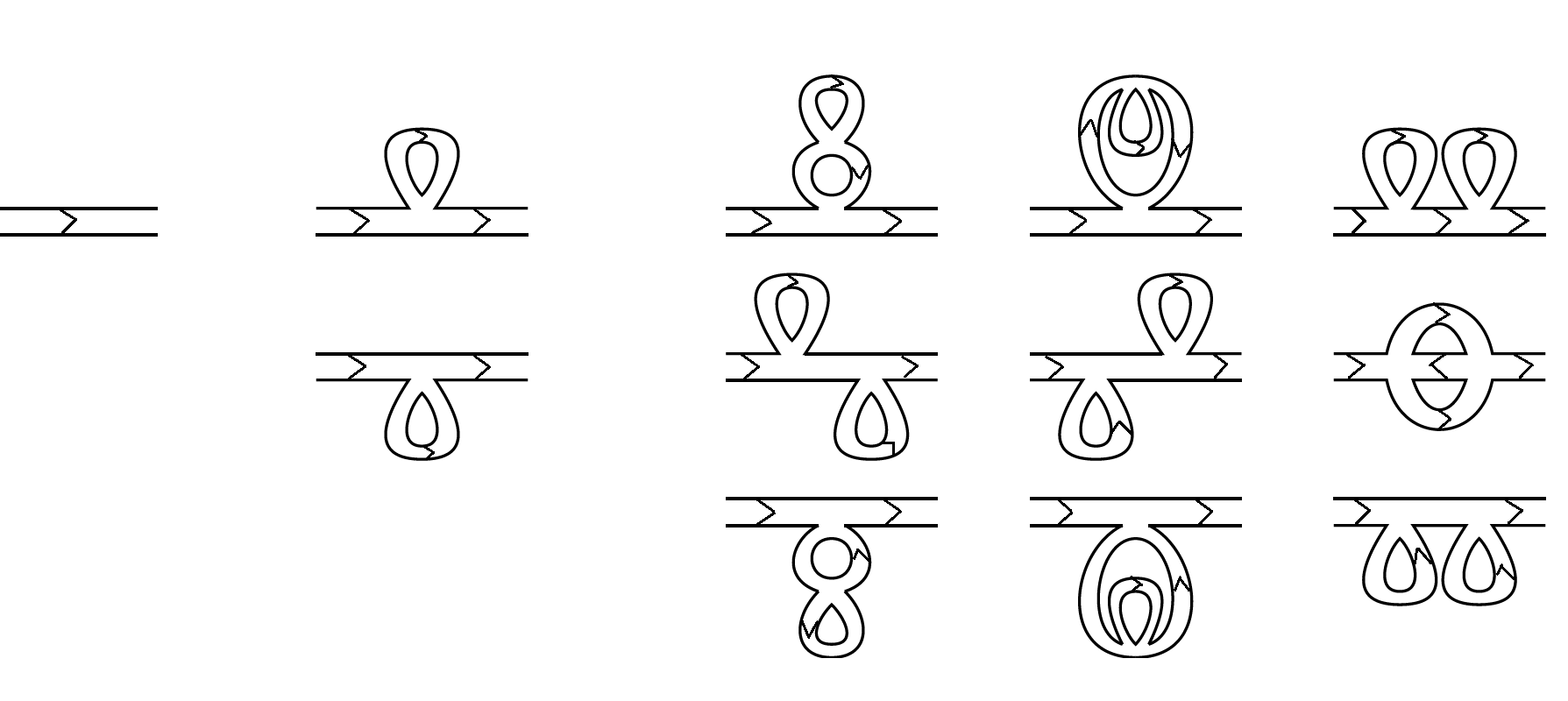} 
	\caption{Ribbon graphs contributing to the planar 2-point function of the quartic LSZ model up to order $\mathcal{O}(\lambda^2)$. At order $\lambda^k$, a total amount of $\frac{2 \cdot 3^k(2k)!}{k!(k+2)!}$ ribbon graphs contributes.
		\label{2p}}
\end{figure}

\begin{figure}[h!]
	\centering
	\includegraphics[width= 0.75\textwidth]{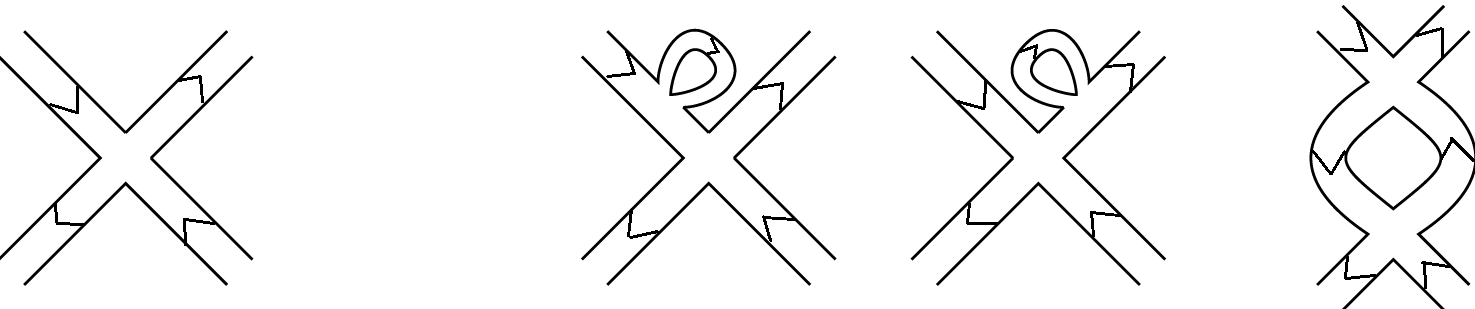} 
	\caption{Ribbon graphs contributing to the planar four-point function of the quartic LSZ model up to order $\mathcal{O}(\lambda^2)$,  where the first two
		graphs of order $\lambda^2$ contribute with four different labellings and the last graph with two different labellings (after probably inverting the orientation).
		\label{4p}}
\end{figure}

Perturbation theory is the formal expansion of the initially introduced cumulants (see eq. \ref{eq:FullExpectation}) by interchanging the integration with the series in the coupling constant $\lambda$:
\begin{align}
	&\langle \Phi^\dagger_{q_1p_1}\Phi_{p_2q_2}... \Phi^\dagger_{q_{n-1}p_{n-1}}\Phi_{p_nq_n}\rangle_c 
	\nonumber
	\\\label{cumulants-1}
	&= \sum_{v=0}^\infty \frac{N^v(-\lambda)^v}{2^vv!}
	\Big[\int_{C_N}\,d\Phi\, d\Phi^\dagger\, e^{-N\mathrm{Tr}(E\Phi^\dagger\Phi +\tilde{E}\Phi\Phi^\dagger)}\\\nonumber
	&\qquad \qquad\qquad  \times
	\Phi^\dagger_{q_1p_1}\Phi_{p_2q_2}... \Phi^\dagger_{q_{n-1}p_{n-1}}\Phi_{p_nq_n} \!\!\!\!\!
	\sum_{j_1,...,m_v=1}^N
	\prod_{i=1}^v\big(\Phi_{j_ik_i}\Phi^\dagger_{k_il_i}\Phi_{l_im_i}\Phi^\dagger_{m_ij_i}\big)\Big]_c,
\end{align}
Performing the integration gives a graphical interpretation in terms of ribbon graphs with labelled strands.
The translation into graphs is performed by substituting a pair of fields by a ribbon (Wick contraction), and the cycle $\Phi_{j_ik_i}\Phi^\dagger_{k_il_i}\Phi_{l_im_i}\Phi^\dagger_{m_ij_i}$ (coming from the expanded interaction $S_{int}$) forming a 4-valent vertex. First recall the free propagator for \textit{hermitian matrix} $\Phi_{ij}\Phi_{ji}$ is a ribbon with strands labelled by $i,j$ ,and the four-valent vertex $\lambda \Phi_{ij}\Phi_{jk}\Phi_{kl}\Phi_{li}$ has four strands labelled by $i,j,k,l$ drawn as (see\cite{Branahl:2020uxs}):
\begin{align*}
	\begin{picture}(30,15)
	\put(0,5){\line(1,0){30}}
	\put(0,9){\line(1,0){30}}
	\put(2,-3){\mbox{\scriptsize$j$}}
	\put(2,11){\mbox{\scriptsize$i$}}
	\put(28,-3){\mbox{\scriptsize$j$}}
	\put(28,11){\mbox{\scriptsize$i$}}
	\end{picture}   \qquad \qquad  \qquad  \begin{picture}(24,19)
	\put(0,6){\line(1,0){10}}
	\put(0,10){\line(1,0){10}}
	\put(24,6){\line(-1,0){10}}
	\put(24,10){\line(-1,0){10}}
	\put(10,-4){\line(0,1){10}}
	\put(14,-4){\line(0,1){10}}
	\put(10,20){\line(0,-1){10}}
	\put(14,20){\line(0,-1){10}}
	\put(5,-9.5){\mbox{\scriptsize$j$}}
	\put(15,-9.5){\mbox{\scriptsize$k$}}
	\put(14,21.5){\mbox{\scriptsize$l$}}
	\put(3,21.5){\mbox{\scriptsize$i$}}
	\put(-10,0){\mbox{\scriptsize$j$}}
	\put(-10,8){\mbox{\scriptsize$i$}}
	\put(26,0){\mbox{\scriptsize$k$}}
	\put(26,8){\mbox{\scriptsize$l$}}
	\end{picture} 
\end{align*}
Looking instead at \textit{complex matrices} with $\Phi, \Phi^\dagger$, the ribbons have additional structure in terms of an direction (depicted by arrows giving an direction to the strands). We will call these graphs \textit{directed ribbon graphs}. $\Phi^\dagger$ will represent an ingoing arrow, whereas the adjoint matrix $\Phi$ will represent an outgoing arrow. Thus, they always have to appear in pairs (otherwise the integation will give zero) like the free propagator $\Phi^\dagger_{ji}\Phi_{ij}$ and four-valent vertex $\lambda \Phi_{ij}\Phi_{jk}^\dagger\Phi_{kl}\Phi_{li}^\dagger$:
\begin{align*}
	\begin{picture}(30,15)
	\put(0,5){\line(1,0){30}}
	\put(0,9){\line(1,0){30}}
	\put(8,7){\vector(1,0){15}}
	\put(2,-3){\mbox{\scriptsize$j$}}
	\put(2,11){\mbox{\scriptsize$i$}}
	\put(28,-3){\mbox{\scriptsize$j$}}
	\put(28,11){\mbox{\scriptsize$i$}}
	\end{picture}   \qquad \qquad  \qquad  \begin{picture}(24,19)
	\put(0,6){\line(1,0){10}}
	\put(0,10){\line(1,0){10}}
	\put(24,6){\line(-1,0){10}}
	\put(24,10){\line(-1,0){10}}
	\put(10,-4){\line(0,1){10}}
	\put(14,-4){\line(0,1){10}}
	\put(10,20){\line(0,-1){10}}
	\put(14,20){\line(0,-1){10}}
	\put(5,-9.5){\mbox{\scriptsize$j$}}
	\put(15,-9.5){\mbox{\scriptsize$k$}}
	\put(14,21.5){\mbox{\scriptsize$l$}}
	\put(3,21.5){\mbox{\scriptsize$i$}}
	\put(-10,0){\mbox{\scriptsize$j$}}
	\put(-10,8){\mbox{\scriptsize$i$}}
	\put(26,0){\mbox{\scriptsize$k$}}
	\put(26,8){\mbox{\scriptsize$l$}}
	\put(25,8){\vector(-1,0){9}}
	\put(0,8){\vector(1,0){9}}
	\put(12,12){\vector(0,1){9}}
	\put(12,4){\vector(0,-1){9}}
	\end{picture}  
\end{align*}
All directed ribbon graphs of the LSZ model have either 1- or 4-valent vertices. For the cumulant \eqref{cumulants-1}, we would have $n$ 1-valent vertices associated to the $\Phi_{p_iq_i},\Phi^\dagger_{q_ip_i}$, where the structure and the orientation is fixed by  $\Phi^\dagger_{q_{2i+1}p_{2i+1}}$ giving ingoing arrows and $\Phi_{p_{2i}q_{2i}}$ giving outgoing arrows. Consequently, all strands are either clockwise or counter-clockwise directed due to the arrows. We associate to a ribbon with counter-clockwise directed strand labelled by $i$ and clockwise directed strand labelled by $j$ the free propagator $\frac{1}{E_i+\tilde{E}_j}$ arising from the integration of a pair $\Phi^{\dagger}_{ji}\Phi_{ij}$. Letting now $\pi\in \mathcal{S}_n$ be a permutation $p_{2i+1}=\pi(p_{2i+1})$  and $q_{2i}=\pi(q_{2i})$, we state the fairly well-known (Feynman-)expansion with the additional structure entailed by the LSZ model (including multiplicities $r_i,\tilde{r}_j$ of eigenvalues $e_i,\tilde{e}_j$)
\begin{proposition}\label{prop:weight}
	Let $p_1,q_1,...,p_n,q_n$ be pairwise different and
	$\mathfrak{G}^{v,\pi}_{p_1,q_1,...,p_n,q_n}$ be the set of labelled
	connected and directed ribbon graphs with $v$ four-valent vertices and $n$
	one-valent vertices labelled $(q_1,\pi(p_1)), (p_1,\pi(q_2)) \dots,
	(q_{n-1},\pi(p_{n-1})),(p_{n},\pi(q_{n}))$, where $p_i$ corresponds to the counter-clockwise directed strands and $q_i$ to clockwise. Then for $n$ even, the integral \eqref{cumulants-1} evaluates to
	\begin{align}
	N^n  \langle \Phi^\dagger_{q_1\pi(p_1)}\Phi_{p_2\pi(q_2)}... \Phi^\dagger_{q_{n-1}\pi(p_{n-1})}\Phi_{p_n\pi(q_n)}\rangle_c 
	=\sum_{v=0}^\infty \sum_{\Gamma \in \mathfrak{G}^{v,\pi}_{p_1,q_1,...,p_n,q_n}}
	N^{v-r+n+s(\Gamma)} \varpi(\Gamma)\;,
	\label{cumulants-2}
	\end{align}
	where $r=2v+n/2$ is the number of ribbons,
	$s_1(\Gamma)$ the number of counter-clockwise directed loops
	and $s_2(\Gamma)$ the number of clockwise directed loops in $\Gamma$
	 and
	the weight $\varpi(\Gamma)$ is derived from the following Feynman rules:
	\begin{itemize} 
		\item label the $s_1=s_1(\Gamma)$ counter-clockwise directed loops by $k_1,...,k_{s_1}$, and the $s_2=s_2(\Gamma)$ clockwise directed loops by $l_1,...,l_{s_2}$;
		\item associate a factor $-\lambda$ to a 4-valent ribbon-vertex;
		\item associate the factor $\frac{1}{e_p+\tilde{e}_q}$ to a ribbon with counter-clockwise directed
		strand labelled by $p$ and clockwise directed strand labelled by $q$;
		\item multiply all factors and apply the summation 
		operator
		\begin{align*}
			\frac{1}{N^{s_1+s_2}}\sum_{k_1,..,k_{s_1},l_1,...,l_{s_2}=1}^{d,\tilde{d}}r_{k_1}..r_{k_{s_1}}\tilde{r}_{l_1}...\tilde{r}_{l_{s_2}}.
		\end{align*}
	\end{itemize}
\end{proposition}
The permutation $\pi\in \mathcal{S}_n$ defines uniquely a cycle-type (see \cite{Branahl:2020uxs} for the cycle-type reconstruction from $\pi$). Thus, Proposition \ref{prop:weight} gives via equation \eqref{defG} the graphical expansion of $G_{|...|}$ as a generating function of directed ribbon graphs. 
\begin{example}\label{ex:2ppert}
	The first  ribbon graphs contributing to the 2-point and 4-point function are shown in Fig. \ref{2p} and \ref{4p}. Label the counter-clockwise directed strands by $p_i$ and clockwise directed by $q_i$. Up to order $\lambda^1$, we get with Proposition \ref{prop:weight}
	\begin{align}\nonumber
		G_{|pq|}= & \frac{1}{N}\frac{\partial^2}{\partial J_{pq}\partial J^\dagger_{qp}}\log \mathcal{Z}(J,J^\dagger)\vert_{J=0}=N\langle \Phi^\dagger_{qp}\Phi_{pq}\rangle\\
		=&\frac{1}{e_p+\tilde{e}_q}- \frac{\lambda}{(e_p+\tilde{e}_q)^2}\frac{1}{N}\bigg(\sum_{l=1}^{\tilde{d}}\frac{\tilde{r}_l}{e_p+\tilde{e}_l}+\sum_{k=1}^{d}\frac{r_k}{\tilde{e}_q+e_k}\bigg)+\mathcal{O}(\lambda^2),\\
		G_{|p_1q_1p_2q_2|}=&\frac{-\lambda}{(e_{p_1}+\tilde{e}_{q_1})(e_{p_2}+\tilde{e}_{q_1})(e_{p_2}+\tilde{e}_{q_2})(e_{p_1}+\tilde{e}_{q_2})}+\mathcal{O}(\lambda^2).
	\end{align}
	The next order can easily read off the figures.
\end{example}

Throughout this article, the main insight of the algebraic structure of the exact solutions is that the correlators $\Omega_{p_1,...,p_n}$ are much simpler from analytic perspective. However, from physics perspective the correlation functions $G_{|...|}$ have the most natural significations and the most common perturbative expansion. We focus on the difference between $\Omega$ and $G$ on the perturbative level and demonstrate the perturbative interpretion of $\Omega$.
In doing so, we illustrate the action of the boundary creation operator (see Definition \ref{def:gDSE}), which is by considering multiplicities $r_i$
\begin{align*}
-\frac{N}{ r_{p_{n+1}} }\frac{\partial}{\partial e_{p_{n+1}}} \Omega_{p_1,...,p_n}= \Omega_{p_1,...,p_n,p_{n+1}}+\frac{\delta_{n,1}}{(e_{p_1}-e_{p_2})^2}.
\end{align*}
To show that the boundary creation operator for the LSZ model was properly chosen, we first need to prove the following correspondence:

\begin{proposition}
\label{prop:GOM}
$\Omega_{p_1,...,p_n}$ has for $n={1,2,3}$ and distinct $p_i$ the following representations in terms of correlation functions:
\begin{align*}
&\Omega^{(g)}_{p_1}=\frac{1}{N} \sum_l \tilde{r}_l G^{(g)}_{|p_1l|} \\
&\Omega^{(g)}_{p_1,p_2}= \frac{\delta_{g,0}}{(e_{p_1}-e_{p_2})^2} + \frac{1}{N^2} \sum_{l_1,l_2} \tilde{r}_{l_1}\tilde{r}_{l_2} G^{(g)}_{|p_1l_1|p_2l_2|}  + \frac{1}{N} \sum_{l}\tilde{r}_l G^{(g)}_{|q_1lq_2l|} \\
&\Omega^{(g)}_{p_1,p_2,p_3}=  \frac{1}{N^3} \sum_{l_1,l_2,l_3}\tilde{r}_{l_1}\tilde{r}_{l_2}\tilde{r}_{l_3} G^{(g)}_{|p_1l_1|p_2l_2|p_3l_3|} + \frac{1}{N^2} \sum_{l_1,l_2}\tilde{r}_{l_1}\tilde{r}_{l_2} \bigg (G^{(g)}_{|p_1l_1p_2l_1|p_3l_2|} + \mathrm{cycl.} \biggl )\\
&\qquad \qquad + \frac{1}{N} \sum_{l}\tilde{r}_{l}  G^{(g)}_{|p_1lp_2lp_3l|} 
\end{align*}
\begin{proof}
	The expression for $\Omega^{(g)}_{p_1}$ is already known. We write
	\begin{align*}
		&\Omega^{(g)}_{p_1,p_2}= [N^{2-2g}]\sum_{l_1,l_2=1}^N \frac{\partial^4}{\partial J_{q_1k}\partial J^\dagger_{kq_1}\partial J_{q_2l}\partial J^\dagger_{lq_2}} \log [\mathcal{Z}(J,J^\dagger)]_{J,J^\dagger =0}
	\end{align*}
	For generic $l_1,l_2$, the derivatives produce two boundaries, yielding $ \frac{1}{N^2} \sum_{l_1,l_2} G^{(g)}_{|p_1l_1|p_2l_2|} $. For $l_1=l_2=l$, we additionally produce $\frac{1}{N} \sum_{l} G^{(g)}_{|p_1lp_2l|} $ as the derivatives commute and we rearrange them to a cycle of four indices, yielding a 4-point function. Allowing multiplicities leads to expression of $\Omega^{(g)}_{p_1,p_2}$
	 Analogous techniques give the expression for $\Omega^{(g)}_{p_1,p_2,p_3}$.
\end{proof}
\end{proposition}

\begin{remark}
A comparison to the Quartic Kontsevich model (hermitian matrices) shows the enormous simplification. For $\Omega^{(g)}_{p_1,p_2}$ e.g., the following analogue to Proposition \ref{prop:GOM} was proved in \cite[Prop. 4.7]{Branahl:2020uxs}:
  \begin{align}
\Omega^{(g)}_{p_1,p_2}   &=
\textcolor{red}{\frac{\delta_{g,0}}{(E_{p_1}-E_{p_2})^2}}
+\sum_{g_1+g_2=g} G^{(g_1)}_{|p_1p_2|} G^{(g_2)}_{|p_1p_2|}
\nonumber
\\
&
+  \textcolor{red}{  \frac{1}{N^2}\sum_{l_1,l_2=1}^dr_{l_1} r_{l_2} G^{(g)}_{|p_1l_1|p_2l_2|}}
+\frac{1}{N}\sum_{l=1}^d r_l \Big(G^{(g)}_{|p_1lp_1p_2|}+G^{(g)}_{|p_2lp_2p_1|}+\textcolor{red}{G^{(g)}_{|p_1lp_2l|}}
\Big)
\nonumber
\\
& +\frac{1}{N} \sum_{l=1}^d r_l \Big(G^{(g-1)}_{|p_1l|p_2|p_2|}+G^{(g-1)}_{|p_2l|p_1|p_1|}\Big)
+G^{(g-1)}_{|p_1p_2p_2|p_2|}+G^{(g-1)}_{|p_2p_1p_1|p_1|}
\nonumber
\\
&+\sum_{g_1+g_2=g-1} G^{(g_1)}_{|p_1|p_2|} G^{(g_2)}_{|p_1|p_2|}
+G^{(g-2)}_{|p_1|p_1|p_2|p_2|}\;.
\end{align}  
Besides odd boundary lengths, also products of correlation functions leading to disconnected ribbon graphs that contribute to $\Omega^{(g)}$ are allowed, as well as chains of external indices $q_i$ within one boundary. This is possible since much more derivatives with respect to the source $J$ can be commuted and rearranged to different chains of indices. Hence, it is not surprising that several surplus structures (in terms of blobs) arise from the analytic perspective forming bobbed topological recursion. 
\end{remark}

Combining the perturbative expansion given by Proposition \ref{prop:weight} together with Proposition \ref{prop:GOM}, we get the perturbative expansion of $\Omega^{(g)}_{p_1,...,p_n}$ in general, which can be compared with the exact results achieved for instance from Theorem \ref{th:main}. 

We illustrate this at the first orders for $\Omega^{(0)}_p$ and $\Omega^{(0)}_{p_1,p_2}$:
\begin{example}
	From the perturbative expansion of the 2- and 4-point function from Example \ref{ex:2ppert}, we find 
	\begin{align}\label{OM1}
		\Omega^{(0)}_{p}=&\frac{1}{N}\sum_{q=1}^{\tilde{d}}\tilde{r}_q\bigg[\frac{1}{e_p+\tilde{e}_q}- \frac{\lambda}{(e_p+\tilde{e}_q)^2}\frac{1}{N}\bigg(\sum_{l=1}^{\tilde{d}}\frac{\tilde{r}_l}{e_p+\tilde{e}_l}+\sum_{k=1}^{d}\frac{r_k}{\tilde{e}_q+e_k}\bigg)\bigg]+\mathcal{O}(\lambda^2)\\\label{OM2}
		\Omega^{(0)}_{p_1,p_2}=&\frac{1}{(e_{p_1}-e_{p_2})^2}-\frac{\lambda}{N}\sum_{q=1}^{\tilde{d}}\frac{ \tilde{r}_q}{(e_{p_1}+\tilde{e}_{q})^2(e_{p_2}+\tilde{e}_{q})^2}+\mathcal{O}(\lambda^2),
	\end{align}
	since the $2+2$-point function starts at order $\mathcal{O}(\lambda^2)$. Note that \eqref{OM2} is also achieved by the boundary creation operator $-\frac{N}{r_{p_2}}\frac{\partial}{\partial e_{p_2}}$ on \eqref{OM1}, as it should by Definition \ref{def:gDSE}. 
	
	On the other hand, the initial data for $x$ and $y$ provided in Theorem \ref{Thm:InitialData} expands $\varepsilon_p,\tilde{\varepsilon}_q$ with $e_p=x(\varepsilon_p)$ and $\tilde{e}_q=y(\tilde{\varepsilon}_q)$ into (note that both expansions are highly coupled)
	\begin{align*}
	\varepsilon_p =& e_p + \frac{\lambda}{N} \sum_{k=1}^{\tilde{d}} \frac{\tilde{r}_k}{e_p+\tilde{e}_k}  -\frac{\lambda^2}{N^2}\sum_{k=1}^{\tilde{d}}\frac{\tilde{r}_k}{e_p+\tilde{e}_k}\bigg(\sum_{n=1}^d\frac{r_n}{(\tilde{e}_k+e_n)^2}+\frac{r_n}{(e_p+\tilde{e}_k)(\tilde{e}_k+e_n)}\\
	&\qquad \qquad\qquad \qquad\qquad \qquad\qquad \qquad +\sum_{l=1}^{\tilde{d}}\frac{\tilde{r}_l}{(e_p+\tilde{e}_k)(\tilde{e}_l+e_p)} \bigg)+ \mathcal{O}(\lambda^3),\\
	 x'(\varepsilon_p) =& 1  + \frac{\lambda}{N} \sum_{k=1}^{\tilde{d}} \frac{\tilde{r}_k}{(e_p+\tilde{e}_k)^2}  + \mathcal{O}(\lambda^2),\\
	 \tilde{\varepsilon}_q =& -\tilde{e}_q - \frac{\lambda}{N} \sum_{n=1}^{d} \frac{r_n}{\tilde{e}_q+e_n}  + \mathcal{O}(\lambda^2), \quad y'(\tilde{\varepsilon}_q) = -1  + \frac{\lambda}{N} \sum_{n=1}^{d} \frac{r_n}{(\tilde{e}_q+e_n)^2}  + \mathcal{O}(\lambda^2).
	\end{align*}
	Inserting this into the solution of $\Omega^{(0)}_p$ yields
	\begin{align*}
		\Omega^{(0)}_p=&-\frac{y(\varepsilon_p)+V'(e_p)}{\lambda}\\
		=&\frac{\varepsilon_p-e_p}{\lambda}+\frac{1}{N}\sum_{n=1}^dr_n\bigg(\frac{1}{e_p-e_n}-\frac{1}{x'(\varepsilon_n)(\varepsilon_p-\varepsilon_n)}\bigg)\\
		=&\frac{1}{N}\sum_{k=1}^{\tilde{d}}\frac{\tilde{r}_k}{e_p+\tilde{e}_k}-\frac{\lambda}{N^2}\sum_{k=1}^{\tilde{d}}\sum_{n=1}^dr_n\tilde{r}_k\bigg(\frac{1}{(\tilde{e}_l+e_n)^2(e_n-e_p)}+\frac{\frac{1}{(\tilde{e}_l+e_n) }-\frac{1}{(\tilde{e}_l+e_p)}}{(e_n-e_p)^2}\bigg)\\
		-&\frac{\lambda}{N^2}\sum_{k=1}^{\tilde{d}}\tilde{r}_k\bigg[\sum_{n=1}^d\frac{r_n}{(e_p+\tilde{e}_k)(\tilde{e}_k+e_n)^2}+\frac{r_n}{(e_p+\tilde{e}_k)^2(\tilde{e}_k+e_n)}
		+\sum_{l=1}^{\tilde{d}}\frac{\tilde{r}_l}{(e_p+\tilde{e}_k)^2(\tilde{e}_l+e_p)}\bigg]\\
		&+ \mathcal{O}(\lambda^2),
	\end{align*}
	where the double sum in second last line cancels the first term in the last line, and therefore it is in full compliance with the perturbative result \eqref{OM1}. Similarly, we find from Proposition \ref{pro:om02}
	\begin{align*}
	&\Omega^{(0)}_{p_1,p_2}-\frac{1}{(e_{p_1}-e_{p_2})^2}
	=\frac{1}{x'(\varepsilon_{p_1})x'(\varepsilon_{p_2})(\varepsilon_{p_1}-\varepsilon_{p_2})^2}-\frac{1}{(e_{p_1}-e_{p_2})^2}
	\\
	&=-\frac{\lambda}{N}\sum_{k=1}^{\tilde{d}} \tilde{r}_k\bigg(2\frac{\frac{1}{\tilde{e}_k+e_{p_1}}
		-\frac{1}{\tilde{e}_k+e_{p_2}}}{(e_{p_1}-e_{p_2})^3}+\frac{\frac{1}{(e_{p_1}+\tilde{e}_k)^2}+\frac{1}{(e_{p_2}+\tilde{e}_k)^2}}{(e_{p_1}-e_{p_2})^2}
	\bigg)+\mathcal{O}(\lambda^2),
	\end{align*}
	which is also after trivial rearrangements in full compliance with the perturbative result \eqref{OM2}.
	
	An interested reader may also check the next orders, where surprising cancellations show up. The perturbative and the analytic result is represented in completely different partial franction decompositions very similar to \cite{Branahl:2020uxs}, but with more structure due to the distinction between $e_n$ and $\tilde{e}_k$.
\end{example}

\subsection{The Combinatorial Limit}
Finally, we want to focus on an extreme limit for the LSZ model: The case where the external fields become a multiple of the identity matrix, meaning $d=\tilde{d}=1$ or $N$-fold degenerate single eigenvalues of $E$ and $\tilde{E}$. For a quartic interaction, this particular model aroused interest already in the 90's, but in the field of combinatorics/enumeration of certain objects. For this reason, the aforementioned limit case shall be called the \textit{combinatorial limit}. In particular, Morris aimed to count ways of checkering a genus $g$-surface like a chess board in \cite{Morris:1990cq} whereas Zinn-Justin and Zuber used the same partition function for the enumeration of virtual tangles and links \cite{Zinn-Justin:2003ecd}. Moreover, the complex 1-matrix model is known to give rise to generating functions of \textit{bipartite maps}, whereas hermitian matrices yield in the same manner generating functions for the enumeration of \textit{ordinary maps} (as they are today often called) - one of the first successes of TR \cite{Eynard:2016yaa}. The spectral curve generating bipartite maps was recently worked out in \cite{Branahl:2022}. 

Interestingly, these two kinds of maps show up again in the comparison between the QKM and the LSZ model: The evaluation of the corresponding $\Omega_{g,1}(z)$ at $z=\varepsilon$ of the QKM (LSZ model) gives rise to generating functions of the number of rooted (bipartite) quadrangulations of a genus-$g$ surface, where the boundary length is 2. This access to generating functions of the enumeration of maps is fairly different compared with earlier techniques - although in the combinatorial limit the same partition functions occur. Within this framework of the complex/hermitian 1-matrix model, topological recursion gives rise to resolvents containing an infinite sum all possible boundary lengths. To evaluate the meromorphic forms $\omega_{g.n}(z_1,...,z_n)$ of these 1-matrix models, one takes residues at $z_i = \infty$ to filter the generating function of interest out of the Laurent series/resolvent in powers of $x(z)$ (see e.g. again \cite{Eynard:2016yaa,Branahl:2022}). Below, we give some exemplary numbers up to $\mathcal{O}(\lambda^5)$ in the LSZ model in the combinatorial limit that coincide with \cite{Branahl:2022}:
\begin{table}[h]
\centerline{\begin{tabular}[h!b]{|c|c|c|c|c|c|c|c|c|c|}
\hline
Order &$\Omega_{0,1}$&$\Omega_{0,2}$ &$\Omega_{0,3}$  &$\Omega_{1,1}$&$\Omega_{1,2}$ &$\Omega_{2,1}$\\
\hline
$\lambda^0$  & 1 & 0  & 0   & 0 &0 &  0   \\
\hline
$\lambda^1$   & 2 &1  & 0 & 0 &0 &0 \\
\hline
$\lambda^2$   & 9 & 13  & 6    & 1  & 0&0 \\
\hline
$\lambda^3$   & 54 &144 &172   &    20 & 0 &0\\
\hline
$\lambda^4$   & 378 &1539  & 3294   & 307 & 21 & 21 \\
\hline
$\lambda^5$   & 2916 & 16335 &53136 & 4280  & 734 & 966 \\
\hline
\end{tabular}}
\hspace*{1ex}
\caption{These numbers are generated by Theorem \ref{th:main} for $d=\tilde{d}=1$, where $\Omega^{(g)}_n(z_1,...,z_n)$ is evaluated at $z_i=\varepsilon$ and then expanded in the coupling constant $\lambda$. $\Omega_{g,1}$ coincides with OEIS no. A000168 (g=0), no. A006300 (g = 1) and no. A006301 (g = 2)  }
\label{tab1}
\end{table}

Moreover, we can give concrete formulae for all planar correlation functions emerging from the LSZ model by exploiting two facts. On the one hand, we know that the correlation functions $G_{...}$ of the QKM count in the combinatorial limit \textit{fully simple maps} in the sense of \cite{Borot:2017agy}. Their generating functions were already worked out in \cite{bernardi2017bijections}. On the other hand, taking only the bipartite ones is an easy task in the planar sector: Since the following relation 
 \begin{align*}
  (G^{(0)}_{|p_1^1q_1^1...p_{N_1}^1q_{N_1}^1|...|p_1^bq_1^b...p_{N_b}^bq_{N_b}^b|} )^{LSZ} = 2^{1-b}  (G^{(0)}_{|p_1^1q_1^1...p_{N_1}^1q_{N_1}^1|...|p_1^bq_1^b...p_{N_b}^bq_{N_b}^b|} )^{QKM}
\end{align*}
holds for $e=\tilde{e}$, in analogy to the discussion in \cite[Sec. 1]{Branahl:2022}, we conclude in the combinatrial limit using \cite{bernardi2017bijections}:
\begin{corollary}
 \begin{align*}
   \bigg ( G^{(0)}_{|k_1^1...k_{n_1}^1|...|k_1^b...k_{n_b}^b|} \bigg )^{LSZ}
   \Big|_{d=1}
   =\sum_{n=0}^{\infty} \frac{3^{b+n-2}(\#n_e-1)!}{n!2^{b-1}(3l_h+b+n-2)!}
   \prod_{i=1}^b n_i \binom{\frac{3n_i}{2}}{\frac{n_i}{2}}
   \cdot \frac{(-\lambda)^{n+l_h+b-2}}{(e+\tilde{e})^{2(n+l_h+b-1)}}\;.
\end{align*}
Here $l_h:=\frac{1}{2}\sum_i n_i$ is the half boundary length and
$\#n_e:=3l_h+2b+2n-4$ the number of edges.  For  $b=1$, $n_1=2$ and thus $l_h= 1$, one recovers  the famous
result of Tutte \cite{Tuttbij} for the number of rooted planar quadrangulations
$\frac{2 \cdot 3^n(2n)!}{n!(n+2)!}$, obtained as coefficient of
$\frac{(-\lambda)^n}{(e+\tilde{e})^{2n+2}}$ in the 2-point
function. 
\end{corollary}

In the combinatorial limit it is possible to write the solution of the implicitly defined variable transforms $x(z)$ and $y(z)$. With the notation $\frac{\tilde{r}_k}{y'(\tilde{\varepsilon}_k)} = \tilde{\rho}_k$ and  $\frac{r_k}{x'(\varepsilon_k)} = \rho_k$, the combinatorial limit gives rise to an exact solution of the set $\{\varepsilon,\tilde{\varepsilon},\rho,\tilde{\rho} \}$: The system of equations
\begin{align*}
 & \biggl \{ e= \varepsilon-\frac{\lambda \tilde{\rho}}{N(\varepsilon-\tilde{\varepsilon})} \,\, ,  \,\,\, \tilde{e}= -\tilde{\varepsilon}-\frac{\lambda \rho}{N(\varepsilon-\tilde{\varepsilon})}  \,\, ,  \,\,\,\frac{N}{\tilde{\rho}} - \frac{\lambda \rho}{N(\varepsilon-\tilde{\varepsilon})^2}=\frac{N}{\rho} - \frac{\lambda \tilde{\rho}}{N(\varepsilon-\tilde{\varepsilon})^2}  \biggl \}
\end{align*}
has the solutions
\begin{align*}
 & \varepsilon= \frac{5e-\tilde{e}+\sqrt{(e+\tilde{e})^2+12 \lambda}}{6} \qquad  \tilde{\varepsilon} =  \frac{e-5\tilde{e}-\sqrt{(e+\tilde{e})^2+12 \lambda}}{6} \\
& \rho= \tilde{\rho} =N\cdot \frac{-(e+\tilde{e})^2+12\lambda +(\tilde{e}+e)\sqrt{(e+\tilde{e})^2+12 \lambda} }{18 \lambda}  
\end{align*}
with $ \varepsilon+\tilde{\varepsilon}=e-\tilde{e}$ and $\lim_{\lambda \to 0} \varepsilon(\lambda)=e$ and $\lim_{\lambda \to 0} \tilde{\varepsilon}(\lambda)=-\tilde{e}$. Setting $e=\tilde{e}$, we get back the solutions of $\{\varepsilon,\rho\}$ in the combinatorial limit of the QKM.

\section{Discussion and Outlook}
\label{ch:concl}
In this article we have generalised the exact solvability of the Langmann-Szabo-Zarembo model with a quartic potential by identifying suitable meromorphic forms $\omega_{g,n}$ that satisfy topological recursion and shed new light on the discussion of its solutions. It was a long way to these objects: Starting with the naturally occurring correlation functions in quantum field theory, that satisfy closed Dyson-Schwinger equations after applying a system of Ward identities, we identified a differential operator later creating boundaries in $\omega_{g,n}$. This, together with the previous success of this method in the quartic analogue of the Kontsevich model, suggested the introduction of the first new set of correlators - the generalised correlation functions $T$. The creation operator gave birth to a second family of correlators $\Omega$ which is eventually governed by topological recursion itself. From the solution of the 2-point function after analytic continuation and two important variable transforms we could directly read off the initial data of the recursion. Complexification of the Dyson-Schwinger equation luckily led to known loop equation for which the recipe to show the fulfilment of the two abstract loop equations was well-known already and applied in a couple of examples. 

The journey to this solution started when finding an answer to the question, for which reasons the analogue of the LSZ model taking hermitian quantum fields into consideration is so much more complicated than the complex case. We analysed the exact and perturbative differences in these two models. 

After a concrete proof of the exact solution of the LSZ model, we are left with the insight that the 2-point function of quantum field theory is undoubtedly a necessary step towards topological recursion to read off the initial data and plays an important role in perturbation theory, but only a smart rearrangement of the correlation functions $G$ to the family of $\omega_{g,n}$ allows for compact and concrete solutions with a simple recursive pattern. This insight might be crucial when turning to exact solutions of more complicated, physical quantum field theories.

\begin{landscape}
\begin{figure}[h!]
  \centering
    \includegraphics[width= 1.45\textwidth]{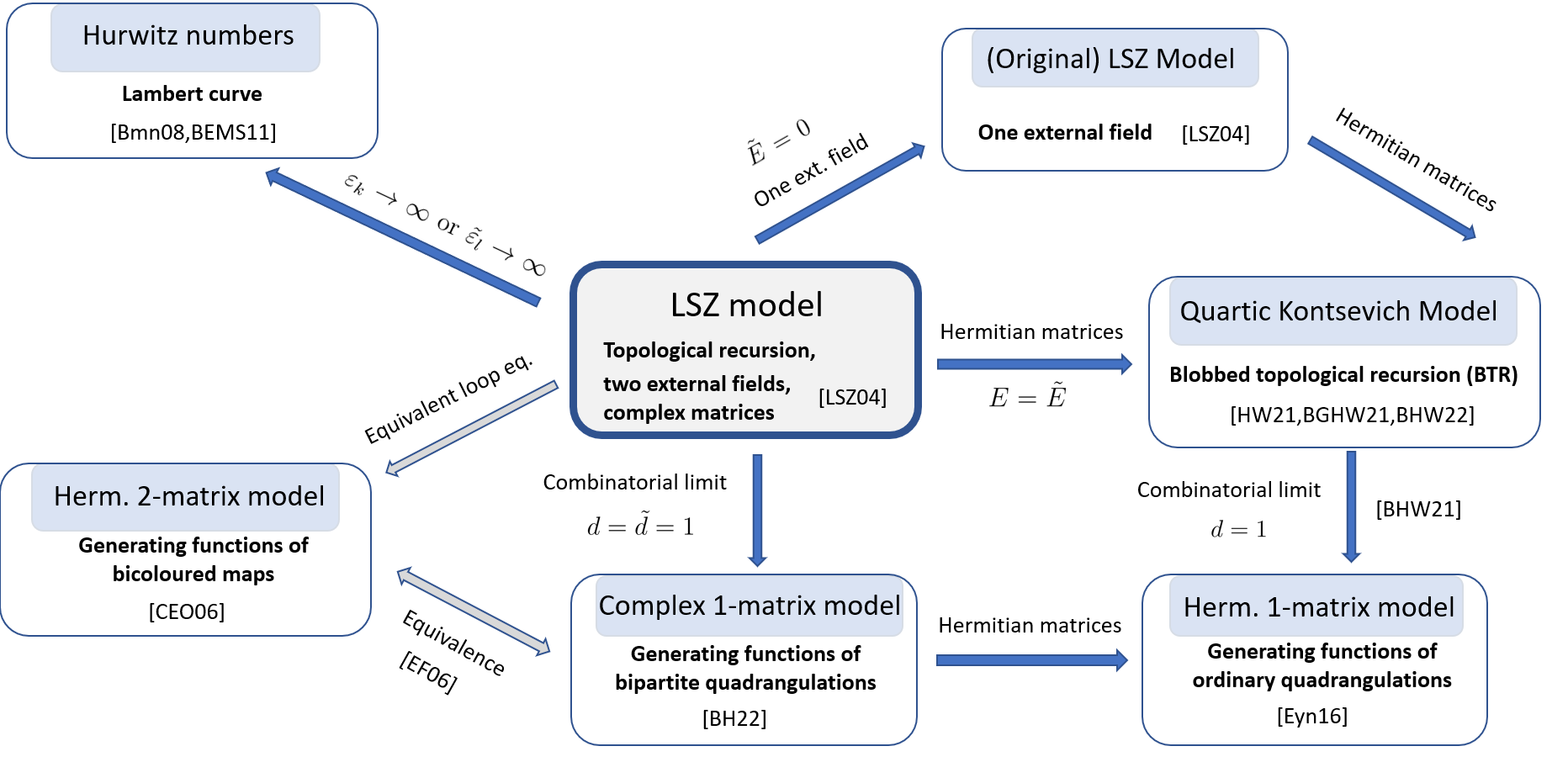} 
    \caption{There exist various connections of the LSZ model to other matrix models in different limits.
\label{mm}}
\end{figure}
\end{landscape}

Finally, this discussion and outlook shall be dedicated to three interesting perspectives for the future. We refresh the discussion of integrability of the LSZ model, the triviality problem on the four-dimensional Moyal space and explain how topological recursion builds a bridge to Hurwitz numbers.
\begin{enumerate}
\item \textit{The triviality problem.} In this article we did not pay much attention to the noncommutative Moyal space, where the actual LSZ model lives. Instead, we directly approximated the quantum fields by finite $N\times N$ matrices and obtained a partition function being a matrix model (this procedure can be retraced in detail e.g. in \cite[Ch. 2.5]{Hock:2020rje}) that could be treated with established methods. However, this is far away from quantum field theory in its original sense. It would be worth a detailed investigation to perform the $N \to \infty$ limit and to see which parts of topological recursion inherit to the Moyal plane in two dimensions or, most importantly, the four-dimensional Moyal space. This limit contains several technical procedures, as a regularisation of the integral equations (occurring by sending $N \to \infty$ in the Dyson-Schwinger equations) by a cut-off $\Lambda^2$ (ratio of the largest eigenvalue and $N$ in the limit $N\to \infty$) and renormalisation of some physical quantities like mass, field and coupling constant (see \cite{Grosse:2019jnv} for the hermitian model). 

There is a severe danger for any quantum field theory: It has to be consistent with the scaling limit $\Lambda^2 \to \infty$ (see e.g. \cite{Landau:1954??} for the Landau ghost problem, a serious hindrance towards renormalisability in the childhood of QFT). The authors of \cite{Langmann:2003if} remark that their scaling limit leads to only trivial Green's functions/connected correlation functions, but also admit that the subtle choice of a proper scaling limit in noncommutative QFT may have meaningful consequences for the triviality problem. On commutative geometries, triviality of the $\lambda \phi^4$ model in $D=4 + \varepsilon$ dimensions is known since the 80's \cite{Aizenman:1981du,Frohlich:1982tw}, whereas in exactly $D=4$ dimensions, (marginal) triviality was recently proved \cite{Aizenman:2019yuo}. But there is hope coming from the investigation of the noncommutative $\lambda \phi^4$ theory with hermitian fields on the four-dimensional Moyal space. This is exactly the $N \to \infty$ limit of the previously explained QKM, obtained by a suitable deformed spectral measure. As shown in \cite{Grosse:2019jnv,Grosse:2019qps}, the spectral measure has a dimension drop in the planar case of $D=4-2 \frac{\arcsin(\lambda \pi)}{\pi}$ implying \textit{non-triviality} in four dimensions in the planar sector! We are convinced that a thorough future investigation of the $N \to \infty$ limit of the LSZ model will reveal many similarities to the hermitian model. There might be an adequate scaling limit to rescue the non-triviality of this noncommutative QFT with complex fields.

\item \textit{Integrability.} Integrable systems in Nature are often those with the richest algebraic beauty in their mathematical description. The Kyoto school method (Sato, Hirota, Miwa, Jimbo et al. \cite{Hirota:1971,Sato:1983}) associates a  $\tau$ function to the integrable system
that encodes the majority of its properties and strongly corresponds to the partition function. It obeys
Hirota’s equation. The characterisation of the integrable system can be performed by finding the integrable PDE class the $\tau$ function obeys. On the other hand, topological recursion provides a simple method to characterise the integrable hierarchy by the geometry of the spectral curve. A first success was the to figure out the bridge between the spectral curve for the Kontsevich model to intersection theory, enumeration of trivalent ribbon graphs and the $\tau$ function of the Korteweg-de Vries (KdV) hierarchy \cite{Witten:1990hr}.

 Today, it is known \cite{Eynard:2017} that the entire moduli space of hyperelliptical spectral curves with $\mathrm{deg}(x)=2$ and a polynomial relation $y^2=P(x)$ yields the space of KdV systems. This is a subspace of all Kadomtsev-Petviashvili (KP) systems corresponding to the moduli space of algebraic curves where $x$ maps  to the Riemann sphere and $x$ and $y$ are meromorphic such that there is a polynomial $P(x,y/dx)=0$. This is exactly the type of spectral curve we dealt with in the LSZ model. The authors of \cite{Langmann:2003if} stated that $\mathcal{Z}(J,J^\dagger)$ from eq. (\ref{partfunLSZ}) is a $\tau$ function of the Toda lattice hierarchy. This is indeed true, but with the knowledge from topological recursion we can restrict this statement to the subspace $\mathcal{M}_{KP} \subset \mathcal{M}_{Toda}$ as long as we are dealing with finite matrix approximation. The moduli space $\mathcal{M}_{Toda}$ namely additionally allows for logarithmic singularities in $x$ and $y$. This kind of singularities will occur in the limit $N\to\infty$, changing the matrix model to a model on the $2D$-dimensional Moyal space - comparing with \cite{Grosse:2019jnv}, where e.g. in 2 dimensions, the spectral curve turns into a logarithm. This kind of \textit{Lambert curve} gives the connection to simple Hurwitz numbers.

\item \textit{A matrix model for simple Hurwitz numbers.} In the early days of topological recursion, prime examples were mostly found in the realm of map enumeration. Bouchard and Mariño then formulated a conjecture \cite{Bouchard:2007hi} that the same recursive structure shows up in the calculation of (simple) Hurwitz numbers. These numbers count $n$-fold coverings over $\overline{\mathbb{C}}$ that are connected genus-$g$ curves with one branch point of arbitrary ramification profile and trivial profiles (only transpositions) for $m$ further points. This conjecture was proved in a twofold way: By so-called cut-and-join equations \cite{Eynard:2011xy} and by a quite artificially constructed matrix model with external field \cite{Borot:2009ix}. We will focus in this discussion on the latter one. 

The partition function of this matrix model for simple Hurwitz numbers looks quite technical and is designed exactly for this purpose. However, spectral curves solving matrix models with external fields are always of the same algebraic structure. We cite their result \cite[eq. (13)]{Borot:2009ix}
\begin{align*}
\quad x(z)=z+C-g_s \sum_{i=1}^N \frac{1}{y'(z_i)(z-z_i)} \qquad y(z)= -z + \sum_{j=1}^{\hat{N}} \frac{u_j}{x'(\hat z_j)(z-\hat z_j)}
\end{align*}
and compare with the curve solving the LSZ model ($z_i \leftrightarrow \tilde{\varepsilon}_i$, $\hat z_j \leftrightarrow \varepsilon_j$, $g_s \leftrightarrow \lambda$, the occurrence of $u_j$ is not important). There is in fact one decisive difference: The coupling constant $\lambda$ in the LSZ model is a global prefactor for both $x(z)$ and $y(z)$. The authors of \cite{Borot:2009ix}, however, aim to reconstruct in the limit $N \to \infty$ the \textit{Lambert curve} $e^x=ye^{-y}$ from the Bouchard-Mariño conjecture that can only be obtained if $y(z)$ reduces to $y(z)=z$. For this purpose, they set the coupling $g_s$ to 0, which is needed before for regularisation. An equivalent apprach in the LSZ model might be to send the deformed eigenvalues either $\varepsilon_k \to \infty$ or $\tilde{\varepsilon}_l \to \infty$ for $k \in \{1,..,d\}$ and $l \in \{1,...,\tilde{d}\}$. Investigating this more thoroughly may be combined with the analysis of the four-dimensional case with respect to the triviality problem in some future.
\end{enumerate}
We have seen that using topological recursion to show the exact solvability of the LSZ model with quartic interaction bridges quantum field theory to several interesting mathematical field. As a summary, Fig. \ref{mm}  shall illustrate these connections for different limiting cases. 
Having the last three points of  (non-)triviality, integrability and Hurwitz numbers in mind, there remain fascinating questions for future investigations. Up to now, we have added another model to the wide range of models and theories obeying topological recursion in its original sense. This work strengthens furthermore the connection of noncommutative QFT toy models to topological recursion and thus renews and enriches discussions about QFT-related problems that were debated about two decades ago. \newpage

\newpage

\appendix
\section{Solution of the 2-Point Function}
\label{sec:2p}
Throughout the appendix, we will need the Lagrange interpolation formula that we want recall:
\begin{lemma}\label{lem:interpol}
Let $f$ be a polynomial of degree $d\geq 0$ and 
$x_1,...,x_{d+1}$ be pairwise distinct complex numbers.
Then, for all $x\in \mathbb{C}$,
\begin{align*}
		f(x)=L(x)\sum_{j=1}^{d+1}\frac{f_j}{(x-x_j)L'(x_j)},\qquad 
\text{where } L(x)=\prod_{j=1}^d(x-x_j) \text{ and } f_j=f(x_j).
\end{align*}
\end{lemma}
This section is dedicated to the proof of the spectral curve, encoded in the solution of the analytically continued 2-point function. 
 
\subsection*{Proof of Theorem \ref{Thm:InitialData}} 
Most methods build on the solution strategy in the QKM, \cite{Schurmann:2019mzu}\footnote{Their proof builds on properties of the inverse Cauchy matrix, however it is sufficient to work with the fundamental polynomials of the Lagrange interpolation formula}. We start with the complexified Dyson-Schwinger equation for the 2-point function. We assume that there are undetermined variable transforms $x(z)$ and $y(z)$ that can be uniquely specified by the upper ansatz. After transformation, we have for eq. (\ref{genusexp}), $g=0$:
	\begin{align}
&\bigg (x(z)+y(w)+ \frac{\lambda}{N} \sum_{k=1}^{\tilde d} \tilde{r}_k \mathcal{G}^{(0)}(z,\tilde{\varepsilon}_k) \bigg )\mathcal{G}^{(0)}(z,w) \\
&= 1+ \frac{\lambda}{N} \sum_{k=1}^d r_k\frac{\mathcal{G}^{(0)}(\varepsilon_k,w)-\mathcal{G}^{(0)}(z,w)}{x(\varepsilon_k)-x(z)}
	\end{align}
The ansatz of Theorem \ref{Thm:InitialData} turns this equation into
	\begin{align}
\label{eq:ans1}
 (y(w) -y(z))\mathcal{G}^{(0)}(z,w) = 1+ \frac{\lambda}{N} \sum_{k=1}^d r_k \frac{\mathcal{G}^{(0)}(\varepsilon_k,w)}{x(\varepsilon_k)-x(z)}
	\end{align}
We set $z= \hat{\tilde{w}}^l$ to get rid off the prefactor on the lhs under the finiteness assumption from Theorem \ref{Thm:InitialData} of $\mathcal{G}^{(0)}(z= \hat{\tilde{w}}^l,w)$. This leads to $d$ equations since $y$ was assumed to be of rational degree $d$
	\begin{align*}
\frac{\lambda}{N} \sum_{k=1}^d r_k \frac{\mathcal{G}^{(0)}(\varepsilon_k,w)}{x(\hat{\tilde{w}}^l)-x(\varepsilon_k)}=1. 
	\end{align*}
Define the fundamental Lagrange interpolation polynomials $A(v) = \prod_{j=1}^d (v-x(\hat{\tilde{w}}^j))$ and $B(v) =\prod_{j=1}^d (v-x(\varepsilon_j))$. Then we can read off
	\begin{align}
\label{eq:lagr}
\frac{\lambda r_k}{N} \mathcal{G}^{(0)}(\varepsilon_k,w) =- \frac{\prod_{j=1}^d (x(\varepsilon_k)-x(\hat{\tilde{w}}^j)}{\prod_{j \neq k}^d (x(\varepsilon_k)-x(\varepsilon_j))}
	\end{align}
since the following sum over all residues at $x(\varepsilon_j)$ for the poles in $B(v)$ reads 1:
	\begin{align*}
\sum_{j=1}^d \Res_{v \to x(\varepsilon_k)} \frac{A(v)}{(x(\hat{\tilde{w}}^l)-v)B(v)} = \sum_{j=1}^d \frac{A(x(\varepsilon_k))}{(x(\hat{\tilde{w}}^l)-x(\varepsilon_j))B'(x(\varepsilon_k))}=1
	\end{align*}
With the same trick, we allow for $d+1$ factors in the interpolation formula and include $x(z)$. This gives the possibility to write
\begin{align*}
\frac{\prod_j (x(z)-x(\hat{\tilde{w}}^j))}{\prod_j (x(z)-x(\varepsilon_j))} + \sum_{k=1}^d \frac{\prod_j(x(\varepsilon_k)-x(\hat{\tilde{w}}^j))}{(x(\varepsilon_k)-x(z))\prod_{j \neq k}(x(\varepsilon_k)-x(\varepsilon_j))}=1
	\end{align*}
Inserting eq. (\ref{eq:lagr}) into eq. (\ref{eq:ans1}) gives
	\begin{align}
\label{eq:2p1}
\mathcal{G}^{(0)}(z,w) =  \frac{1}{y(w)-y(z)} \prod_{j=1}^{d} \frac{x(z)-x\big (\widehat{\tilde{w}}^j\big )}{x(z)-x(\varepsilon_j)} 
	\end{align}
and thus, we principally finished the proof of the first representation. However, it is still unclear why $y(z)$ is necessarily given by $ y(z)=-z+\frac{\lambda}{N}\sum_{n=1}^{d}\frac{r_n}{x'(\varepsilon_n)(z-\varepsilon_n)}+C$ where the constant $C$ must vanish in the end, and the same for $x(z)$. This can be  shown using Liouville's theorem. Inserting $z=\varepsilon_k$ into eq. (\ref{eq:2p1}) and comparing with  eq. (\ref{eq:lagr}), $y(z)$ must have $d$ poles at $z= \varepsilon_k$. Moreover, we have a simple pole at $z=\infty$. Remembering the degree $d+1$ of the assumption, we already have the complete set of poles. Thus, $-z+\frac{\lambda}{N}\sum_{n=1}^{d}\frac{r_n}{x'(\varepsilon_n)(z-\varepsilon_n)}+C-y(z)$ is a bounded holomorphic function and by Liouville a constant. Analogously to \cite{Schurmann:2019mzu}, one can show with some technical effort, that the initial ansatz guarantees that $C=0$. 

Now we exchange the role of $x$ and $y$, where we have to take the DSE of Remark \ref{rem:xy}. Carrying out the analytic continuation of Definition \ref{def:compl}, we get a second complexified DSE for the 2-point function
\begin{align}
	\label{eq:ans2}
	(x(z) -x(w))\mathcal{G}^{(0)}(z,w) = 1+ \frac{\lambda}{N} \sum_{k=1}^{\tilde{d}} \tilde{r}_k \frac{\mathcal{G}^{(0)}(z,\tilde{\varepsilon}_k)}{y(\tilde{\varepsilon}_k)-y(w)}.
\end{align}
Applying exactly the same steps yields the second representation, where $x(z)$ is necessarily given by $z-\frac{\lambda}{N}\sum_{k=1}^{\tilde{d}}\frac{\tilde{r}_k}{y'(\tilde{\varepsilon}_k)(z-\tilde{\varepsilon}_k)}$. \qed

We remark that the form of the spectral curve is not surprising - it is the typical interwoven structure between $x$ and $y$ - as implicitly defined equations - that usually appears when solving a matrix model with external field (for instance the matrix model for simple Hurwitz numbers \cite{Borot:2009ix} or the Kontsevich model \cite[Sec. 10.4]{Eynard:2007kz}).

\section{Proof of $\Omega^{(0)}_2$}
\label{sec:om02}
It remains the proof for the second part of the initial data, namely the Bergman kernel. By rational parametrisation of $\omega_{0,1}$, the spectral curve is a planar algebraic curve and thus, the shape of the fundamental form of the second kind is automatically determined, which we will show to coincide with $\Omega_{0,2}$.
\begin{proposition}
	The cylinder amplitude of the LSZ model reads
	\begin{align}
		\Omega_{0,2}(z_1,z_2)=\frac{1}{x'(z_1)x'(z_2)(z_1-z_2)^2}.
	\end{align}
\end{proposition}
The proof is split into two lemmata that follow from techniques of \cite{Branahl:2020yru}. Taking Corollary \ref{Cor:DSE} for $(g,n)=(0,2)$, setting $v=y(z)$ (the first term vanishes since $H$ is rational in $v$) and inserting the DSE of the 2-point function with exchanged $x$ and $y$ (eq. \ref{eq:ans2}), we deduce that $\Omega^{(0)}_{2}$ satisfies the DSE
\begin{align}
&x'(z)\mathfrak{G}_0(z)\Omega^{(0)}_{2}(u,z)
-\frac{\lambda}{N^2}\sum_{k,n=1}^{d,\tilde{d}}
\frac{ r_k \tilde{r}_n \mathcal{T}^{(0)}(u\|\varepsilon_k,\tilde{\varepsilon}_n|)}{
(x(\varepsilon_k)-x(z))(y(\tilde{\varepsilon}_n)-y(z))}=-\frac{\partial}{\partial x(u)}
\mathcal{G}^{(0)}(u,z) \;.
\end{align}
where
$\mathfrak{G}_0(z)=-H_{0,1}(y(z);z)$
and thus $\mathfrak{G}_0(z)=\Res_{u\to z} \mathcal{G}^{(0)}(z,u)du$ by Theorem \ref{Thm:InitialData}. To specify the pole structure of the 2-point function, we show the following decomposition:
\begin{lemma}
\label{prop:Omega0}
The planar 2-point function can be rewritten in the following decomposition of its poles:
\begin{align}
&\mathcal{G}^{(0)}(z,u) = \frac{1}{u-z} \bigg (1+ \frac{\lambda^2}{N^2} \sum_{k,n=1}^d \sum_{l,m=1}^{\tilde{d}} \frac{C_{k,l}^{m,n}}{\Big (z-\widehat{\varepsilon}_k^{\,m}\Big )\Big (u-\widehat{\tilde{\varepsilon_l}}^n\Big )} \biggl) \quad ,
\end{align}
where
\begin{align*}
& C_{k,l}^{m,n} = \frac{\Big (\widehat{\tilde{\varepsilon_l}}^n - \widehat{\varepsilon}_k^{\,m}\Big ) r_k \tilde{r}_l \mathcal{G}^{(0)}(\varepsilon_k,\tilde{\varepsilon}_l)}{x'\big (\widehat{\varepsilon}_k^{\,m} \big ) y' \big ( \widehat{\tilde{\varepsilon_l}}^n\Big ) \Big ( y\big (\widehat{\varepsilon}_k^{\,m} \big ) - y (\tilde{\varepsilon}_l) \Big ) \Big ( x\big (\widehat{\tilde{\varepsilon_l}}^n \big ) - x( \varepsilon_k) \Big ) }
\end{align*}
\begin{proof}
	To figure out all possible poles, we have to look at the two DSE's \eqref{eq:ans1} and \eqref{eq:ans2}. After dividing by $y(w)-y(z)$ (or $x(z)-x(w)$) and the regularity assumption of Theorem \ref{Thm:InitialData}, possible poles are located at $z=u$ as well as at $z=\hat{\varepsilon}_k^m$ and $u=\widehat{\tilde{\varepsilon_l}}^n$. Then, we look at the function $(u-z)
	\mathcal{G}^{(0)}(u,z)$, which approaches 1 for $z,u \to \infty$. Again, the basic relations eq. \eqref{eq:ans1} and \eqref{eq:ans2} give the chance to directly read off the residues at  $z=\widehat{\varepsilon_k}^m$ and $u=\widehat{\tilde{\varepsilon_l}}^n$ as stated. For the regularity discussion at $u=\hat z^k$ and $u=\widehat{\tilde{z}^k}$, one can look at the finite limit of the product representation of $\mathcal{G}(z,u)$ given in Theorem \ref{Thm:InitialData}.
\end{proof}
\end{lemma}
Liouville's theorem now proves:
\begin{lemma}
Assume that (for generic u) the function $\Omega^{(0)}_2(u,z)$ is regular at any zero $z$ of $\mathfrak{G}_0(z)$. Then
	\begin{align}
\label{eq:liou}
		x'(z)\Omega_{0,2}(z,u)-\frac{1}{x'(u)(z-u)^2}=0
	\end{align}
\begin{proof}
	We deduce the representation for $\mathfrak{G}_0(z)=\Res_{u\to z} \mathcal{G}^{(0)}(z,u)du$:
	\begin{align*}
		&\mathfrak{G}_0(z)= 1- \frac{\lambda^2}{N^2} \sum_{k,n}^d \sum_{l,m}^{\tilde{d}} \frac{C_{k,l}^{m,n}}{\Big (z-\widehat{\varepsilon}_k^{\,m}\Big )\Big (z-\widehat{\tilde{\varepsilon_l}}^n\Big )} 
	\end{align*}
	and therefore the partical fraction decomposition
	\begin{align*}
		&\mathcal{G}^{(0)}(u,z) = \frac{\mathfrak{G}_0(z)}{u-z} + \frac{\lambda^2}{N^2} \sum_{k,n}^d \sum_{l,m}^{\tilde{d}} \frac{C_{k,l}^{m,n}}{\Big (z-\widehat{\varepsilon}_k^{\,m}\Big )\Big ( z-\widehat{\tilde{\varepsilon_l}}^n\Big )\Big (u-\widehat{\tilde{\varepsilon_l}}^n\Big )}   \quad ,
	\end{align*}
	Comparing with eq. (\ref{prop:Omega0}), the right hand side of eq. (\ref{eq:liou}) reads with the upper partial fraction decomposition
	\begin{align*}
		& \frac{1}{\mathfrak{G}_0(z)}\bigg [ \frac{\lambda}{N^2}\sum_{k=1}^d  \sum_{n=1}^{\tilde{d}} r_k \tilde{r}_n
		\frac{\mathcal{T}^{(0)}(u\|\varepsilon_k,\tilde{\varepsilon}_n|)}{
			(x(\varepsilon_k)-x(z))(y(\tilde{\varepsilon}_n)-y(z))} \\
		& + \frac{\lambda^2}{N^2 x'(u)} \sum_{k,n}^d \sum_{l,m}^{\tilde{d}} \frac{C_{k,l}^{m,n}}{\Big (z-\widehat{\varepsilon}_k^{\,m}\Big )\Big (z-\widehat{\tilde{\varepsilon_l}}^n\Big )\Big (u-\widehat{\tilde{\varepsilon_l}}^n\Big )^2} \bigg]
	\end{align*}
	The simple poles at $z= \widehat{\varepsilon}_k^{\,m},\widehat{\tilde{\varepsilon_l}}^n$ cancel due to the prefactor $ \frac{1}{\mathfrak{G}_0(z)}$. By assumption, we can exclude further poles on the rhs at any zero $z$ of $\mathfrak{G}_0(z)$. Thus, the lhs and rhs of eq. (\ref{eq:liou})  must be a constant, and sending $z \to \infty$, the constant has to be zero as claimed.
\end{proof}

\end{lemma}

\section{Proofs of Section \ref{sec:compl}}
\label{sec:appproof}
 Section \ref{sec:compl} is completely free of any proofs. This has two reasons: First, the proofs are very technical without greater learning effects. Second, the proofs are carried out in quite similar manner to established results for hermitian fields.  Proposition \ref{prop:GpqJ} and Theorem \ref{thm:complete} show up in a slightly more general form for hermitian fields in \cite{Hock:2020rje}. The other two results are generalisations of results found in \cite[Chapter 4]{Branahl:2020yru}. \\ \\ 
\textbf{Proof of Proposition \ref{prop:GpqJ}} \\ \\ 
Define the differential operator
 \begin{align*}
\hat D = \frac{\partial^{2b-2}}{\partial J_{p^2q^2}\partial J^\dagger_{q^2p^2}...\partial J_{p^bq^b}\partial J^\dagger_{q^bp^b}}
 \end{align*}
The same steps as for Proposition \ref{thm:2p}, namely using Proposition \ref{prop:ward2}, give:
 \begin{align*}
G_{|pq|\J}& =N^{b-2} \hat D \frac{\partial^2}{\partial J_{pq} \partial J^\dagger_{qp}} \log[\mathcal{Z}(J,J^\dagger)]_{J,J^\dagger=0}\\
&=-    \frac{\lambda N^{b-4} \hat D}{E_p+\tilde{E}_q} \frac{\partial}{\partial J_{qp}^\dagger} \frac{1}{ \mathcal{Z}(J,J^\dagger)}\sum_{m,n=1}^N\frac{\partial^3}{\partial J_{pm}\partial J_{mn}^\dagger\partial J_{nq}}   \mathcal{Z}(J,J^\dagger)_{J,J^\dagger =0}\\
& = -  \frac{\lambda N^{b-4} \hat D}{E_p+\tilde{E}_q} \frac{\partial}{\partial J_{qp}^\dagger}  \frac{1}{\mathcal{Z}(J,J^\dagger)} \biggl [ \frac{\partial  }{\partial J_{pq} }  W_p[J,J^\dagger] \mathcal{Z}(J,J^\dagger)\\
&+ \sum_{m,n=1}^N \frac{N}{E_n-E_p} \frac{\partial}{\partial J_{nq}}  \bigg (  J^\dagger_{mp} \frac{\partial}{\partial J^\dagger_{mn}}-J_{nm} \frac{\partial}{\partial J_{pm}} \bigg ) \mathcal{Z}(J,J^\dagger)\bigg ] _{J,J^\dagger =0}
 \end{align*}
Note that we generalised the steps in Example \ref{ex:2p} to the general operator $\hat D$. For the second line, write
 \begin{align*}
  \frac{1}{\mathcal{Z}(J,J^\dagger)}  \frac{\partial  }{\partial J_{pq} }  \big ( W_p[J,J^\dagger] \mathcal{Z}(J,J^\dagger) \big )  = \frac{\partial  }{\partial J_{pq} } W_p[J,J^\dagger]  +  W_p[J,J^\dagger] \frac{\partial  }{\partial J_{pq} } \log[\mathcal{Z}(J,J^\dagger)]\;,
 \end{align*}
with $W_p[J,J^\dagger]$ given by Proposition \ref{prop:ward2}.
Applying all derivatives on $W_p$ gives by definition at $J=J^\dagger=0$:
 \begin{align*}
  N^{b-2} \hat D \frac{\partial^2  }{\partial J_{pq}\partial J_{qp}^\dagger}  W_p[J,J^\dagger] = \frac{1}{N^2} T_{p\|pq|\J|} 
 \end{align*}
The second term, however, gives rise to a partitioned sum of two subsets, dependent on which subset of indices in $\hat D$ hit $ \log[\mathcal{Z}(J,J^\dagger)]$ itself:
 \begin{align*}
N^{b-4}  \hat D \bigg( W_p[J,J^\dagger]   \frac{\partial^2  }{\partial J_{pq}\partial J_{qp}^\dagger} \log[\mathcal{Z}(J,J^\dagger)]\bigg)\bigg\vert_{J=0}=  \sum_{\mathcal{J}'\uplus \mathcal{J}''=\mathcal{J}}
T_{p\|\mathcal{J}'|} G_{|pq|\mathcal{J}''|}
 \end{align*}
Demanding $\mathcal{J}'\neq \emptyset$ gives $\Omega_pG_{|pq|\J}$ in Proposition \ref{prop:GpqJ}. The other terms arise from the part of the Ward identity. Excluding the term $l=p$, we can write $ \frac{1}{N}\sum_{ l=1, l\neq p}^N 
\frac{G_{|lq|\mathcal{J}|}-G_{|pq|\mathcal{J}|}}{E_l-E_p}$ 
with the same argument as in Proposition \ref{thm:2p}. The excluded term can be included in $\frac{1}{N^2} T_{p\|pq|\J|}$ to write it as $\frac{1}{N}\frac{\partial G_{|pq|\J|}}{\partial E_p}$. The other realisation of fixing the indices $n,m$ by $p^s$ and $q^s$ (when $\frac{\partial}{\partial J_{p^sq^s}} \in \hat{D}$ hits $J_{mn}$ and vice versa) gives the very last term in Proposition \ref{prop:GpqJ} where two boundaries merge. \qed \\ \\ 
\textbf{Proof of Proposition \ref{prop2}} \\ \\ 
The second term of the lhs of the DSE of Corollary \ref{cor2+} is
rewritten into    polynomial $f(\,.\,;w|I|\J)$ of degree $d-1$ and a denominator also appearing in $\mathcal{G}^{(0)}(z,w)$.
\begin{align*}
  -\frac{\lambda}{N}\sum_{k=1}^d r_k
  \frac{\mathcal{T}^{(g)}(I\|\varepsilon_k,w|\mathcal{J}|)}{x(\varepsilon_k)-x(z)}
  &=\frac{\frac{\lambda}{N}\sum_{k=1}^d r_k
    \mathcal{T}^{(g)}(I\|\varepsilon_k,w|\mathcal{J}|) \prod_{i\neq k}^d
    (x(z)-x(\varepsilon_i))}{\prod_{j=1}^d(x(z)-x(\varepsilon_j))}
\\
&=:\frac{f(x(z);w|I|\J)}{\prod_{j=1}^d (x(z)-x(\varepsilon_j))},
\end{align*}
 With one of the product representations of the 2-point function, Theorem \ref{Thm:InitialData}, application of
the interpolation formula (see \ref{lem:interpol}) with $L_w(x(z)):=\prod_{j=1}^d (x(z)-x(\hat{\tilde{w}}^j)$, yields :
\begin{align*}
\frac{f(x(z);w|I|\J)}{\prod_{j=1}^d (x(z)-x(\varepsilon_j))}&=\frac{ L_w(x(z))}{\prod_{j=1}^d (x(z)-x(\varepsilon_j))}\sum_{j=1}^d\frac{f(x(\hat{\tilde{w}}^j);w|I|\mathcal{J}|)}{(x(z)-x(\hat{\tilde{w}}^j) 
L_w'(x(\hat{\tilde{w}}^j)}
\\
&=-\lambda(y(w){-}y(z))\mathcal{G}^{(0)}(z,w) \sum_{j=1}^d\Res\displaylimits_{t\to \hat{\tilde{w}}^j}
\frac{f(x(t);w|I|\mathcal{J}|)x'(t)dt}{(x(z)-x(t)) L_w(x(t))},
\end{align*}
where the analyticity of $f(x(z);w|I|\J)$ at $z=\hat{\tilde{w}}^j$ was used.  
Next, insert
the rhs of Corollary \ref{cor2+} for $z\mapsto t$ near $t=\hat{\tilde{w}}^j$
at  which the first term of the lhs vanishes (here it is important
that the integrand has only a simple pole at
$t=\hat{\tilde{w}}^j$). Inserting it for $f(x(t);w|I|\J)$ leads to
\begin{align}
&-\frac{\lambda}{N}\sum_{k=1}^dr_k 
\frac{\mathcal{T}^{(g)}(I\|\varepsilon_k,w|)}{x(\varepsilon_k)-x(z)}
\label{eq:Tsum}
\\
 &=-\lambda(y(w){-}y(z))\mathcal{G}^{(0)}(z,w)
\sum_{j=1}^d \Res\displaylimits_{t\to \hat{\tilde{w}}^j}
\frac{x'(t)dt}{(x(z){-}x(t))(y(w){-}y(t))\mathcal{G}^{(0)}(t,w)}  
\nonumber
\\
&\times\bigg[ \sum_{\substack{I_1\uplus I_2=I\\
g_1+g_2=g,~(g_1,I_1)\neq (0,\emptyset)}} \hspace*{-2em}
\Omega^{(g_1)}_{|I_1|+1}(I_1,t) 
\mathcal{T}^{(g_2)}(I_2\|t,w|\mathcal{J}|)
\nonumber
\\
&+\T^{(g-1)}(I,t\|t,w|\mathcal{J}|)
+
\sum_{\substack{I_1\uplus I_2=I\\
\mathcal{J}_1\uplus \mathcal{J}_2=\mathcal{J},~
\mathcal{J}_1\neq \emptyset\\
g_1+g_2=g}}
\hspace*{-1.5em}
\mathcal{T}^{(g_1)}(I_1,t\|\mathcal{J}_1|) 
\mathcal{T}^{(g_2)}(I_2\|t,w|\mathcal{J}_2|)
\nonumber
\\
& 
+ 
\sum_{i=1}^m \frac{\partial}{\partial x(u_i)}
\Big(\frac{\mathcal{T}^{(g)}(I{\setminus} u_i\|u_i,w|\mathcal{J}|)}{
x(u_i)-x(t)}\Big)
\nonumber
\\
&- \sum_{s=2}^b\frac{
\mathcal{T}^{(g)}(I\|z^s,w^s,z^s,w|\mathcal{J}{\setminus} J^s|)
-
\mathcal{T}^{(g)}(I\|t,w^s,z^s,w|\mathcal{J}{\setminus} J^s|)
}{x(z^s)-x(t)}
  \nonumber
\end{align}
 
Next, compute for the same integrand the residue at $t= z$ (for 
arbitrary $z$):
\begin{align*}
&\lambda(y(w)-y(z))\mathcal{G}^{(0)}(z,w)\Res\displaylimits_{t\to z}\frac{x'(t)dt}{(x(z)-x(t))(y(w)-y(t))\mathcal{G}^{(0)}(t,w)}  \\
&\times\bigg[  \sum_{\substack{I_1\uplus I_2=I\\
g_1+g_2=g,~(g_1,I_1)\neq (0,\emptyset)}} \hspace*{-2em}
\Omega^{(g_1)}_{|I_1|+1}(I_1,t) 
\mathcal{T}^{(g_2)}(I_2\|z,w|\mathcal{J}|)
\nonumber
\\
&+\T^{(g-1)}(I,t\|t,w|\mathcal{J}|)
+
\sum_{\substack{I_1\uplus I_2=I\\
\mathcal{J}_1\uplus \mathcal{J}_2=\mathcal{J},~
\mathcal{J}_1\neq \emptyset\\
g_1+g_2=g}}
\hspace*{-1.5em}
\mathcal{T}^{(g_1)}(I_1,t\|\mathcal{J}_1|) 
\mathcal{T}^{(g_2)}(I_2\|t,w|\mathcal{J}_2|)
\nonumber
\\
& 
+ 
\sum_{i=1}^m \frac{\partial}{\partial x(u_i)}
\Big(\frac{\mathcal{T}^{(g)}(I{\setminus} u_i\|u_i,w|\mathcal{J}|)}{
x(u_i)-x(t)}\Big)
\nonumber
\\
&- \sum_{s=2}^b\frac{
\mathcal{T}^{(g)}(I\|z^s,w^s,z^s,w|\mathcal{J}{\setminus} J^s|)
-
\mathcal{T}^{(g)}(I\|t,w^s,z^s,w|\mathcal{J}{\setminus} J^s|)
}{x(z^s)-x(t)}
\nonumber
\\
&=(y(w)-y(z))\mathcal{T}^{(g)}(I\|z,w|\J|)
-\frac{\lambda}{N}\sum_{k=1}^d r_k 
\frac{\mathcal{T}^{(g)}(I\|\varepsilon_k,w|\J|)}{x(\varepsilon_k)-x(z)}\;.
\end{align*}
 Summing both expressions finishes the proof. \qed \\ \\ 
\textbf{Proof of Corollary \ref{cor2+}} \\ \\ 
First of all, we need to proof analyticity at  certain points:
\begin{lemma}\label{lem:analy}
 Let $\mathcal{J}=\{J^2,...,J^b\}$ for $J^s=[z^s,w^s]$
 and $I=\{u_1,...,u_m\}$.  The generalised correlation functions
 $\T^{(g)}(I\|z,w|\mathcal{J})$  
are analytic at $z=u_i$ and $z=z^s$.
\begin{proof}
In principal, this is clear when returning to the matrix base with a finite limit of coinciding indices. The result can be shown inductively in the Euler characteristic in  DSE \eqref{DSE-cT2}, with analogous considerations to \cite[Lemma 4.1]{Branahl:2020yru}.
%... rhs of the DSE \eqref{DSE-cT2} 
%is analytic .
%The only critical terms for $z\to u_i$ arise in combination
%\begin{align*}
%&\Omega^{(0)}_2(u_i,z)\mathcal{T}^{(g)}(I\backslash
 % u_i\|z,w|\mathcal{J})
%+\frac{\partial}{\partial R(u_i)}\Big(\frac{\T^{(g)}(I\backslash
  %u_i\|u_i,w|\mathcal{J}|)}{R(u_i)-R(z)}\Big)
%\\
%&= \Big(\Omega^{(0)}_2(u_i,z)-\frac{1}{(R(u_i)-R(z))^2}\Big)
%\mathcal{T}^{(g)}(I\backslash
 % u_i\|z,w|\mathcal{J})
%\\
%&+\frac{\partial}{\partial R(u_i)}\Big(\frac{\T^{(g)}(I\backslash
 % u_i\|u_i,w|\mathcal{J})-\T^{(g)}(I\backslash
  %u_i\|z,w|\mathcal{J})}{R(u_i)-R(z)}\Big)\;,
%\end{align*}
%which is analytic for $z\to u_i$. 
\end{proof}
\end{lemma}
The strategy of the proof consists of rewriting a term in Proposition \ref{prop2} in terms of all the others - but taking the residues at different points:
\begin{align*}
&\frac{\partial}{\partial x(u_i)}\sum_{j=1}^d \Res\displaylimits_{t\to z,\hat{\tilde{w}}^j}
\frac{x'(t)dt\prod_{k=1}^d(x(t)-x(\varepsilon_k)) }{(x(z)-x(t)) L_w(x(t))}
\frac{\mathcal{T}^{(g)}(I\backslash u_i\|u_i,w|\J)}{x(u_i)-x(t)}
\\
&=\frac{\partial}{\partial x(u_i)} \sum_{j=1}^d \Res\displaylimits_{v\to x(z),x(\hat{\tilde{w}}^j)}
\frac{dv\prod_{k=1}^d(v-x(\varepsilon_k)) }{(x(z)-v) L_w(v)}
\frac{\mathcal{T}^{(g)}(I\backslash u_i\|u_i,w|\J)}{x(u_i)-v}
\\
&=-\frac{\partial}{\partial x(u_i)}\Res\displaylimits_{v\to x(u_i)}
\frac{dv\prod_{k=1}^d(v-x(\varepsilon_k)) }{(x(z)-v) L_w(v)}
\frac{\mathcal{T}^{(g)}(I\backslash u_i\|u_i,w|\J)}{x(u_i)-v}
\\
&=\frac{1}{x'(u_i)}\frac{\partial}{\partial u_i}
\frac{\prod_{k=1}^d(x(u_i)-x(\varepsilon_k)) }{(x(z)-x(u_i)) 
L_w(x(u_i))}\mathcal{T}^{(g)}(I\backslash u_i\|u_i,w|\J)
\\
&=\Res\displaylimits_{t\to u_i}\frac{x'(t)dt
\prod_{k=1}^d(x(t)-x(\varepsilon_k)) }{(x(z)-x(t))
L_w(x(t))}\Big\{\frac{1}{x'(u_i)x'(t)(t-u_i)^2 }
\mathcal{T}^{(g)}(I\backslash u_i\|t,w|\J)\Big\}\;,
\end{align*}
where we substituted $t\mapsto v=x(t)$, then moved the integration contour 
and finally represented the result in form of a residue formula.
$\Omega^{(0)}_2(z,w)$ is the only correlation function divergent 
on the diagonal so that the terms in $\{~ \}$ extend 
to $\sum_{I_1,I_2,g_1,g_2}
\Omega^{(g_1)}_{|I_1|+1}(I_1;t)\mathcal{T}^{(g_2)}(I_2\|t,w|\J)$ and finally to 
\begin{align}
\label{integrand}
&\sum_{\substack{I_1\uplus I_2=I\\
g_1+g_2=g,~(g_1,I_1)\neq (0,\emptyset)}} \hspace*{-2em}
\Omega^{(g_1)}_{|I_1|+1}(I_1,t) 
\mathcal{T}^{(g_2)}(I_2\|t,w|\mathcal{J}|)
\nonumber
\\
&+\T^{(g-1)}(I,t\|t,w|\mathcal{J}|)
+
\sum_{\substack{I_1\uplus I_2=I\\
\mathcal{J}_1\uplus \mathcal{J}_2=\mathcal{J},~
\mathcal{J}_1\neq \emptyset\\
g_1+g_2=g}}
\hspace*{-1.5em}
\mathcal{T}^{(g_1)}(I_1,t\|\mathcal{J}_1|) 
\mathcal{T}^{(g_2)}(I_2\|t,w|\mathcal{J}_2|)
\nonumber
\\
&+ \sum_{s=2}^b\frac{
\mathcal{T}^{(g)}(I\|t,w^s,z^s,w|\mathcal{J}{\setminus} J^s|)
}{x(z^s)-x(t)}
\end{align}
 where the regularity at $z=u_i$ was exploited - under the residue operation, we added an effective zero. In the same manner, we rewrite the term
\begin{align*}
&\sum_{j=1}^d \Res\displaylimits_{t\to z,\hat{\tilde{w}}^j}
\frac{x'(t)dt\prod_{k=1}^d(x(t)-x(\varepsilon_k)) }{(x(z)-x(t)) L_w(x(t))}
\frac{\mathcal{T}^{(g)}(I\|z^s,w^s,z^s,w|\mathcal{J}{\setminus} J^s|)}{x(z^s)-x(t)}
\end{align*}
for $s \in \{2,...,b\}$ into a residue formula where we take the residue at $t=z^s$ by deforming the contour. Since we have just a pole at order one at $t=z^s$, we set an in the integrand the function 
\begin{align*}
	\mathcal{T}^{(g)}(I\|z^s,w^s,z^s,w|\mathcal{J}{\setminus} J^s|)\quad \to\quad \mathcal{T}^{(g)}(I\|t,w^s,z^s,w|\mathcal{J}{\setminus} J^s|).
\end{align*}
 Again, the regularity argument of the all the other terms for the integrand in eq. (\ref{integrand}) at $t=z^s$ allows for adding another zero.  \qed \\ \\ 
\textbf{Proof of Theorem \ref{thm:complete}} \\ \\
  Assume $p^i_j$ and $q^i_j$ such that all $E_{p^i_j}$ and $\tilde{E}_{q^i_j}$ are pairwise different. Set $a=p^1_1$, $d=q^1_1$ and $c=q^1_{N_1}$ to have a clear distinction between these and the remaining $p^i_j$ and $q^i_j$. Again, we define a differential operator $\hat D_{dc}$:
  \begin{align*}
   \hat{D}_{dc}=\frac{\partial^{2N_1+..+2N_b-2}}{\partial  J^\dagger_{dp^1_2}\partial J_{p^1_2q^1_2}..
   \partial J_{p^1_{N_1}c} \partial J_{p^2_1q^2_1}\partial J^\dagger_{q^2_1p^2_2}..
   \partial J^\dagger_{q^2_{N_2}p^2_1} ..\partial J_{p^b_1q^b_1}..
   \partial J^\dagger_{q^b_{N_b}p^b_1}}.
  \end{align*}
Bringing the global denominator of the theorem to the other side yields by definition of the correlation function
\begin{align*}
 &(\tilde{E}_{d}-\tilde{E}_{c})G_{|p_1^1 q^1_1..q^1_{N_1}|\mathcal{J}|}\\
 =&(\tilde{E}_{d}-\tilde{E}_{c})N^{b-2} \hat{D}_{dc}\frac{\partial^2}{\partial J^\dagger_{ca}\partial J_{ad}}
 \log \Z(J,J^\dagger)\big\vert_{J,J^\dagger=0}\\
 =&N^{b-2} \hat{D}_{dc}\frac{\partial^2}{\partial J^\dagger_{ca}\partial J_{ad}}[E_a+\tilde{E}_{d}-(E_a+\tilde{E}_{c})]\log \Z(J,J^\dagger)\big\vert_{J,J^\dagger=0}
\end{align*}
We generalise the steps in Example \ref{ex:2p} to the general operator $\hat D_{dc}$:
\begin{align*}
 &C \hat{D}_{dc}N^{b-1}\bigg (\frac{\partial}{\partial J^\dagger_{ca}} \frac{\exp(-N S_{int}(\frac{1}{N}\partial_J,\frac{1}{N}\partial_{J^\dagger}))J_{da}}
 {\Z(J,J^\dagger)}\\
&\qquad \quad - \frac{\partial}{\partial J_{ad}} \frac{\exp(-N S_{int}(\frac{1}{N}\partial_J,\frac{1}{N}\partial_{J^\dagger}))J^\dagger_{ac}}{\Z(J,J^\dagger)}\bigg)\Z_{free}(J,J^\dagger)\big\vert_{J,J^\dagger=0}\\
 =&-N^{b-4}\lambda \hat{D}_{dc}\sum_{n,m=1}^N\left(\frac{\partial}{\partial J^\dagger_{ca}}
 \frac{\frac{\partial^3}{\partial J_{an}\partial J^\dagger_{nm}\partial J_{md}}}{\Z(J,J^\dagger)}-\frac{\partial}{\partial J_{ad}}
\frac{ \frac{\partial^3}{\partial J^\dagger_{cm}\partial J_{mn}\partial J^\dagger_{na}}}{\Z(J,J^\dagger)}\right)\Z(J,J^\dagger)\big\vert_{J,J^\dagger=0},
\end{align*}
For $E_m=E_a$, the bracket vanishes for regular and non-regular terms. This comes from the fact that $\frac{\partial}{\partial J^\dagger_{ca}}$ and 
$\frac{\partial}{\partial J_{ad}}$ do not act on $\frac{1}{\Z(J,J^\dagger)}$ because it gives 0 after taking $J=0$ (no cycle in $a$), then setting $m=a$ yields the same four derivatives.

Therefore, we can assume $E_m\neq E_a$ and apply the simpler Ward identity of Proposition \ref{prop:ward}.
\begin{align*}
 =&-\lambda \hat{D}_{dc}N^{b-3}\sum_{\substack{n,m \\ m\neq a} }\bigg(\frac{\partial}{\partial J^\dagger_{ca}}
 \frac{\frac{\partial}{\partial J_{md}}}{\Z(J,J^\dagger)(E_m-E_a)}
 \big(J_{mn}\frac{\partial}{\partial J_{an}}-
 J^\dagger_{na}\frac{\partial}{\partial J^\dagger_{nm}}\big)\\
 &\qquad +\frac{\partial}{\partial J_{ad}}\frac{
 \frac{\partial}{\partial J^\dagger_{cm}}}{\Z(J,J^\dagger)(E_m-E_a)}\big(J_{an}\frac{\partial}{\partial J_{mn}}-
 J^\dagger_{nm}\frac{\partial}{\partial J^\dagger_{na}}\big)\bigg)\Z(J,J^\dagger)\big\vert_{J,J^\dagger=0}.
\end{align*}
The following derivatives lead to cancellations:
\begin{itemize}
\item in the first line $\frac{\partial}{\partial J^\dagger_{ca}}$ on $J^\dagger_{na}$ is cancelled by $\frac{\partial}{\partial J_{ad}}$ on $J_{an}$ in the second line 
\item if
$\frac{\partial}{\partial J_{md}}$ acts on $J_{mn}$ is cancelled by
$\frac{\partial}{\partial J^\dagger_{cm}}$ on $J^\dagger_{nm}$ in the second line
\end{itemize}
We end up with the surviving terms
\begin{align*}
&-\lambda \hat{D}_{dc}N^{b-3}\sum_{\substack{n,m \\ m\neq a} }\frac{1}{E_m-E_a}
 \biggl(\frac{\partial}{\partial J^\dagger_{ca}}J_{mn}\frac{
 \frac{\partial^2}{\partial J_{an}\partial J_{md}}}{\Z(J,J^\dagger)}\\
&
\qquad\qquad\qquad\qquad\qquad\qquad -\frac{\partial}{\partial J_{ad}}J^\dagger_{nm}\frac{
 \frac{\partial^2}{\partial J^\dagger_{cm}\partial J^\dagger_{na}}}{\Z(J,J^\dagger)}\biggl)\Z(J,J^\dagger)\big\vert_{J,J^\dagger=0}.
\end{align*}
We manipulate this equation using the general identity
 \begin{align*}
\frac{\partial_x \partial_yf(x,y)}{f(x,y)}=\partial_x\partial_y\log[f(x,y)]+\partial_x\log[f(x,y)] \partial_y\log[f(x,y)]
\end{align*}
 yields
\begin{align*}
 &-\lambda \hat{D}_{dc}N^{b-3}\sum_{\substack{n,m \\ m\neq a} }\frac{1}{E_m-E_a}
 \bigg\{\frac{\partial}{\partial J^\dagger_{ca}}J_{mn}
 \frac{\partial^2}{\partial J_{an}\partial J_{md}}\log[\Z(J,J^\dagger)]\\
& \qquad\qquad\qquad\qquad\qquad\qquad-\frac{\partial}{\partial J_{ad}}J^\dagger_{nm}
 \frac{\partial^2}{\partial J^\dagger_{cm}\partial J^\dagger_{na}}\log[\Z(J,J^\dagger)]\\
 &\qquad\qquad\qquad\qquad\qquad\qquad +
 \frac{\partial}{\partial J^\dagger_{ca}}J_{mn}
 \bigg(\frac{\partial}{\partial J_{an}}\log[\Z(J,J^\dagger)]\bigg)\bigg(\frac{\partial}{\partial J_{md}}\log[\Z(J,J^\dagger)]\bigg)\\
& \qquad\qquad\qquad\qquad\qquad\qquad-\frac{\partial}{\partial J_{ad}}J^\dagger_{nm}
 \bigg(\frac{\partial}{\partial J^\dagger_{cm}}\log[\Z(J,J^\dagger)]\bigg)
 \bigg(\frac{\partial}{\partial J^\dagger_{na}}\log[\Z(J,J^\dagger)]\bigg)\bigg\}\bigg\vert_{J,J^\dagger=0}.
\end{align*}
The $n,m$ are fixed by a derivative acting on $J_{mn}$ (or $J^\dagger_{nm}$). We collect the following terms:
\begin{itemize}
  \item either a derivative of the form $\frac{\partial}{\partial J^\dagger_{q^1_k p^1_{k}}}$ and $\frac{\partial}{\partial J_{p^1_k q^1_{k}}}$ , respectively, fixes the $n,m$ in the first two lines, which produces 
  separated cycles
\item  or a derivative of the form $\frac{\partial}{\partial J^\dagger_{q^\beta_k p^\beta_{k}}}$ and $\frac{\partial}{\partial J_{p^\beta_k q^\beta_{k}}}$ , respectively, with 
  $\beta>1$ fixes the $n,m$, which merges the first cycle with the $\beta^{\text{th}}$-cycle. 
\item in the last two lines it is only possible that 
  a derivative of the form$\frac{\partial}{\partial J^\dagger_{q^1_k p^1_{k}}}$ and $\frac{\partial}{\partial J_{p^1_k q^1_{k}}}$ , respectively, fixes the $n,m$, otherwise setting $J=0$ 
 leads to vanishing contributions. Acting with the remaining derivatives of $\hat{D}_{dc}$ on the 
  product of the logarithms by considering 
  the Leibniz rule leads to the assertion for pairwise different $E_{p^i_j}$. 
\end{itemize}  
  The expression stays true for coinciding $E_{p^i_j}$ since the lhs is regular which induces a well-defined limit of the rhs
  by continuation to differentiable functions. A genus expansion and application of the boundary creation $-N\frac{\partial}{\partial E_{p_i}}$ operator yields the assertion. \qed
 
\bibliographystyle{halpha-abbrv}
\bibliography{cqkm_bib}
 
\end{document}